\documentclass[twocolumn,prd,amsmath,amssymb,nofootinbib,superscriptaddress,floatfix]{revtex4-2}
\usepackage{graphicx}
\usepackage{booktabs}
\usepackage{xcolor}
\usepackage[colorlinks=true,allcolors=blue]{hyperref}

\newcommand{\pvec}{\Lambda} \newcommand{\svec}{\lambda}  \newcommand{\comp}{\boldsymbol{\rho}}
\newcommand{\svecz}{\boldsymbol{\lambda}} \newcommand{\pvecz}{\boldsymbol{\Lambda}}

\newcommand{\Msun}{M_\odot}
\newcommand{\chieff}{\chi_{\rm eff}}
\newcommand{\mc}{\mathcal{M}}
\newcommand{\Rapster}{\textsc{Rapster}}
\newcommand{\gwkokab}{\textsc{gwkokab}}
\newcommand{\fgc}{f_{\rm GC}}
\newcommand{\hatfgc}{\hat{f}_{\rm GC}}
\newcommand{\Nexp}{N_{\rm exp}}

\providecommand{\apj}{ApJ}\providecommand{\apjl}{ApJL}\providecommand{\apjs}{ApJS}
\providecommand{\mnras}{MNRAS}\providecommand{\araa}{ARA\&A}\providecommand{\aap}{A\&A}
\providecommand{\prd}{Phys.~Rev.~D}\providecommand{\prl}{Phys.~Rev.~Lett.}
\providecommand{\prx}{Phys.~Rev.~X}

\providecommand{\aj}{AJ}\providecommand{\aapr}{A\&A~Rev.}

\providecommand\result[1]{  \ifcsname #1\endcsname\csname #1\endcsname\else\textbf{??#1??}\fi}

\expandafter\providecommand\csname ChiIsoChiMax\endcsname{}\expandafter\renewcommand\csname ChiIsoChiMax\endcsname{\ensuremath{+0.50}}
\expandafter\providecommand\csname ChiIsoChiMin\endcsname{}\expandafter\renewcommand\csname ChiIsoChiMin\endcsname{\ensuremath{-0.39}}
\expandafter\providecommand\csname ChiIsoChiMinSig\endcsname{}\expandafter\renewcommand\csname ChiIsoChiMinSig\endcsname{\ensuremath{0.11}}
\expandafter\providecommand\csname ChiIsoLoudChi\endcsname{}\expandafter\renewcommand\csname ChiIsoLoudChi\endcsname{\ensuremath{+0.50}}
\expandafter\providecommand\csname ChiIsoLoudSig\endcsname{}\expandafter\renewcommand\csname ChiIsoLoudSig\endcsname{\ensuremath{0.036}}
\expandafter\providecommand\csname ChiIsoMu\endcsname{}\expandafter\renewcommand\csname ChiIsoMu\endcsname{\ensuremath{+0.09}}
\expandafter\providecommand\csname ChiIsoMuErr\endcsname{}\expandafter\renewcommand\csname ChiIsoMuErr\endcsname{\ensuremath{0.04}}
\expandafter\providecommand\csname ChiIsoN\endcsname{}\expandafter\renewcommand\csname ChiIsoN\endcsname{\ensuremath{34}}
\expandafter\providecommand\csname ChiIsoNpos\endcsname{}\expandafter\renewcommand\csname ChiIsoNpos\endcsname{\ensuremath{22}}
\expandafter\providecommand\csname ChiIsoNsigNeg\endcsname{}\expandafter\renewcommand\csname ChiIsoNsigNeg\endcsname{\ensuremath{1}}
\expandafter\providecommand\csname ChiIsoNsigPos\endcsname{}\expandafter\renewcommand\csname ChiIsoNsigPos\endcsname{\ensuremath{8}}
\expandafter\providecommand\csname ChiIsoP\endcsname{}\expandafter\renewcommand\csname ChiIsoP\endcsname{\ensuremath{5\times10^{-56}}}
\expandafter\providecommand\csname ChiIsoPiso\endcsname{}\expandafter\renewcommand\csname ChiIsoPiso\endcsname{\ensuremath{0.02}}
\expandafter\providecommand\csname ChiIsoSigmaInt\endcsname{}\expandafter\renewcommand\csname ChiIsoSigmaInt\endcsname{\ensuremath{0.17}}
\expandafter\providecommand\csname ChiIsoSigmaIntIso\endcsname{}\expandafter\renewcommand\csname ChiIsoSigmaIntIso\endcsname{\ensuremath{0.20}}
\expandafter\providecommand\csname ChiIsoSigmaIntSumm\endcsname{}\expandafter\renewcommand\csname ChiIsoSigmaIntSumm\endcsname{\ensuremath{0.16}}
\expandafter\providecommand\csname ChiIsoT\endcsname{}\expandafter\renewcommand\csname ChiIsoT\endcsname{\ensuremath{359}}
\expandafter\providecommand\csname ClusterFracTotal\endcsname{}\expandafter\renewcommand\csname ClusterFracTotal\endcsname{\ensuremath{36\%}}
\expandafter\providecommand\csname ClusterFracTotalHi\endcsname{}\expandafter\renewcommand\csname ClusterFracTotalHi\endcsname{\ensuremath{44}}
\expandafter\providecommand\csname ClusterFracTotalLo\endcsname{}\expandafter\renewcommand\csname ClusterFracTotalLo\endcsname{\ensuremath{26}}
\expandafter\providecommand\csname ClusterReachChi\endcsname{}\expandafter\renewcommand\csname ClusterReachChi\endcsname{\ensuremath{0.67}}
\expandafter\providecommand\csname DelayMedianMyr\endcsname{}\expandafter\renewcommand\csname DelayMedianMyr\endcsname{\ensuremath{10\,{\rm Myr}}}
\expandafter\providecommand\csname DenseEndMonotonic\endcsname{}\expandafter\renewcommand\csname DenseEndMonotonic\endcsname{\ensuremath{0}}
\expandafter\providecommand\csname DenseEndSpread\endcsname{}\expandafter\renewcommand\csname DenseEndSpread\endcsname{\ensuremath{15}}
\expandafter\providecommand\csname FiducialRh\endcsname{}\expandafter\renewcommand\csname FiducialRh\endcsname{\ensuremath{0.25}}
\expandafter\providecommand\csname FiducialS\endcsname{}\expandafter\renewcommand\csname FiducialS\endcsname{\ensuremath{0}}
\expandafter\providecommand\csname FieldDetFrac\endcsname{}\expandafter\renewcommand\csname FieldDetFrac\endcsname{\ensuremath{0.20}}
\expandafter\providecommand\csname FieldDetFracHi\endcsname{}\expandafter\renewcommand\csname FieldDetFracHi\endcsname{\ensuremath{0.22}}
\expandafter\providecommand\csname FieldDetFracLo\endcsname{}\expandafter\renewcommand\csname FieldDetFracLo\endcsname{\ensuremath{0.16}}
\expandafter\providecommand\csname FieldFedBandExcess\endcsname{}\expandafter\renewcommand\csname FieldFedBandExcess\endcsname{\ensuremath{-11\%}}
\expandafter\providecommand\csname FieldFedDlnL\endcsname{}\expandafter\renewcommand\csname FieldFedDlnL\endcsname{\ensuremath{3}}
\expandafter\providecommand\csname FieldFedDlnLBound\endcsname{}\expandafter\renewcommand\csname FieldFedDlnLBound\endcsname{\ensuremath{32}}
\expandafter\providecommand\csname FieldFedDlnLHi\endcsname{}\expandafter\renewcommand\csname FieldFedDlnLHi\endcsname{\ensuremath{13}}
\expandafter\providecommand\csname FieldFedDlnLLo\endcsname{}\expandafter\renewcommand\csname FieldFedDlnLLo\endcsname{\ensuremath{0}}
\expandafter\providecommand\csname FieldFedDlnLSmallNHi\endcsname{}\expandafter\renewcommand\csname FieldFedDlnLSmallNHi\endcsname{\ensuremath{1}}
\expandafter\providecommand\csname FieldFedDlnLSmallNLo\endcsname{}\expandafter\renewcommand\csname FieldFedDlnLSmallNLo\endcsname{\ensuremath{1}}
\expandafter\providecommand\csname FieldFedDriverAlphaChi\endcsname{}\expandafter\renewcommand\csname FieldFedDriverAlphaChi\endcsname{\ensuremath{0.53}}
\expandafter\providecommand\csname FieldFedDriverAlphaDlnL\endcsname{}\expandafter\renewcommand\csname FieldFedDriverAlphaDlnL\endcsname{\ensuremath{2.5}}
\expandafter\providecommand\csname FieldFedDriverAlphaMone\endcsname{}\expandafter\renewcommand\csname FieldFedDriverAlphaMone\endcsname{\ensuremath{24}}
\expandafter\providecommand\csname FieldFedDriverAlphaQ\endcsname{}\expandafter\renewcommand\csname FieldFedDriverAlphaQ\endcsname{\ensuremath{0.57}}
\expandafter\providecommand\csname FieldFedDriverBetaChi\endcsname{}\expandafter\renewcommand\csname FieldFedDriverBetaChi\endcsname{\ensuremath{0.42}}
\expandafter\providecommand\csname FieldFedDriverBetaDlnL\endcsname{}\expandafter\renewcommand\csname FieldFedDriverBetaDlnL\endcsname{\ensuremath{1.2}}
\expandafter\providecommand\csname FieldFedDriverBetaMone\endcsname{}\expandafter\renewcommand\csname FieldFedDriverBetaMone\endcsname{\ensuremath{25}}
\expandafter\providecommand\csname FieldFedDriverBetaQ\endcsname{}\expandafter\renewcommand\csname FieldFedDriverBetaQ\endcsname{\ensuremath{0.41}}
\expandafter\providecommand\csname FieldFedEpsilon\endcsname{}\expandafter\renewcommand\csname FieldFedEpsilon\endcsname{\ensuremath{7\%}}
\expandafter\providecommand\csname FieldFedEssLowBound\endcsname{}\expandafter\renewcommand\csname FieldFedEssLowBound\endcsname{\ensuremath{18}}
\expandafter\providecommand\csname FieldFedFieldR\endcsname{}\expandafter\renewcommand\csname FieldFedFieldR\endcsname{\ensuremath{15}}
\expandafter\providecommand\csname FieldFedNtwoG\endcsname{}\expandafter\renewcommand\csname FieldFedNtwoG\endcsname{\ensuremath{7}}
\expandafter\providecommand\csname FieldFedNtwoGSamp\endcsname{}\expandafter\renewcommand\csname FieldFedNtwoGSamp\endcsname{\ensuremath{10}}
\expandafter\providecommand\csname FieldFedPosSum\endcsname{}\expandafter\renewcommand\csname FieldFedPosSum\endcsname{\ensuremath{7.4}}
\expandafter\providecommand\csname FieldFedRateFrac\endcsname{}\expandafter\renewcommand\csname FieldFedRateFrac\endcsname{\ensuremath{3\%}}
\expandafter\providecommand\csname FieldFedTwoGPeak\endcsname{}\expandafter\renewcommand\csname FieldFedTwoGPeak\endcsname{\ensuremath{0.06}}
\expandafter\providecommand\csname FieldFedTwoGPeakMone\endcsname{}\expandafter\renewcommand\csname FieldFedTwoGPeakMone\endcsname{\ensuremath{20\,\Msun}}
\expandafter\providecommand\csname FieldFedTwoGR\endcsname{}\expandafter\renewcommand\csname FieldFedTwoGR\endcsname{\ensuremath{0.5}}
\expandafter\providecommand\csname FieldFloorDecades\endcsname{}\expandafter\renewcommand\csname FieldFloorDecades\endcsname{\ensuremath{10}}
\expandafter\providecommand\csname FieldFloorFracShift\endcsname{}\expandafter\renewcommand\csname FieldFloorFracShift\endcsname{\ensuremath{0.006}}
\expandafter\providecommand\csname FieldGridBestRh\endcsname{}\expandafter\renewcommand\csname FieldGridBestRh\endcsname{\ensuremath{0.10}}
\expandafter\providecommand\csname FieldGridDenserDlnL\endcsname{}\expandafter\renewcommand\csname FieldGridDenserDlnL\endcsname{\ensuremath{32}}
\expandafter\providecommand\csname FieldGridDenserRh\endcsname{}\expandafter\renewcommand\csname FieldGridDenserRh\endcsname{\ensuremath{0.07}}
\expandafter\providecommand\csname FieldGridDensestDlnL\endcsname{}\expandafter\renewcommand\csname FieldGridDensestDlnL\endcsname{\ensuremath{17}}
\expandafter\providecommand\csname FieldGridDiffuserDlnL\endcsname{}\expandafter\renewcommand\csname FieldGridDiffuserDlnL\endcsname{\ensuremath{26}}
\expandafter\providecommand\csname FieldGridFlatDlnL\endcsname{}\expandafter\renewcommand\csname FieldGridFlatDlnL\endcsname{\ensuremath{0.0}}
\expandafter\providecommand\csname FieldGridInterior\endcsname{}\expandafter\renewcommand\csname FieldGridInterior\endcsname{\ensuremath{1}}
\expandafter\providecommand\csname FieldGridRhFlatHi\endcsname{}\expandafter\renewcommand\csname FieldGridRhFlatHi\endcsname{\ensuremath{0.05}}
\expandafter\providecommand\csname FieldGridRhHi\endcsname{}\expandafter\renewcommand\csname FieldGridRhHi\endcsname{\ensuremath{0.43}}
\expandafter\providecommand\csname FieldGridRhLo\endcsname{}\expandafter\renewcommand\csname FieldGridRhLo\endcsname{\ensuremath{0.05}}
\expandafter\providecommand\csname FieldGridRhSpan\endcsname{}\expandafter\renewcommand\csname FieldGridRhSpan\endcsname{\ensuremath{42}}
\expandafter\providecommand\csname FieldGridRhSpanNoField\endcsname{}\expandafter\renewcommand\csname FieldGridRhSpanNoField\endcsname{\ensuremath{182}}
\expandafter\providecommand\csname FieldRefChi\endcsname{}\expandafter\renewcommand\csname FieldRefChi\endcsname{\ensuremath{0.075}}
\expandafter\providecommand\csname FieldRefChiSig\endcsname{}\expandafter\renewcommand\csname FieldRefChiSig\endcsname{\ensuremath{0.05}}
\expandafter\providecommand\csname FieldRefMone\endcsname{}\expandafter\renewcommand\csname FieldRefMone\endcsname{\ensuremath{11.0}}
\expandafter\providecommand\csname FieldRefMoneSig\endcsname{}\expandafter\renewcommand\csname FieldRefMoneSig\endcsname{\ensuremath{1.0}}
\expandafter\providecommand\csname FieldSplitDlnL\endcsname{}\expandafter\renewcommand\csname FieldSplitDlnL\endcsname{\ensuremath{99}}
\expandafter\providecommand\csname FieldSplitFgc\endcsname{}\expandafter\renewcommand\csname FieldSplitFgc\endcsname{\ensuremath{0.39\%}}
\expandafter\providecommand\csname FieldSplitFrac\endcsname{}\expandafter\renewcommand\csname FieldSplitFrac\endcsname{\ensuremath{0.19}}
\expandafter\providecommand\csname FieldSplitFracErr\endcsname{}\expandafter\renewcommand\csname FieldSplitFracErr\endcsname{\ensuremath{0.03}}
\expandafter\providecommand\csname FieldSplitFracLowMass\endcsname{}\expandafter\renewcommand\csname FieldSplitFracLowMass\endcsname{\ensuremath{62}}
\expandafter\providecommand\csname FieldTunedChi\endcsname{}\expandafter\renewcommand\csname FieldTunedChi\endcsname{\ensuremath{0.050}}
\expandafter\providecommand\csname FieldTunedChiSig\endcsname{}\expandafter\renewcommand\csname FieldTunedChiSig\endcsname{\ensuremath{0.05}}
\expandafter\providecommand\csname FieldTunedDlnL\endcsname{}\expandafter\renewcommand\csname FieldTunedDlnL\endcsname{\ensuremath{112}}
\expandafter\providecommand\csname FieldTunedMone\endcsname{}\expandafter\renewcommand\csname FieldTunedMone\endcsname{\ensuremath{9.5}}
\expandafter\providecommand\csname FieldTunedMoneSig\endcsname{}\expandafter\renewcommand\csname FieldTunedMoneSig\endcsname{\ensuremath{1.0}}
\expandafter\providecommand\csname FitNcluster\endcsname{}\expandafter\renewcommand\csname FitNcluster\endcsname{\ensuremath{136}}
\expandafter\providecommand\csname FitNfield\endcsname{}\expandafter\renewcommand\csname FitNfield\endcsname{\ensuremath{34}}
\expandafter\providecommand\csname FitNreproc\endcsname{}\expandafter\renewcommand\csname FitNreproc\endcsname{\ensuremath{9}}
\expandafter\providecommand\csname FitSigchiThree\endcsname{}\expandafter\renewcommand\csname FitSigchiThree\endcsname{\ensuremath{0.14}}
\expandafter\providecommand\csname FitSigchiTwo\endcsname{}\expandafter\renewcommand\csname FitSigchiTwo\endcsname{\ensuremath{0.07}}
\expandafter\providecommand\csname FollowupRatio\endcsname{}\expandafter\renewcommand\csname FollowupRatio\endcsname{\ensuremath{0.88}}
\expandafter\providecommand\csname FracLVK\endcsname{}\expandafter\renewcommand\csname FracLVK\endcsname{\ensuremath{41\%}}
\expandafter\providecommand\csname FracLVKHi\endcsname{}\expandafter\renewcommand\csname FracLVKHi\endcsname{\ensuremath{73}}
\expandafter\providecommand\csname FracLVKLo\endcsname{}\expandafter\renewcommand\csname FracLVKLo\endcsname{\ensuremath{29}}
\expandafter\providecommand\csname GapChiMin\endcsname{}\expandafter\renewcommand\csname GapChiMin\endcsname{\ensuremath{0.30}}
\expandafter\providecommand\csname GapFieldSigChi\endcsname{}\expandafter\renewcommand\csname GapFieldSigChi\endcsname{\ensuremath{5}}
\expandafter\providecommand\csname GapFieldSigMone\endcsname{}\expandafter\renewcommand\csname GapFieldSigMone\endcsname{\ensuremath{10}}
\expandafter\providecommand\csname GapMoneHi\endcsname{}\expandafter\renewcommand\csname GapMoneHi\endcsname{\ensuremath{30}}
\expandafter\providecommand\csname GapMoneLo\endcsname{}\expandafter\renewcommand\csname GapMoneLo\endcsname{\ensuremath{15}}
\expandafter\providecommand\csname GapN\endcsname{}\expandafter\renewcommand\csname GapN\endcsname{\ensuremath{4}}
\expandafter\providecommand\csname GapNslice\endcsname{}\expandafter\renewcommand\csname GapNslice\endcsname{\ensuremath{32}}
\expandafter\providecommand\csname GapShapeDecadesMax\endcsname{}\expandafter\renewcommand\csname GapShapeDecadesMax\endcsname{\ensuremath{21}}
\expandafter\providecommand\csname GapShapeDecadesMin\endcsname{}\expandafter\renewcommand\csname GapShapeDecadesMin\endcsname{\ensuremath{3}}
\expandafter\providecommand\csname HatFgc\endcsname{}\expandafter\renewcommand\csname HatFgc\endcsname{\ensuremath{0.5\%}}
\expandafter\providecommand\csname HatFgcHi\endcsname{}\expandafter\renewcommand\csname HatFgcHi\endcsname{\ensuremath{1.1}}
\expandafter\providecommand\csname HatFgcLo\endcsname{}\expandafter\renewcommand\csname HatFgcLo\endcsname{\ensuremath{0.2}}
\expandafter\providecommand\csname HierFracDet\endcsname{}\expandafter\renewcommand\csname HierFracDet\endcsname{\ensuremath{0.24}}
\expandafter\providecommand\csname HierFracDetHi\endcsname{}\expandafter\renewcommand\csname HierFracDetHi\endcsname{\ensuremath{0.36}}
\expandafter\providecommand\csname HierFracDetLo\endcsname{}\expandafter\renewcommand\csname HierFracDetLo\endcsname{\ensuremath{0.11}}
\expandafter\providecommand\csname HierFracIntr\endcsname{}\expandafter\renewcommand\csname HierFracIntr\endcsname{\ensuremath{0.11}}
\expandafter\providecommand\csname HierFracIntrHi\endcsname{}\expandafter\renewcommand\csname HierFracIntrHi\endcsname{\ensuremath{0.13}}
\expandafter\providecommand\csname HierFracIntrLo\endcsname{}\expandafter\renewcommand\csname HierFracIntrLo\endcsname{\ensuremath{0.05}}
\expandafter\providecommand\csname HighMassRate\endcsname{}\expandafter\renewcommand\csname HighMassRate\endcsname{\ensuremath{2.1^{+2.6}_{-0.9}}}
\expandafter\providecommand\csname JointBestRh\endcsname{}\expandafter\renewcommand\csname JointBestRh\endcsname{\ensuremath{0.15\,{\rm pc}}}
\expandafter\providecommand\csname JointBestS\endcsname{}\expandafter\renewcommand\csname JointBestS\endcsname{\ensuremath{0.2}}
\expandafter\providecommand\csname JsDelayMax\endcsname{}\expandafter\renewcommand\csname JsDelayMax\endcsname{\ensuremath{0.04\,{\rm bits}}}
\expandafter\providecommand\csname LvkTotalRate\endcsname{}\expandafter\renewcommand\csname LvkTotalRate\endcsname{\ensuremath{24}}
\expandafter\providecommand\csname MdfQuadDev\endcsname{}\expandafter\renewcommand\csname MdfQuadDev\endcsname{\ensuremath{0.9\%}}
\expandafter\providecommand\csname MdfQuadDlnL\endcsname{}\expandafter\renewcommand\csname MdfQuadDlnL\endcsname{\ensuremath{0.1}}
\expandafter\providecommand\csname NatalMdfDlnL\endcsname{}\expandafter\renewcommand\csname NatalMdfDlnL\endcsname{\ensuremath{21}}
\expandafter\providecommand\csname NatalMdfDlnLOFour\endcsname{}\expandafter\renewcommand\csname NatalMdfDlnLOFour\endcsname{\ensuremath{18}}
\expandafter\providecommand\csname NatalSpinDlnL\endcsname{}\expandafter\renewcommand\csname NatalSpinDlnL\endcsname{\ensuremath{20}}
\expandafter\providecommand\csname NatalSpinSeedSd\endcsname{}\expandafter\renewcommand\csname NatalSpinSeedSd\endcsname{\ensuremath{18}}
\expandafter\providecommand\csname NatalSpinSeeds\endcsname{}\expandafter\renewcommand\csname NatalSpinSeeds\endcsname{\ensuremath{9}}
\expandafter\providecommand\csname Nclu\endcsname{}\expandafter\renewcommand\csname Nclu\endcsname{\ensuremath{144}}
\expandafter\providecommand\csname NcluHi\endcsname{}\expandafter\renewcommand\csname NcluHi\endcsname{\ensuremath{150}}
\expandafter\providecommand\csname NcluLo\endcsname{}\expandafter\renewcommand\csname NcluLo\endcsname{\ensuremath{140}}
\expandafter\providecommand\csname NevScored\endcsname{}\expandafter\renewcommand\csname NevScored\endcsname{\ensuremath{179}}
\expandafter\providecommand\csname Nfld\endcsname{}\expandafter\renewcommand\csname Nfld\endcsname{\ensuremath{35}}
\expandafter\providecommand\csname NfldHi\endcsname{}\expandafter\renewcommand\csname NfldHi\endcsname{\ensuremath{39}}
\expandafter\providecommand\csname NfldLo\endcsname{}\expandafter\renewcommand\csname NfldLo\endcsname{\ensuremath{29}}
\expandafter\providecommand\csname NfollowupValid\endcsname{}\expandafter\renewcommand\csname NfollowupValid\endcsname{\ensuremath{22}}
\expandafter\providecommand\csname NodeSdPeak\endcsname{}\expandafter\renewcommand\csname NodeSdPeak\endcsname{\ensuremath{14}}
\expandafter\providecommand\csname NodeSdTypical\endcsname{}\expandafter\renewcommand\csname NodeSdTypical\endcsname{\ensuremath{3.6}}
\expandafter\providecommand\csname NodeStableRh\endcsname{}\expandafter\renewcommand\csname NodeStableRh\endcsname{\ensuremath{0.05}}
\expandafter\providecommand\csname NodeStableS\endcsname{}\expandafter\renewcommand\csname NodeStableS\endcsname{\ensuremath{0.0}}
\expandafter\providecommand\csname NodeStableSd\endcsname{}\expandafter\renewcommand\csname NodeStableSd\endcsname{\ensuremath{3.6}}
\expandafter\providecommand\csname NodeSubDraws\endcsname{}\expandafter\renewcommand\csname NodeSubDraws\endcsname{\ensuremath{24}}
\expandafter\providecommand\csname NodeSubFrac\endcsname{}\expandafter\renewcommand\csname NodeSubFrac\endcsname{\ensuremath{70\%}}
\expandafter\providecommand\csname NtwoG\endcsname{}\expandafter\renewcommand\csname NtwoG\endcsname{\ensuremath{9}}
\expandafter\providecommand\csname PisnEdgeLL\endcsname{}\expandafter\renewcommand\csname PisnEdgeLL\endcsname{\ensuremath{47\,\Msun}}
\expandafter\providecommand\csname PisnEdgeLLtight\endcsname{}\expandafter\renewcommand\csname PisnEdgeLLtight\endcsname{\ensuremath{50\,\Msun}}
\expandafter\providecommand\csname PisnLowEdgeDlnL\endcsname{}\expandafter\renewcommand\csname PisnLowEdgeDlnL\endcsname{\ensuremath{3.0}}
\expandafter\providecommand\csname PisnPlateau\endcsname{}\expandafter\renewcommand\csname PisnPlateau\endcsname{\ensuremath{55\,\Msun}}
\expandafter\providecommand\csname PromptFracDet\endcsname{}\expandafter\renewcommand\csname PromptFracDet\endcsname{\ensuremath{73\%}}
\expandafter\providecommand\csname PromptFracInt\endcsname{}\expandafter\renewcommand\csname PromptFracInt\endcsname{\ensuremath{82\%}}
\expandafter\providecommand\csname RateLocal\endcsname{}\expandafter\renewcommand\csname RateLocal\endcsname{\ensuremath{10^{+10}_{-5}}}
\expandafter\providecommand\csname RateMatched\endcsname{}\expandafter\renewcommand\csname RateMatched\endcsname{\ensuremath{10}}
\expandafter\providecommand\csname RateMatchedHi\endcsname{}\expandafter\renewcommand\csname RateMatchedHi\endcsname{\ensuremath{18}}
\expandafter\providecommand\csname RateMatchedLo\endcsname{}\expandafter\renewcommand\csname RateMatchedLo\endcsname{\ensuremath{7}}
\expandafter\providecommand\csname Rcl\endcsname{}\expandafter\renewcommand\csname Rcl\endcsname{\ensuremath{9.1}}
\expandafter\providecommand\csname RclHi\endcsname{}\expandafter\renewcommand\csname RclHi\endcsname{\ensuremath{10.3}}
\expandafter\providecommand\csname RclLo\endcsname{}\expandafter\renewcommand\csname RclLo\endcsname{\ensuremath{6.1}}
\expandafter\providecommand\csname Rfield\endcsname{}\expandafter\renewcommand\csname Rfield\endcsname{\ensuremath{16.0}}
\expandafter\providecommand\csname RfieldHi\endcsname{}\expandafter\renewcommand\csname RfieldHi\endcsname{\ensuremath{17.9}}
\expandafter\providecommand\csname RfieldLo\endcsname{}\expandafter\renewcommand\csname RfieldLo\endcsname{\ensuremath{13.0}}
\expandafter\providecommand\csname RhMixDenseFrac\endcsname{}\expandafter\renewcommand\csname RhMixDenseFrac\endcsname{\ensuremath{35\%}}
\expandafter\providecommand\csname RhMixDenseRh\endcsname{}\expandafter\renewcommand\csname RhMixDenseRh\endcsname{\ensuremath{0.05}}
\expandafter\providecommand\csname RhMixDiffuseFrac\endcsname{}\expandafter\renewcommand\csname RhMixDiffuseFrac\endcsname{\ensuremath{53\%}}
\expandafter\providecommand\csname RhMixDiffuseRh\endcsname{}\expandafter\renewcommand\csname RhMixDiffuseRh\endcsname{\ensuremath{0.43}}
\expandafter\providecommand\csname RhMixNbins\endcsname{}\expandafter\renewcommand\csname RhMixNbins\endcsname{\ensuremath{6}}
\expandafter\providecommand\csname RhMixSigInt\endcsname{}\expandafter\renewcommand\csname RhMixSigInt\endcsname{\ensuremath{0.10}}
\expandafter\providecommand\csname SelPlaceDistortion\endcsname{}\expandafter\renewcommand\csname SelPlaceDistortion\endcsname{\ensuremath{28}}
\expandafter\providecommand\csname SelPlaceGapRatio\endcsname{}\expandafter\renewcommand\csname SelPlaceGapRatio\endcsname{\ensuremath{2.0}}
\expandafter\providecommand\csname SelPlaceSpearman\endcsname{}\expandafter\renewcommand\csname SelPlaceSpearman\endcsname{\ensuremath{0.61}}
\expandafter\providecommand\csname SelPlaceSpearmanOFour\endcsname{}\expandafter\renewcommand\csname SelPlaceSpearmanOFour\endcsname{\ensuremath{0.42}}
\expandafter\providecommand\csname ShapeDropAtLargest\endcsname{}\expandafter\renewcommand\csname ShapeDropAtLargest\endcsname{\ensuremath{52}}
\expandafter\providecommand\csname ShapeGridRhHi\endcsname{}\expandafter\renewcommand\csname ShapeGridRhHi\endcsname{\ensuremath{0.43}}
\expandafter\providecommand\csname ShapeGridRhLo\endcsname{}\expandafter\renewcommand\csname ShapeGridRhLo\endcsname{\ensuremath{0.10}}
\expandafter\providecommand\csname ShapeHalvingHi\endcsname{}\expandafter\renewcommand\csname ShapeHalvingHi\endcsname{\ensuremath{26}}
\expandafter\providecommand\csname ShapeHalvingLo\endcsname{}\expandafter\renewcommand\csname ShapeHalvingLo\endcsname{\ensuremath{15}}
\expandafter\providecommand\csname SigmaChiDet\endcsname{}\expandafter\renewcommand\csname SigmaChiDet\endcsname{\ensuremath{0.13}}
\expandafter\providecommand\csname SigmaChiDetHi\endcsname{}\expandafter\renewcommand\csname SigmaChiDetHi\endcsname{\ensuremath{0.15}}
\expandafter\providecommand\csname SigmaChiDetLo\endcsname{}\expandafter\renewcommand\csname SigmaChiDetLo\endcsname{\ensuremath{0.09}}
\expandafter\providecommand\csname SigmaChiIntr\endcsname{}\expandafter\renewcommand\csname SigmaChiIntr\endcsname{\ensuremath{0.09}}
\expandafter\providecommand\csname SigmaChiIntrHi\endcsname{}\expandafter\renewcommand\csname SigmaChiIntrHi\endcsname{\ensuremath{0.09}}
\expandafter\providecommand\csname SigmaChiIntrLo\endcsname{}\expandafter\renewcommand\csname SigmaChiIntrLo\endcsname{\ensuremath{0.07}}
\expandafter\providecommand\csname SigmaChiThreeComp\endcsname{}\expandafter\renewcommand\csname SigmaChiThreeComp\endcsname{\ensuremath{0.14}}
\expandafter\providecommand\csname SigmaChiThreeCompHi\endcsname{}\expandafter\renewcommand\csname SigmaChiThreeCompHi\endcsname{\ensuremath{0.14}}
\expandafter\providecommand\csname SigmaChiThreeCompLo\endcsname{}\expandafter\renewcommand\csname SigmaChiThreeCompLo\endcsname{\ensuremath{0.11}}
\expandafter\providecommand\csname SigmaChiThreeCompSpread\endcsname{}\expandafter\renewcommand\csname SigmaChiThreeCompSpread\endcsname{\ensuremath{0.02}}
\expandafter\providecommand\csname SigmaChiTwoComp\endcsname{}\expandafter\renewcommand\csname SigmaChiTwoComp\endcsname{\ensuremath{0.07}}
\expandafter\providecommand\csname SmearChiStdIntr\endcsname{}\expandafter\renewcommand\csname SmearChiStdIntr\endcsname{\ensuremath{0.11}}
\expandafter\providecommand\csname SmearChiStdObs\endcsname{}\expandafter\renewcommand\csname SmearChiStdObs\endcsname{\ensuremath{0.17}}
\expandafter\providecommand\csname SmearChiStdSmear\endcsname{}\expandafter\renewcommand\csname SmearChiStdSmear\endcsname{\ensuremath{0.18}}
\expandafter\providecommand\csname SmearLgmStdIntr\endcsname{}\expandafter\renewcommand\csname SmearLgmStdIntr\endcsname{\ensuremath{0.13}}
\expandafter\providecommand\csname SmearLgmStdObs\endcsname{}\expandafter\renewcommand\csname SmearLgmStdObs\endcsname{\ensuremath{0.15}}
\expandafter\providecommand\csname SmearLgmStdSmear\endcsname{}\expandafter\renewcommand\csname SmearLgmStdSmear\endcsname{\ensuremath{0.14}}
\expandafter\providecommand\csname SmearSigmaChiMed\endcsname{}\expandafter\renewcommand\csname SmearSigmaChiMed\endcsname{\ensuremath{0.15}}
\expandafter\providecommand\csname SmearSigmaMoneHigh\endcsname{}\expandafter\renewcommand\csname SmearSigmaMoneHigh\endcsname{\ensuremath{9.2}}
\expandafter\providecommand\csname SmearSigmaMoneLow\endcsname{}\expandafter\renewcommand\csname SmearSigmaMoneLow\endcsname{\ensuremath{1.8}}
\expandafter\providecommand\csname SpinTradeoffFhighSpin\endcsname{}\expandafter\renewcommand\csname SpinTradeoffFhighSpin\endcsname{\ensuremath{1.6\%}}
\expandafter\providecommand\csname SpinTradeoffFhighZero\endcsname{}\expandafter\renewcommand\csname SpinTradeoffFhighZero\endcsname{\ensuremath{6.5\%}}
\expandafter\providecommand\csname SpinTradeoffHierSuppression\endcsname{}\expandafter\renewcommand\csname SpinTradeoffHierSuppression\endcsname{\ensuremath{4.3}}
\expandafter\providecommand\csname SpinTradeoffSigmaMax\endcsname{}\expandafter\renewcommand\csname SpinTradeoffSigmaMax\endcsname{\ensuremath{0.14}}
\expandafter\providecommand\csname SpinTradeoffSuppression\endcsname{}\expandafter\renewcommand\csname SpinTradeoffSuppression\endcsname{\ensuremath{4.2}}
\expandafter\providecommand\csname TwoChanPostRhBest\endcsname{}\expandafter\renewcommand\csname TwoChanPostRhBest\endcsname{\ensuremath{0.10}}
\expandafter\providecommand\csname TwoChanPostRhHi\endcsname{}\expandafter\renewcommand\csname TwoChanPostRhHi\endcsname{\ensuremath{0.10}}
\expandafter\providecommand\csname TwoChanPostRhLo\endcsname{}\expandafter\renewcommand\csname TwoChanPostRhLo\endcsname{\ensuremath{0.10}}
\expandafter\providecommand\csname TwoChannelDlnL\endcsname{}\expandafter\renewcommand\csname TwoChannelDlnL\endcsname{\ensuremath{99}}
\expandafter\providecommand\csname TwoChannelDlnLLo\endcsname{}\expandafter\renewcommand\csname TwoChannelDlnLLo\endcsname{\ensuremath{32}}
\expandafter\providecommand\csname TwocompClusterR\endcsname{}\expandafter\renewcommand\csname TwocompClusterR\endcsname{\ensuremath{9.3}}
\expandafter\providecommand\csname TwocompFieldR\endcsname{}\expandafter\renewcommand\csname TwocompFieldR\endcsname{\ensuremath{24}}
\expandafter\providecommand\csname TwocompHierFrac\endcsname{}\expandafter\renewcommand\csname TwocompHierFrac\endcsname{\ensuremath{9\%}}
\expandafter\providecommand\csname TwocompNmergers\endcsname{}\expandafter\renewcommand\csname TwocompNmergers\endcsname{\ensuremath{575714}}
\expandafter\providecommand\csname TwocompOneGR\endcsname{}\expandafter\renewcommand\csname TwocompOneGR\endcsname{\ensuremath{8.4}}
\expandafter\providecommand\csname TwocompTwoGR\endcsname{}\expandafter\renewcommand\csname TwocompTwoGR\endcsname{\ensuremath{0.9}}
\expandafter\providecommand\csname ZKernOffsetVar\endcsname{}\expandafter\renewcommand\csname ZKernOffsetVar\endcsname{\ensuremath{9}}
\expandafter\providecommand\csname ZKernOffsetVarOFour\endcsname{}\expandafter\renewcommand\csname ZKernOffsetVarOFour\endcsname{\ensuremath{3}}

\begin{document}

\title{Physics-based phenomenological modeling of binary black hole hierarchical formation 1: Synthetic universes from
  globular cluster simulations for GWTC}

\author{R.~O'Shaughnessy}
\affiliation{Center for Computational Relativity and Gravitation, Rochester
Institute of Technology, Rochester, NY 14623, USA}
\author{R.~Mechum}
\affiliation{Center for Computational Relativity and Gravitation, Rochester
Institute of Technology, Rochester, NY 14623, USA}
\author{M.~Qazalbash}
\affiliation{Center for Computational Relativity and Gravitation, Rochester
Institute of Technology, Rochester, NY 14623, USA}
\author{Z.~Rosenberg}
\affiliation{Center for Computational Relativity and Gravitation, Rochester
Institute of Technology, Rochester, NY 14623, USA}
\author{M.~Zeeshan}
\affiliation{Center for Computational Relativity and Gravitation, Rochester
Institute of Technology, Rochester, NY 14623, USA}

\date{\today}

\begin{abstract}
  We iteratively construct a physical model of the GWTC-5.0 binary-black-hole
  census by adding three components. First, we introduce a globular-cluster
  population built from physically normalized \Rapster{} simulations spanning
  cluster mass, metallicity, formation redshift, compactness, and natal
  black-hole spin. This component reproduces the higher-mass events and their
  larger effective-spin dispersion through hierarchical mergers, provided that
  black holes are born with zero natal spin. Second, because clusters
  cannot reproduce the low-mass, preferentially aligned population, we add a
  phenomenological field (isolated-binary) channel. Once this channel supplies
  the low-mass peak, the cluster compactness likelihood flattens: the
  field-inclusive inference permits ordinary dense globular-cluster birth radii
  and does not require nuclear-cluster-like conditions. Finally, motivated by
  residual tension between this two-component mixture and the census at
  $\result{GapMoneLo}$--$\result{GapMoneHi}\,\Msun$, we include an
  intermediate-mass isotropic component motivated by field remnants reprocessed
  in clusters. We treat this third component as a hypothesis, not a detection,
  because present numerical support does not permit a reliable evidence
  comparison. Physical normalization converts the observed high-mass rate into
  $\hatfgc\simeq\result{FieldSplitFgc}$ and local rates
  $R_{\rm cl}\simeq\result{Rcl}$ and
  $R_{\rm field}\simeq\result{Rfield}\,$Gpc$^{-3}$yr$^{-1}$. The forward
  model further predicts linked mass-spectrum breaks near $35$ and
  $70\,\Msun$, a $q\simeq0.5$ feature from first-plus-second-generation
  pairings, and a symmetric effective-spin distribution that broadens sharply
  above $45\,\Msun$. These correlated, mass-resolved predictions can be tested
  directly as the gravitational-wave census grows.
\end{abstract}

\maketitle

\section{Introduction}
\label{sec:intro}
The growing binary-black-hole (BBH) census from LIGO, Virgo, and KAGRA
\cite{DiscoveryPaper,LIGO-O2-Catalog,LIGO-O3-O3b-catalog,LIGO-O4a-cbc-catalog_results,LIGO-O4b-CBC-catalog,2015CQGra..32g4001L,2015CQGra..32b4001A,2020LRR....23....3A,2021PTEP.2021eA101A,2025PhRvD.111f2002L,2024arXiv240902831S}
now supports population-level tests of compact-binary formation across cosmic time
\cite{LIGO-O2-Rates,LIGO-O3-O3a-RP,LIGO-O4a-cbc-population,LIGO-GWTC5-populations-2026}.
These tests use parametric models
\cite{gwastro-PopulationReconstruct-Parametric-Wysocki2018,LIGO-O2-Rates,LIGO-O3-O3a-RP,LIGO-O3-O3bpop,LIGO-O4a-cbc-population,LIGO-GWTC5-populations-2026,2026PhRvL.137b1404P,2026ApJ..1005L..55R},
nonparametric reconstructions
\cite{dcc-Tong-Hierarchical-2025,2026arXiv260614472F,2026ApJ..1006L...5A,2025PhRvD.111f1305H,2025PhRvD.111f3043H,2023ApJ...957...37R,2024PhRvX..14b1005C,2024PhRvD.109j3006H,2025PhRvD.112l3054S,2025arXiv251025579T,2025ApJ...994L..52T,2023MNRAS.524.5844T,2026A&A...706A.361R},
and direct comparisons with simulation libraries
\cite{gwastro-ConstrainChannels-KickRatePaper-2017,popsyn-gwastro-STInterp-Vera2020,gwastro-PopulationReconstruct-Hierarchical-WysockiDoctor2019,gwastro-agndisk-GayathriPopModels2022,popsyn-gwastro-STInterpFinal-Vera2023,gwastro-wd-DelfaveroCosmic-2024,2021ApJ...910..152Z,2026PhRvD.113h3006T,gwastro-agndisk-GayathriPopModels2025,2024MNRAS.534.3506R,2025arXiv250819336C,2024ApJ...967...62Y}.
The reconstructed population can therefore distinguish among formation mechanisms
\cite{2010CQGra..27k4007M,gwastro-popsynVclusters-Rodriguez2016,LIGO-O3-O3a-RP,LIGO-O4a-cbc-population,LIGO-GWTC5-populations-2026}.

Despite their differences, parametric and nonparametric analyses recover a common broad picture of BBH masses, spins,
and merger redshifts \cite{LIGO-O4a-cbc-population,LIGO-GWTC5-populations-2026}. At low primary mass
($<15\,\Msun$), binaries tend toward equal masses and a narrow, positive effective-spin distribution, consistent with
small, preferentially aligned spins. At higher mass, the inferred spin distribution is broader and more nearly
isotropic
\cite{LIGO-GWTC5-populations-2026,2026arXiv260614472F,2026PhRvL.137b1404P,dcc-Tong-Hierarchical-2025,gwastro-pop-Zeeshan-O4aMixture,2026arXiv260612205A}.
The detailed mass dependence of mass ratio and effective spin remains model dependent
\cite{LIGO-GWTC5-populations-2026,gwastro-pop-Zeeshan-O4aMixture}; some analyses, for example, favor partial alignment
among higher-mass binaries \cite{2026PhRvL.137b1407L,2026arXiv260612205A,2026arXiv260623305R}.
Together with individual systems that have distinctive masses, large spins, or misaligned spins
\cite{LIGO-O4-HierarchicalPair-2025,LIGO-O3-O3b-catalog,LIGO-O4a-cbc-catalog_results,LIGO-O4a-cbc-population,LIGO-GWTC5-populations-2026},
these population results suggest that some observed BBHs form hierarchically through repeated mergers
\cite{2021NatAs...5..749G,LIGO-O4-GW231123,gwastro-agndisk-VeraGW231123McFacts,2025arXiv251113820L,2026ApJ...999..127L,LIGO-O4-HierarchicalPair-2025,2026PhRvL.137b1407L,2024arXiv241107304A,2025arXiv250909123A,2026arXiv260600234P},
as long expected for dynamically assembled populations and inferred from earlier observations
\cite{gwastro-mergers-hierarchical-smallerBHs-Fishbach2017,2017PhRvD..95l4046G,gwastro-PopulationReconstruct-Hierarchical-WysockiDoctor2019,2021ApJ...915L..35K,2025ApJ...981..177L,gwastro-agndisk-GayathriPopModels2025,2022ApJ...927..231F,2025PhRvD.112f3034X,2022PhRvD.106j3013M}.

Physics tightly constrains hierarchical formation. A previous merger fixes the remnant mass and, barring extreme
alternatives \cite{gwastro-pop-Zeeshan-O4aMixture}, produces a rapidly spinning black hole
\cite{2003ApJ...585L.101H,gwastro-mergers-hierarchical-smallerBHs-Fishbach2017,2021NatAs...5..749G}.
Anisotropic gravitational-wave emission also kicks the remnant, often ejecting it from
low-velocity-dispersion environments like globular clusters \cite{1973ApJ...183..657B,1983MNRAS.203.1049F,2019PhRvD.100d3027R,2021NatAs...5..749G,2025ApJ...987..146B}. 
The processes that assemble new binaries are likewise mass dependent
\cite{2016ApJ...831..187A,2018ApJ...866...66M,2021NatAs...5..749G}. Hierarchical models must therefore correlate
merger rates, masses, and spins across mass scales---correlations that ad hoc population models need not preserve.

We test these correlations by fitting the GW census directly with Monte Carlo simulations of hierarchical formation.
Building on earlier simulation-to-catalog comparisons
\cite{gwastro-ConstrainChannels-KickRatePaper-2017,popsyn-gwastro-STInterp-Vera2020,2021ApJ...910..152Z}, we construct
physically normalized synthetic universes from \Rapster{} globular-cluster simulations
\cite{2024PhRvD.110d3023K} spanning cluster masses, metallicities, formation redshifts, compactness, and natal spin.
We combine this cluster channel with an analytic, preferentially aligned low-mass field (isolated-binary) component motivated by previous
studies \cite{2026arXiv260700565F,gwastro-pop-Zeeshan-O4aMixture}. A companion paper applies the same inference strategy
to a coagulation model of hierarchical growth
\cite{gwastro-PopulationReconstruct-Hierarchical-WysockiDoctor2019,gwastro-PopulationReconstruct-Hierarchical-PhysicalParam2-Coagulation}.

The physical forward model adds constraints that a channel assignment alone
cannot provide: it maps the observed population shape into natal spin and
cluster compactness, converts the merger rate into a cluster-formation
efficiency, and predicts linked mass, mass-ratio, and spin features. Its
high-mass interpretation is consistent with earlier work
\cite{2026ApJ...997..267Y,2024PhRvL.133e1401L}. Conversely, our simplified
event likelihood does not sharply constrain the pair-instability boundary,
unlike analyses that use more event-level information
\cite{2026NatAs.tmp..111A,2025PhRvD.112f3040A}. These model-specific
contributions and non-results organize the analysis below.

Section \ref{sec:methods} describes the hierarchical likelihood, the \Rapster{} simulation library, our synthetic
universes, the field ansatz, and the event-likelihood approximation. Section \ref{sec:results} presents the cluster-only,
field-plus-cluster, and diagnostic three-component fits. Section \ref{sec:discussion} interprets these results and their
limitations, and Sec.~\ref{sec:conclusions} summarizes the main conclusions. The Appendices develop the
simulation-to-observation response formalism, validate the prompt-merger approximation, and document supporting tests.

\section{Methods}
\label{sec:methods}

\subsection{Adaptive hierarchical inference  }
\label{sec:methods:hyperpipe}
We infer the cluster-formation hyperparameters $\Lambda$ (natal spin, cluster
compactness, formation efficiency) by hierarchical Bayesian inference over the
observed binary--black-hole population. For a population whose intrinsic
differential merger rate (per comoving volume, per source-frame time) is
$\mathcal{R}(\svecz\,|\,\Lambda)$ in single-event parameters
$\svecz=(m_1,m_2,z,\chieff,\dots)$, the inhomogeneous-Poisson likelihood of the
observed catalog $\{d_i\}_{i=1}^{N_{\rm obs}}$ is, following the notation of the
\gwkokab{} framework~\cite{gwastro-mergers-zeeshan-gwkokab},
\begin{equation}
    \label{eq:poisson_like}
    \mathcal{L}(\pvecz) \propto
    e^{-\mu{(\pvecz)}}
    \prod_{j=1}^{N_{\rm obs}}
    \int\ell_j(\svecz) \cdot \comp(\svecz\mid\pvecz)
     \sqrt{ g_{\svecz}}
    d \svecz,
\end{equation}
Here $\sqrt{g_{\svecz}}d\svecz$ contains $T_{\mathrm{obs}}\,dz\,(1+z)^{-1}
(dV_c/dz)\,dm_1dm_2$ and the coordinate-dependent spin and eccentricity measures;
$\comp(\svecz\mid\pvecz)$ is the population rate density in this measure, and $\ell_i(\svecz)$ is the per-event
likelihood (Sec.~\ref{sec:methods:likelihood}). The expected number of
\emph{detections} is
\begin{equation}
  \mu(\Lambda) = T_{\rm obs}\!\int\! d\svecz\; \mathcal{R}(\svecz\,|\,\Lambda)\,
                 \frac{dV_c}{dz}\frac{1}{1+z}\, p_{\rm det}(\svecz),
  \label{eq:mu}
\end{equation}
with $p_{\rm det}$ the detection probability (Sec.~\ref{sec:methods:likelihood}).

We combine an analytic field rate with Monte Carlo cluster rates,
${\cal R}={\cal R}_f+{\cal R}_c$. Both the expected count and the single-event integrals therefore contain field and
cluster contributions:
\begin{align}
{\cal L} = e^{-\mu_c -\mu_f} \prod_{j=1}^{N_{\rm obs}} \int d\svecz  \sqrt{g_{\svecz}} \ell_j(\svecz) [{\cal R}_f+{\cal R}_c]
\label{eq:twocomp}
 \end{align}
We can likewise treat the simulation grid as a mixture, with weights $\lambda_\alpha$ satisfying
$\sum_\alpha\lambda_\alpha=1$ \cite{gwastro-ConstrainChannels-KickRatePaper-2017}. This promotes discrete simulations
to an inferred distribution over natal spin and cluster compactness.

Following Refs.
 \cite{gwastro-ConstrainChannels-KickRatePaper-2017,gwastro-mergers-GaussianLikelihoods-Delfavero2021,gwastro-mergers-GaussianLikelihoods-Delfavero2022,gwastro-mergers-NarrowPopAnalyticSophiya},
we use lower-dimensional analytic approximations to $\ell(\svecz)$ so that the likelihood can act directly on the
Monte Carlo catalogs.

Each likelihood evaluation requires an expensive cluster ensemble. After an initial grid, we therefore use the
single-cluster likelihood to place new simulations adaptively \cite{gwastro-wd-DelfaveroCosmic-2024}.
RIFT's \texttt{simulation\_manager} organizes this content-addressed library and reuses completed simulations
\cite{code-RIFT-research-projects-RIT}.
Following the RIFT hyperpipeline strategy \cite{gwastro-bns-eos-Atul2024,gwastro-RIFT-Update}, we evaluate
Eq.~\eqref{eq:poisson_like} at selected $\Lambda$ points, fit an approximate likelihood to those evaluations, and
resample until the simulations cover the hyperparameter posterior. This yields a posterior over cluster and field
properties rather than only a maximum-likelihood population.

\subsection{The \Rapster{} cluster backend and adaptively-placed simulation library}
\label{sec:methods:rapster}
We generate cluster BBHs with \Rapster~\cite{2024PhRvD.110d3023K}, a fast semi-analytic model of dynamical BBH
formation. Given particle number $N$, half-mass radius $r_h$, metallicity $Z$, formation redshift, binary fraction,
and first-generation natal spin $s$, \Rapster{} returns component masses and spins, $\chieff$, mass ratio, merger
redshift, formation channel, component generations $g_1,g_2$, and remnant recoil. Cluster stars follow the
\citet{2001MNRAS.322..231K} broken-power-law initial mass function over
$m_\star\in[0.08,150]\,\Msun$, with $dN_\star/dm_\star\propto m_\star^{-2.3}$ above $1\,\Msun$.
Compact-remnant masses follow one of three metallicity-dependent prescriptions: the \emph{delayed} or \emph{rapid}
models of Ref.~\cite{2012ApJ...749...91F}, or the Spera--Mapelli \textsc{sevn} tables
\cite{2015MNRAS.451.4086S}. In default \Rapster{} output, pulsational
pair instability removes first-generation black holes above a prescription-dependent lower edge and forbids remnants
up to $\simeq120\,\Msun$ \cite{2019ApJ...882...36M}. The lower edge is $55\,\Msun$ for \textsc{sevn} and
$45\,\Msun$ for the two Fryer engines.

\noindent \emph{A tunable pair-instability edge.}
The lower pair-instability edge---the maximum first-generation black-hole mass---is uncertain at the tens-of-percent
level: the ${}^{12}{\rm C}(\alpha,\gamma){}^{16}{\rm O}$ reaction rate alone places it anywhere over
$\simeq40$--$65\,\Msun$~\cite{2019ApJ...887...53F,2020ApJ...902L..36F}, and the two built-in
prescriptions already straddle this ($45\,\Msun$ Fryer, $55\,\Msun$ \textsc{sevn}).
We modify \Rapster{} to make both gap edges and their treatment configurable. Our added ``pile-up'' mode places
would-be gap remnants at the lower edge, approximating pulsational pair-instability mass loss.
All scored cluster ensembles use the fiducial \emph{delayed} prescription
\cite{2012ApJ...749...91F}, \Rapster's native default with a lower edge near $45\,\Msun$; Sec.~\ref{sec:disc:remnant}
tests the high-mass robustness. Only the pair-instability scan (Sec.~\ref{sec:results:pisn}) uses \textsc{sevn}.
Because an edge can be scanned only downward from the native remnant curve, \textsc{sevn}'s $\simeq55\,\Msun$
ceiling permits the full $40$--$55\,\Msun$ range. Appendix~\ref{app:pisn_truncate} compares the pile-up model with
delete-truncation.

\subsection{Synthetic cluster universe }
\label{sec:methods:universe}
We construct a synthetic universe by drawing a population of clusters, evolving each cluster with \Rapster{}, and
pooling the mergers with physical weights. A cluster initial-mass function, cosmic formation history, and metallicity distribution set the population;
the cluster mass-formation efficiency $\fgc$ fixes its absolute rate.
Each merger is placed at its simulated merger redshift and weighted by the cosmic metallicity distribution at its
\emph{formation} redshift [Eqs.~\eqref{eq:wk}--\eqref{eq:ri}]. This exactly retains the formation-epoch--metallicity
coupling and simulated delay times within the discrete sample. Appendix~\ref{app:delay} quantifies the remaining
placement approximation: the population is prompt dominated (formation-weighted median delay
$\simeq\result{DelayMedianMyr}$; $\simeq\result{PromptFracInt}$ within $1\,$Gyr), and its $m_1$ and $\chieff$ shapes
are nearly delay independent (Jensen--Shannon divergence $<\result{JsDelayMax}$). Placing mergers at formation rather
than merger redshift changes $\Nexp$ and $R_0$ by tens of percent but detected shapes by only $10^{-3}$ bits.

\paragraph*{Cluster initial-mass function.}
Each cluster is labeled by its particle number $N$ (initial stellar mass
$M_{\rm cl}\simeq N\bar m$, $\bar m=0.6\,\Msun$ for a Kroupa IMF). We draw $N$ from a
truncated power law,
\begin{equation}
  \frac{dN_{\rm cl}}{dN}\propto N^{\alpha},\quad N\in[10^{5.3},10^{6.0}],\quad
  \alpha=-2,
  \label{eq:icmf}
\end{equation}
consistent with observed young-massive-cluster mass functions and the CMC
suite~\cite{2020ApJS..247...48K}, with a Schechter cutoff $\exp(-N/N_\star)$ applied
in the fiducial runs at $N_\star=10^{6.48}$.

\paragraph*{Formation history.}
The cluster formation rate per comoving volume follows the cosmic star-formation-rate density
\cite{2017ApJ...840...39M},
\begin{equation}
  \psi_{\rm SFR}(z)=0.01\,\frac{(1+z)^{2.6}}{1+[(1+z)/3.2]^{6.2}}
  \quad[\Msun\,{\rm yr^{-1}\,Mpc^{-3}}],
  \label{eq:madau}
\end{equation}
We draw cluster formation redshift from
\begin{equation}
  p(z_f)\propto\psi_{\rm SFR}(z_f)\left|\frac{dt}{dz}\right|_{z_f},
    \label{eq:pzf}
\end{equation}
using flat-$\Lambda$CDM parameters
$(\Omega_m,H_0)=(0.31,67.7\,{\rm km\,s^{-1}Mpc^{-1}})$~\cite{2016A&A...594A..13P}. Weighting by
$\psi_{\rm SFR}|dt/dz|$ gives cluster mass formed per comoving volume per redshift, as required by the absolute
normalization below. The simulations and normalization use the same $\psi_{\rm SFR}$.

\paragraph*{Metallicity ensemble and absolute normalization.}
To represent the cosmic metallicity distribution,
we run \Rapster{} on a discrete metallicity grid and assume a lognormal distribution at each redshift,
\begin{equation}
  p(\log_{10}Z\,|\,z)=\mathcal{N}\!\big(\mu_Z(z),\sigma_Z\big),\quad
  \sigma_Z=0.5\,{\rm dex},
  \label{eq:pZz}
\end{equation}
centered on the mean-metallicity evolution~\cite{2017ApJ...840...39M,2026ApJ...997..267Y}
\begin{equation}
  \mu_Z(z)=\log_{10}Z_\odot+0.153-0.074\,z^{1.34}.
  \label{eq:muZ}
\end{equation}
The total cluster stellar mass formed per comoving volume over cosmic history is
\begin{equation}
  \Sigma_\star(\fgc)=\fgc\!\int_0^{z_{\max}}\!\!\psi_{\rm SFR}(z)
     \left|\frac{dt}{dz}\right|dz,
  \label{eq:sigmastar}
\end{equation}
where $\fgc$ is the fraction of cosmic star formation in surviving dense clusters. It connects cosmic star formation
to the absolute GW rate.

For fixed metallicity bins $Z_k$, the cluster-formation mass fraction is the exact MDF mass in each bin integrated
over formation history:
\begin{equation}
  w_k=\frac{1}{\mathcal{Z}}\!
  \int_0^{z_{\max}}\!\!\psi_{\rm SFR}(z)\!\left|\frac{dt}{dz}\right|
      [  F(e_{k+1})-F(e_k)]
      dz,
  \label{eq:wk}
\end{equation}
Here $F(x)=\Phi((x-\mu_Z(z))/\sigma_Z)$, $\Phi$ is the unit-normal CDF, $e_k$ are the $\log_{10}Z$ bin edges
(adjacent midpoints, with end bins extended by half a spacing), $z_{\max}=10$, and $\mathcal{Z}$ enforces
$\sum_k w_k=1$.
Within each $Z_k$ universe we draw cluster mass [Eq.~\eqref{eq:icmf}] and formation redshift
[Eq.~\eqref{eq:pzf}], fix a characteristic $r_h$, pool $N_{\rm seed}$ seeds, and record the simulated stellar mass
$M_{{\rm sim},k}$. Metal-rich clusters dominate the formed mass and supply lower-mass black holes; rarer metal-poor
clusters supply the high-mass tail. Their superposition produces the broad $dN/dm_1$ spectrum in
Sec.~\ref{sec:results}.
Each metallicity bin then contributes the comoving number density of
clusters-like-the-simulated-ones
\begin{equation}
  \kappa_k = w_k\,\frac{\Sigma_\star(\fgc{=}1)}{M_{{\rm sim},k}}\,\fgc
  \qquad[{\rm Mpc^{-3}}],
  \label{eq:kappak}
\end{equation}
Because every simulated merger records its formation redshift $z_{f,i}$, we evaluate the MDF at that natal epoch.
Merger $i$ in bin $k$ carries weight $\kappa_k r_i$, where
\begin{equation}
  r_{i}=\frac{\mathcal{N}\big(\log_{10}Z_k;\,\mu_Z(z_{f,i}),\sigma_Z\big)}
  {\big\langle\mathcal{N}\big(\log_{10}Z_k;\,\mu_Z(z_{f}),\sigma_Z\big)
  \big\rangle_k},
  \label{eq:ri}
\end{equation}
and the mean runs over the bin's simulated mergers so that its total weight remains $w_k$. Thus early mergers in a
metal-poor bin, where the cosmic MDF favors low $Z$, receive more weight than late mergers. Replacing this natal-epoch
weight with one bin-global factor shifts the joint $(s,r_h)$ surface by up to
$\result{NatalMdfDlnL}\,\ln\mathcal{L}$
($\result{NatalMdfDlnLOFour}$ for O4 only), although shape-only rankings are insensitive.

We use fixed-metallicity universes log-spaced over $0.007$--$1.4\,Z_\odot$ ($Z_\odot=0.0142$), extending the
metal-poor floor below Refs.~\cite{2026ApJ...997..267Y,2020ApJS..247...48K} to capture the high-mass tail. The
likelihood uses $N_Z=8$ bins, sufficient for spin--mass trends; the mass spectrum and merger rate use $N_Z=29$ to
avoid spurious metallicity structure.

\paragraph*{Reconstructing rate observables.}
Every merger $i$ in bin $k$ inherits rate weight $\kappa_k$. The expected number of detections is
\begin{equation}
  \Nexp = T_{\rm obs}\sum_k \kappa_k \sum_{i\in k} p_{{\rm det},i}\,
          \left[\frac{dV_c/dz}{(1+z)\,|dt/dz|}\right]_{z_{m,i}},
  \label{eq:Nexp}
\end{equation}
where $z_{m,i}$ is the merger redshift and $dV_c/dz$ is the comoving volume element. The
intrinsic comoving rate density $R(z)=\sum_{i:z_{m,i}\in\Delta z}\kappa_i/
[|dt/dz|\,\Delta z]$ gives $dR/dm_1=R(z_{\rm ref})\,p(m_1)$ at
$z_{\rm ref}=0.2$.

\paragraph*{Compactness mixture.}
The construction above fixes $r_h$, whereas real clusters span diffuse globulars through dense young and nuclear
cores, and their hierarchical yield depends steeply on compactness. We therefore also fit a mixture over an $r_h$ grid,
with weights $g(r_h)\ge0$ summing to unity. Equation~\eqref{eq:kappak} becomes
$\kappa_{k,j}=w_k\,g_j\,\Sigma_\star(\fgc{=}1)/M_{{\rm sim},k,j}\,\fgc$ over
metallicity bins $k$ and compactness bins $j$; $g_j=\delta_{jj_0}$ recovers a single $r_h$. Because $g_j$ enters each
event evidence linearly, it can be fit cheaply with the channel amplitudes (Sec.~\ref{sec:methods:fit}).

\paragraph*{Relation to galaxy-simulation cluster suites.}
Our synthetic universe is population averaged. More elaborate models inherit metallicity, formation time, and tidal
environment from galaxy or cosmological simulations, including the GAMESH-coupled Milky-Way GC model of
\cite{2026A&A...708A.364A} and the
semi-analytic GC-on-merger-tree models of~\cite{2019MNRAS.482.4528E}. Our metallicity
ensemble approximates this joint $(z_f,Z)$ structure with a separable MDF
[Eq.~\eqref{eq:wk}]; we discuss the resulting simplifications in
Sec.~\ref{sec:discussion}. The same \Rapster/cBHBd/CMC/\textsc{FastCluster}
backends~\cite{2021MNRAS.505..339M} (with \textsc{sevn} remnant
masses~\cite{2017MNRAS.470.4739S}) underlie recent cluster-channel inferences
against the GW catalog~\cite{2025PhRvL.134a1401A,2020PhRvD.102l3016A,
2023MNRAS.522..466A,2024A&A...688A.148T,2026ApJ...998..138M,2025ApJ...994L..54P}, to
which our physically-normalized, $\fgc$-marginalized analysis is complementary.

\begin{figure}
  \centering
  \includegraphics[width=\columnwidth]{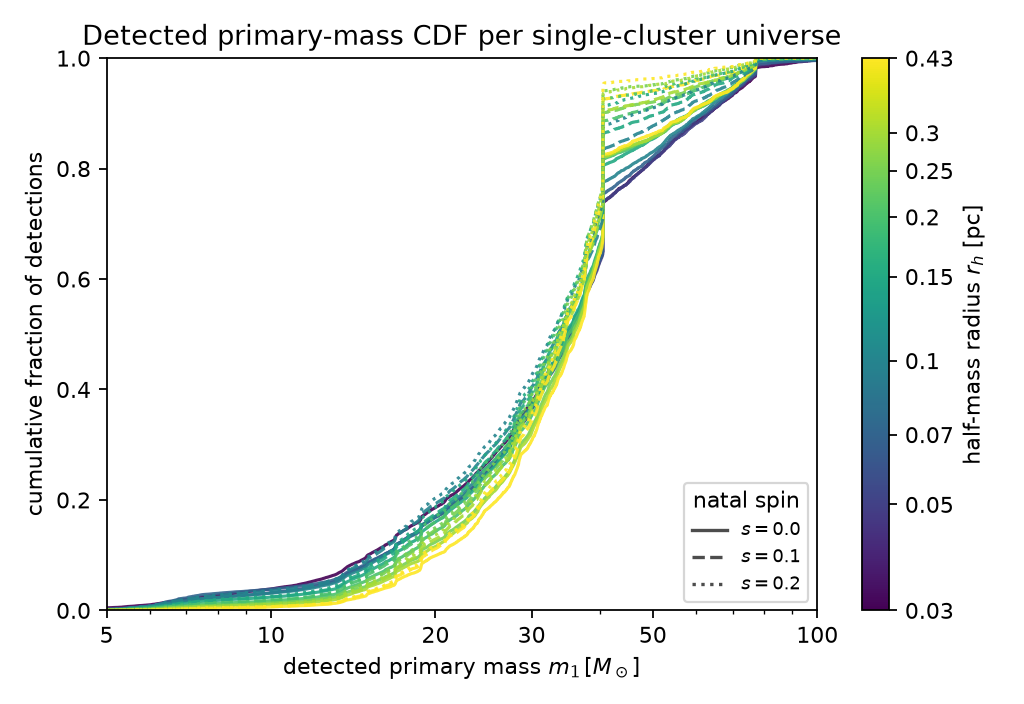}
  \caption{Cumulative distribution of \emph{detected} primary mass for each
  single-cluster-universe configuration---natal spin $s$ by line style, half-mass
  radius $r_h$ by color---Z-averaged (MDF weight) and seed-averaged. The distributions
  are normalized and so independent of $\fgc$. Detections concentrate at the
  $\simeq30$--$45\,\Msun$ hierarchical population; denser clusters extend to higher mass.}
  \label{fig:cluster_cdf}
\end{figure}

\begin{figure}
  \centering
  \includegraphics[width=\columnwidth]{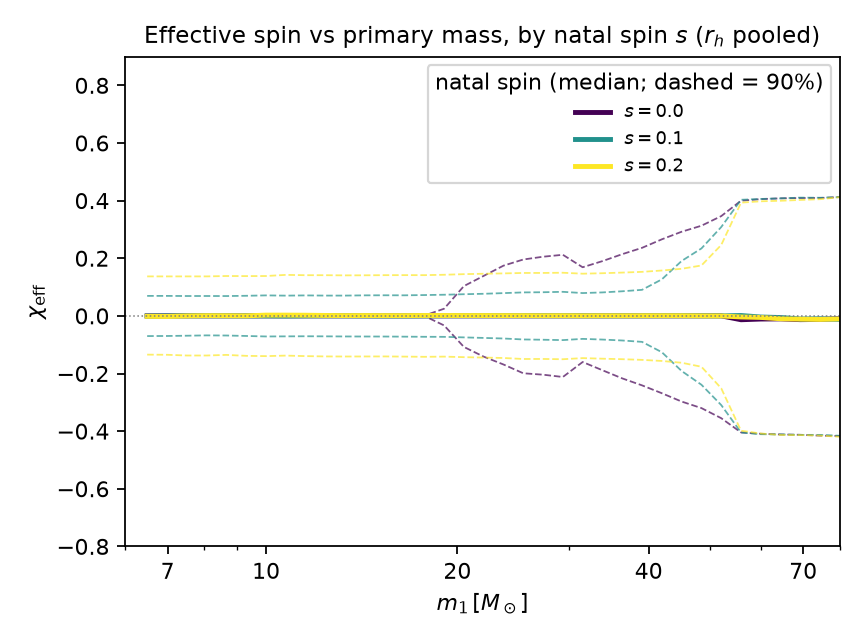}
  \caption{Effective-spin distribution versus primary mass per single-cluster universe,
  overlaid for all natal-spin values $s$ (one color each, $r_h$ pooled within each $s$):
  median (solid) with the $90\%$ contour edges (dashed, same color). Because $\chieff$ is
  bounded, contours rather than a dispersion are shown. The natal spin sets the low-mass,
  first-generation width---pinned at zero for $s=0$, widening with $s$---while
  hierarchical mergers broaden $\chieff$ symmetrically above $\simeq40\,\Msun$, regardless
  of natal spin.}
  \label{fig:cluster_chieff}
\end{figure}

\paragraph*{Single-cluster-universe predictions.}
Figures~\ref{fig:cluster_cdf} and~\ref{fig:cluster_chieff} show the cluster-only detected populations before fitting an
amplitude or adding the field. For each $(s,r_h)$, we average over MDF-weighted metallicities and seeds, weighting
mergers by $p_{\rm det}$ and the comoving volume--time element. The normalized distributions are independent of
$\fgc$. In Fig.~\ref{fig:cluster_cdf}, the low-mass rise traces the metallicity-dependent
first-generation peak---the Kroupa IMF folded through the \cite{2012ApJ...749...91F}
mass--metallicity relation, with metal-rich clusters peaking near $\simeq15\,\Msun$; the distribution then steepens through $\simeq30$--$45\,\Msun$ and shows a sharp break
at the $\simeq45\,\Msun$ pair-instability edge, above which \emph{no} first-generation
black holes form. The entire tail beyond $45\,\Msun$ is hierarchical (second-generation
and higher), so denser clusters---which retain and re-merge more remnants---reach
systematically higher masses. The intrinsic $dR/dm_1$ (Fig.~\ref{fig:cmp_dRdm1}) has the same broadened
first-generation peak and sharp $45\,\Msun$ truncation. In Fig.~\ref{fig:cluster_chieff}, quantile contours show the
bounded $\chieff$ distribution: natal spin sets the low-mass first-generation width, while the $\simeq0.7$ remnant
spin makes hierarchical mergers broad and symmetric. All natal-spin cases therefore converge above
$\simeq40$--$50\,\Msun$.

\subsection{Rate posterior marginalization and reconstruction}
\label{sec:methods:rate}

Because $\kappa_k\propto\fgc$ [Eq.~\eqref{eq:kappak}], $\fgc$ is a single multiplicative scale. For simulated
parameters $\theta=(r_h,s)$, write the cluster mean as
$\mu_c=\fgc\,\bar\mu_c(\theta)$, with $\bar\mu_c\equiv\Nexp(\fgc{=}1)$, and each event integral as
$\fgc c_j(\theta)$. A log-flat prior, $\pi(\fgc)d\fgc=d\fgc/\fgc$, then permits analytic marginalization:
\begin{align}
  Z(\theta) &\equiv \int_0^\infty \frac{d\fgc}{\fgc}\,
  e^{-\fgc\bar\mu_c(\theta)} \prod_{j=1}^{N_{\rm obs}} \fgc\,c_j(\theta)
  \nonumber\\
  &= \Gamma(N_{\rm obs})\prod_{j=1}^{N_{\rm obs}}
     \frac{c_j(\theta)}{\bar\mu_c(\theta)}.
  \label{eq:fgc_marg}
\end{align}
The count factor locks the amplitude scale: the integrand peaks at
\begin{equation}
  \hat{\fgc}=N_{\rm obs}/\Nexp(\fgc{=}1),
  \label{eq:fgchat}
\end{equation}
so the fitted detected count equals the observed count. Equation~\eqref{eq:fgc_marg} therefore tests only
\emph{shape}, through $c_j/\bar\mu_c$. Physical bounds on $\fgc$, from the total star formation and observed local
cluster mass density, remain an independent check rather than a prior input.

The same integral gives the exact fixed-$\theta$ amplitude posterior,
\begin{equation}
  p(\fgc\,|\,d,\theta)
  = \frac{\bar\mu_c^{N_{\rm obs}}}{\Gamma(N_{\rm obs})}\,
    \fgc^{\,N_{\rm obs}-1}\,e^{-\fgc\,\bar\mu_c(\theta)},
  \label{eq:fgc_post}
\end{equation}
with mean $\hatfgc$ and fractional width $N_{\rm obs}^{-1/2}\simeq7\%$. The deterministic map
$R_0=\fgc R_{\rm tot}(\theta)$ rescales this Gamma density into the local-rate posterior, so the cluster rate remains
a finite-width posterior. With a field channel, write its mean as
$\mu_f=R_{\rm field}\,\bar\mu_f$ and the field per-event evidence as $d_j$
[Eq.~\eqref{eq:twocomp}], the joint amplitude posterior
$p(\fgc,R_{\rm field}\,|\,d,\theta)\propto
e^{-\fgc\bar\mu_c-R_{\rm field}\bar\mu_f}
\prod_j\big(\fgc\,c_j+R_{\rm field}\,d_j\big)\,\pi(\fgc,R_{\rm field})$
is log-concave and evaluated on a two-dimensional grid. Profiling instead of marginalizing adds only a
$\theta$-independent Stirling constant in the one-amplitude case and leaves relative weights across $\theta$ unchanged.

\paragraph*{Monte-Carlo seed uncertainty.}
At each $\theta$ we run $N_{\rm seed}$ independent universes per metallicity bin. Seed $a$ supplies a complete rate
estimate, normalized by its simulated mass,
$\kappa_k^{(a)}=w_k\,\Sigma_\star(\fgc{=}1)/M_{{\rm sim},k,a}$
[cf.\ Eq.~\eqref{eq:kappak}]. For any rate functional $X$, we assign the pooled estimate the standard error
\begin{equation}
  \sigma^2_{\ln X}(\theta)
  = \frac{1}{N_{\rm seed}}\,
    {\rm Var}_a\!\left[\ln X^{(a)}(\theta)\right].
  \label{eq:seedvar}
\end{equation}
Per-seed rescoring likewise sets the Monte-Carlo noise floor on the $(r_h,s)$ likelihood surface
(Sec.~\ref{sec:methods:fit}). We convolve the fixed-$\theta$ rate posterior with an independent lognormal factor
$e^{\sigma_{\ln X}\xi}$, $\xi\sim\mathcal{N}(0,1)$, to propagate this seed variance.

\paragraph*{Marginalizing the formation parameters.}
The $\theta$ dependence of the marginalized likelihood separates into a
\emph{number} term and a \emph{shape} term,
\begin{equation}
  \ln Z(\theta)=\ln\Gamma(N_{\rm obs})
  -N_{\rm obs}\ln\bar\mu_c(\theta)
  +\sum_{j=1}^{N_{\rm obs}}\ln c_j(\theta),
  \label{eq:lnZsplit}
\end{equation}
and we define the surface statistic used for formation-parameter inference,
\begin{equation}
  \Delta\ln\mathcal{L}(\theta)\equiv \ln Z(\theta)-\max_{\theta'}\ln Z(\theta'),
  \label{eq:dlnl}
\end{equation}
For the two-channel model, $Z(\theta)$ is the amplitude-marginalized joint evidence scored with the tuned field
(Sec.~\ref{sec:methods:fit}); this field flattens the cluster-only compactness rail (Sec.~\ref{sec:compactness}). A flat
prior gives $p(\theta\mid d)\propto e^{\Delta\ln\mathcal{L}(\theta)}$. When comparing models instead, such as
cluster-only and two-channel in Sec.~\ref{sec:results:split}, $\Delta\ln\mathcal{L}$ denotes their $\ln Z$ difference
at fixed $\theta$. We interpolate the seed-averaged number and shape terms in Eq.~\eqref{eq:lnZsplit} separately, then
marginalize the fixed-$\theta$ rate posterior:
\begin{equation}
  p(R_0\,|\,d)=\int d\theta\;p(\theta\,|\,d)\;p(R_0\,|\,d,\theta),
  \label{eq:ratepost}
\end{equation}
The reported bands evaluate Eq.~\eqref{eq:ratepost} over the simulated nodes; structure below the seed noise floor is
not informative. Every simulation-derived total curve uses this convention. Intrinsic rates include seed uncertainty;
count-anchored detected totals include the $\theta$ and Poisson terms. Component curves are fiducial point predictions
for readability, and only totals are banded.

\paragraph*{External rate anchor and validation.}
The likelihood above anchors $\fgc$ internally to the observed count. For catalog-level rate comparisons we also use
an external normalization: choose $\fgc$ so that the cluster rate above $m_1=30\,\Msun$ matches the LVK high-mass
rate. We call this case \emph{rate matched} ($m_1>30$) and plot it in goldenrod. Its local rate factorizes as
\begin{equation}
  R_0(\theta)=\fgc(\theta)\,R_{\rm tot}(\theta)
  = R_{\rm hm}^{\rm LVK}\,\frac{R_{\rm tot}(\theta)}{R_{\rm hm}(\theta)}
  \equiv R_{\rm hm}^{\rm LVK}\,\rho(\theta),
  \label{eq:ratefac}
\end{equation}
Here $R_{\rm hm}^{\rm LVK}$ carries a factor-$\simeq2$ $90\%$ uncertainty, obtained by integrating the LVK
$dR/dm_1$ band above $30\,\Msun$ and modeled as lognormal. It dominates the $R_0$ uncertainty over the
$\theta$-posterior, Poisson, and seed terms; the shape ratio $\rho=R_{\rm tot}/R_{\rm hm}$ depends only on $\theta$
after MDF marginalization. Monte Carlo propagation through Eq.~\eqref{eq:ratepost} gives the reported $R_0$, $\fgc$,
and $dR/dm_1$ bands. Fresh \Rapster{} universes drawn from the joint $(r_h,s,Z)$ posterior$\times$MDF validate the
cache-interpolated result.

\subsection{A parametric field channel and a second generation}
\label{sec:methods:field}
The cluster channel produces neither the sharp low-mass peak nor the preferential spin alignment seen in the data
(Sec.~\ref{sec:results}). We add a parametric isolated-binary, or ``field,'' population:
\begin{multline}
  p_{\rm field}(m_1,q,\chieff,z)=
  \mathcal{N}_{[5,45]}(m_1;\mu_m,\sigma_m)\,p(q)\\
  \times\,\mathcal{N}(\chieff;\mu_\chi,\sigma_\chi)\,
  \frac{dV_c/dz}{1+z}\,(1+z)^{\kappa_z},
  \label{eq:field}
\end{multline}
with $\mathcal{N}_{[a,b]}$ a normal truncated to $[a,b]\,\Msun$ and $m_2\ge3\,\Msun$.
We fix $p(q)\propto q^{\beta_q}$ at $\beta_q=3$, consistent with LVK fits to the full BBH population
($\beta_q\simeq3$--$4$ over GWTC-3 through
GWTC-5.0~\cite{LIGO-O3-O3bpop,LIGO-GWTC5-populations-2026}) and with the near-equal-mass preference expected after
mass transfer and two supernovae. Varying $\beta_q$ over $[0,6]$ moves the inferred field fraction only from $0.171$
to $0.185$.
We set $\kappa_z=2.7$ to $z_{\max}=1.5$ and apply the same measured selection function as for clusters. The field
therefore requires no additional simulations. Its sharp low-mass peak and mildly aligned $\chieff$ target the residual
that the dynamical channel cannot reproduce.

In each fit we fix $(\mu_m,\sigma_m,\mu_\chi,\sigma_\chi)$ and vary only $R_{\rm field}$. The fiducial two-channel
decomposition (Sec.~\ref{sec:results:split}) uses the \emph{a priori} low-mass shape
$\mu_m=\result{FieldRefMone}\,\Msun$, $\sigma_m=\result{FieldRefMoneSig}\,\Msun$,
$\mu_\chi=+\result{FieldRefChi}$, $\sigma_\chi=\result{FieldRefChiSig}$, adopted from published low-mass results rather
than fitted here. Appendix~\ref{app:field} instead tunes these parameters on the all-mass catalog, necessarily giving a
larger in-sample gain; Fig.~\ref{fig:posterior_field} uses that tuned shape. Neither analysis propagates field-shape
uncertainty.

\subsection{GW observations and the data likelihood}
\label{sec:methods:likelihood}
Each observed event $i$ enters Eqs.~\eqref{eq:poisson_like}--\eqref{eq:mu}
through a Normal Approximate Likelihood (NAL) kernel: a single multivariate
Gaussian in $\boldsymbol{x}=(\mc,\eta,\chieff,z)$,
\begin{equation}
  \ell_i(\svecz)=\mathcal{N}\big(\boldsymbol{x}(\svecz)\,;\,
  \boldsymbol{x}_i,\boldsymbol{\Sigma}_i\big),
  \label{eq:nal_kernel}
\end{equation}
with per-event mean $\boldsymbol{x}_i$ and covariance $\boldsymbol{\Sigma}_i$.

\paragraph*{The per-event likelihood representation.}
For O1--O3 we use published prior-divided NAL fits in $(\mc,\eta,\chieff)$
\cite{popsyn-gwastro-STInterpFinal-Vera2023}. For O4a--O4b we fit the full parameter-estimation samples
\cite{data-GWTC4-PE,data-GWTC5-PE}, dividing out the sampling prior and the
$(m_1,m_2)\!\to\!(\mc,\eta)$ measure. Retaining this equal-mass-divergent Jacobian would bias $\eta$ upward. Each event
keeps its full $3\times3$ covariance (median $|\rho(\mc,\eta)|\simeq0.38$). The companion paper describes and tests the
O4 fits \cite{gwastro-PopulationReconstruct-Hierarchical-PhysicalParam2-Coagulation}. We append $z$ from each event's
GWOSC median and $90\%$ interval \cite{2023ApJS..267...29A,data-GWOSC-eventapi}, treating it as independent because
the NAL fits omit mass--redshift covariance. The Gaussian approximation is cruder in $z$, but here $z$ only localizes
the kernel where $p_{\rm det}$ changes.

\paragraph*{Evaluating the kernel in its own coordinates.}
We evaluate Eq.~\eqref{eq:nal_kernel} in its fitted coordinates. Each weighted model merger $(m_1,m_2)$ maps to
$\mc=(m_1m_2)^{3/5}(m_1+m_2)^{-1/5}$ and $\eta=m_1m_2(m_1+m_2)^{-2}$, and the likelihood
is evaluated there. This direction is exact and needs no Jacobian. Reducing to $(m_1,\chieff)$ would both approximate
a nonlinear transformation and discard mass-ratio information. The extra dimension reduces the number of model samples
under each kernel, so we monitor every evidence estimate's effective sample size and size the field template accordingly.

\paragraph*{The event list.}
We use confident BBHs from GWTC-1 through GWTC-5.0, requiring $m_2\ge3\,\Msun$ and FAR $<1\,$yr$^{-1}$
\cite{LIGO-O2-Catalog,2024PhRvD.109b2001A,LIGO-O3-O3b-catalog,LIGO-O4a-cbc-catalog_results,data-GWOSC-eventapi}.
This excludes the NSBH-like GW190814 and GW190917\_114630, whose $\simeq2.6$ and $\simeq2.1\,\Msun$ secondaries lie
outside our BBH channels and the GWTC-5.0 comparison products
\cite{2020ApJ...896L..44A,2024PhRvD.109b2001A}.

\paragraph*{Which events enter which fit.}
We use two overlapping---not nested---samples for different questions. All results involving the field use the
\emph{all-mass} O4a$+$O4b sample: $188$ confident BBHs, $179$ with the required $(m_1,\chieff)$. These include the count anchor, two-channel
decomposition, field tuning, and field-fed component. Intermediate-mass events near $20$--$25\,\Msun$, especially
GW241113, drive the last test. Only the cluster shape fit uses the \emph{high-mass} subsample: $160$ events with
$m_1>30\,\Msun$ ($51$ O1--O3, $53$ O4a, and $56$ O4b). There the cluster dominates and the field is negligible.
Restricting this fit prevents the unmodeled low-mass cluster mismatch from controlling $(r_h,s)$; low-mass events
remain in the analysis elsewhere.

\paragraph*{High-mass sample selection.}
This high-mass selection acts in data space: membership uses the catalog median $m_1$, whereas the scored model is
truncated at true $m_1>30\,\Msun$. An event straddling the threshold loses sub-threshold model support, and a source
that scatters upward can be scored against a population that cannot contain its true mass. Equation~\eqref{eq:poisson_like}
is normalized consistently for the truncated model but omits the probability of passing the catalog cut. With median
high-mass uncertainty $\simeq\result{SmearSigmaMoneHigh}\,\Msun$ (App.~\ref{app:smear}), threshold migration may matter.
Related work on the $35\,\Msun$ peak instead required each selected event to have at least $50\%$ posterior probability
of satisfying its mass cuts \cite{2025CQGra..42v5008R}. We have not propagated an analogous probabilistic selection and
therefore treat threshold migration as an unquantified systematic on the $(r_h,s)$ shape inference.

We score these kernels with Eq.~\eqref{eq:poisson_like} and $\mu=\Nexp$. Selection enters only $\mu$, because each
$\ell_i$ already conditions on detection
\cite{2004AIPC..735..195L,LIGO-P1600187-Farr-SelectionBiasaAndMonteCarlo}.

\paragraph*{Why a Gaussian kernel and not a kernel-density estimate?}
The Gaussian NAL is stable and tractable; it need not be exact. Tabulated KDEs preserve the tested $(s,r_h)$ ordering
and best-fit point, but their likelihood differences depend appreciably on bandwidth and they cost roughly $10^3$
times more to contract against the simulations. We therefore use the NAL and treat KDE agreement as a robustness check.
Tabulated likelihoods would be preferable for scientifically consequential multimodality once bandwidth and finite-sample
convergence are controlled.

\paragraph*{Kernel tails and a floor on the per-event evidence.}
A typical cluster evidence receives only $\simeq0.6\%$ of its weight beyond $3\sigma$, so its resolved kernel core
dominates. The minimal field template peaks near $\result{FieldTunedMone}\,\Msun$; for high-mass events, its tiny
evidence instead comes almost entirely from an unconstrained Gaussian tail. This does not affect channel assignment,
because the field is negligible there (Sec.~\ref{sec:results:split}), but the tail is a modeling artifact. The template
omits high-mass isolated pathways such as chemically homogeneous evolution and mass-ratio reversal, so its bare tail
cannot justify exponentially small field probabilities (as low as $10^{-296}$). A defensible analysis would floor the
field evidence to represent this uncertainty. Flooring it across $\result{FieldFloorDecades}$ decades around its median
moves the detected field fraction by only $\result{FieldFloorFracShift}$, well below
$\pm\result{FieldSplitFracErr}$. We therefore quote the unfloored fit but do not interpret its high-mass tail physically.

Including $z$ keeps the kernel narrow where $p_{\rm det}$ varies most. Appendix~\ref{app:response} shows that this
selection-free placement and the historical detected-weight convention differ only by a grid-constant offset below seed
scatter, whereas omitting redshift distorts model rankings. For detected-space plots only, we smear model densities by
the catalog widths $\boldsymbol{\Sigma}_i$ (App.~\ref{app:smear}); the likelihood is unchanged.

\begin{figure}
  \centering
  \includegraphics[width=\columnwidth]{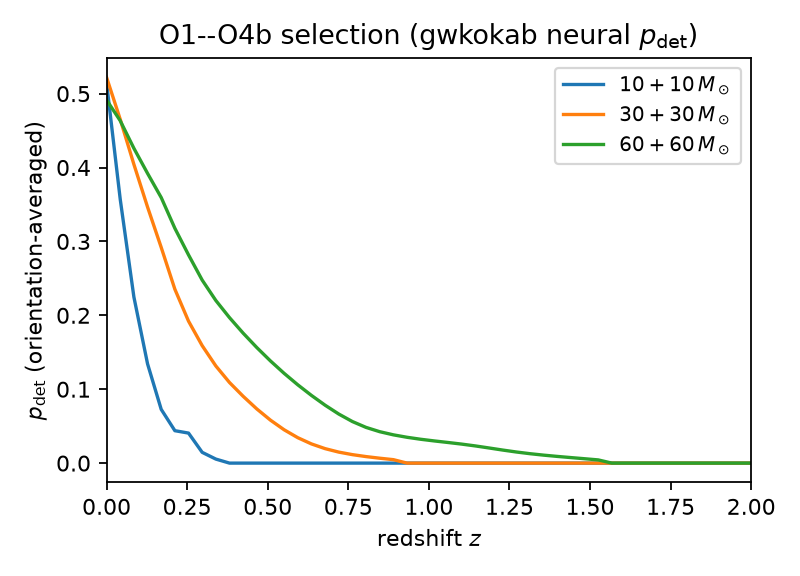}
  \caption{Combined O1--O4b detection probability $p_{\rm det}(m,m,z)$ (gwkokab
  neural emulator trained on the LVK sensitivity injections), orientation-averaged,
  for equal-mass binaries. Detectability falls off sharply with redshift---a
  $30+30\,\Msun$ source is undetectable beyond $z\simeq1.5$, a $60+60\,\Msun$
  source reaches farther---so cluster hierarchical mergers, which form
  preferentially at high $z$, are largely beyond the horizon.}
  \label{fig:pdet}
\end{figure}

We measure $p_{\rm det}(m_1,m_2,z)$ by training a neural-network emulator on the LVK O1--O4b
sensitivity-injection campaigns~\cite{2025PhRvD.112j2001E,data-O4-injections} using the \gwkokab{}
population-inference framework~\cite{gwastro-mergers-zeeshan-gwkokab,code-Zeeshan-gwkokab},
marginalizing spins and extrinsic parameters over the injection distribution. It captures the high-redshift falloff and
reproduces the raw-injection sensitive volume to $\sim10\%$. The injections and catalog span the same observing period,
making the rate term internally consistent (Fig.~\ref{fig:pdet}).

\subsection{Fitting the models}
\label{sec:methods:fit}
We fit four configurations. (a) A one-component cluster model uses one $(r_h,s)$ ensemble and marginalizes $\fgc$
analytically. (b) A cluster$+$field model adds Eq.~\eqref{eq:twocomp} with one compactness
($g_r=\delta_{rr_0}$) and free $R_{\rm field}$. (c) A mixture cluster$+$field model instead fits compactness weights
$\{g_r\}$ on the unit simplex. Thus (a)$\to$(b) adds a physical channel, whereas (b)$\to$(c) changes only the cluster
compactness distribution. Configuration (d) adds the diagnostic field-fed component with its own profiled rate. We
treat it separately in Sec.~\ref{sec:results:fieldfed} because we do not quote its likelihood ratio.

The simulated formation parameters $(s,r_h)$ lie on a discrete grid with one \Rapster{} ensemble per node; we maximize
over this grid rather than claim a smooth posterior. Per-seed rescoring gives a noise floor of a few
$\ln\mathcal{L}$ at dense nodes and more at diffuse nodes with sparse high-mass mergers. The one- versus two-channel
differences are tens of $\ln\mathcal{L}$ and exceed this floor. Natal-spin and compactness trends keep the same sign
across ensembles, but some node contrasts do not: at the densest node, $s=0$ versus $s=0.2$ is comparable to seed
scatter. We therefore quote a preference direction and natal-spin upper limit, not a precise likelihood ratio or
compactness point estimate, and marginalize $r_h$ through the mixture. Continuous parameters are treated more fully:
$\fgc$ is marginalized analytically, $(\fgc,R_{\rm field})$ is a two-dimensional convex problem, and $\{g_r\}$ spans
a cheaply mapped simplex. A continuous $(r_h,s)$ posterior and Dirichlet treatment of $\{g_r\}$ require a deeper
simulation library with seed noise below the interpolated surface structure.
Appendix~\ref{app:response} summarizes the response formalism: archived weighted populations, frozen-array
contractions, and mixture weights entering through Eq.~\eqref{eq:mixlnL}.

\paragraph*{A field-fed second generation.}
Real clusters can host both dynamical and primordial binaries \cite{2026arXiv260614846O}. Retained primordial-binary
remnants may be dynamically re-paired, producing isotropic second-generation mergers near twice the natal mass. As a
diagnostic rather than a self-consistent simulation, we add an optional isotropic-spin Gaussian with
$m_1\sim\mathcal{N}(20,3)\,\Msun$, remnant spin $\chi\simeq0.7$ with isotropic tilts ($\chieff$ width $\simeq0.25$),
and the field $q$ and $(1+z)^{2.7}$ distributions. Its rate is profiled with $\fgc$ and $R_{\rm field}$
(Sec.~\ref{sec:results:fieldfed}).

\subsection{Published population reconstructions used for comparison}
\label{sec:methods:lvkproducts}
Our figures use specific LIGO--Virgo--KAGRA GWTC-5.0 population products
\cite{LIGO-GWTC5-populations-2026,data-GWTC5-pop}. Because the released models differ in flexibility, every caption
labeled ``LVK GWTC-5.0'' refers to the following fixed mapping.
\emph{(i) Primary-mass rate $dR/dm_1$ and effective-spin density $p(\chieff)$}
(Fig.~\ref{fig:cmp_dRdm1}, and the high-mass rate anchor
$R_{\rm hm}^{\rm LVK}$ of Eq.~\eqref{eq:ratefac}): the BBH analysis with the
skew-normal $(\chieff,\chi_p)$ spin model, variance cut $1$
(release product \texttt{skewNormalChiEffChiP\_varcut1}). We multiply each normalized 1D density draw by its
local-rate parameter $R_0$ (median
$\simeq\result{LvkTotalRate}\,$Gpc$^{-3}$yr$^{-1}$) to form $dR/dm_1$, and
$R_{\rm hm}^{\rm LVK}$ is its integral over $m_1>30\,\Msun$.
\emph{(ii) The joint $\chieff$--$m_1$ plane}
(Fig.~\ref{fig:chieff_m1}): the nonparametric \textsc{PixelPop} joint $(m_1,\chieff)$ rate model, variance cut $1$
(\texttt{m1chieff\_varcut1}), reduced to $\chieff$ quantiles in each log-mass column. Figure~\ref{fig:chieff_consistency}
instead uses event-level prior-divided likelihoods (App.~\ref{app:chieff_iso}).
\emph{(iii) Mass ratio $p(q)$ and rate evolution $R(z)$}
(Fig.~\ref{fig:mixed_q}): the GWTC-5.0 updated default parametric model
(\texttt{TwoPeakBrokenPowerLaw} smoothed mass distribution,
\texttt{PowerLawRedshift} evolution, iid Gaussian spin magnitudes and tilts).
\emph{(iv) The joint mass plane $dR/dm_1dm_2$} (Fig.~\ref{fig:mixed_m1m2}):
the \texttt{NotchFilter} binned-pairing mass model (joint posterior-median rate). We use the less-flexible parametric
default in (iii) only for band-level $q$ and $z$ comparisons.

\section{Results}
\label{sec:results}

We add channels only when the data expose a specific failure. The cluster-only model explains the broad, hierarchical
high-mass population but not the aligned low-mass peak. Adding a field channel repairs the mass spectrum and yields a
field-dominated local rate, while weakening the apparent cluster-only preference for extreme compactness. A remaining
intermediate-mass spin mismatch motivates a diagnostic third component.

All cluster predictions include redshift-dependent measurement error and selection, cosmic star formation, simulated
formation-to-merger delays [Eqs.~\eqref{eq:ri} and \eqref{eq:Nexp}], and the stellar and cluster metallicity distributions.
Physical normalization turns the observed high-mass rate into a constraint on $f_{\rm GC}$ while the population shape
constrains the cluster parameters.
We compare these fits with the released GWTC-5.0 posterior-predictive populations
\cite{LIGO-GWTC5-populations-2026,data-GWTC5-pop}.

\subsection{Cluster-only constraints}
\label{sec:results:forward}
We first compare cluster-only universes with the $160$ O1--O4b events having $m_1>30\,\Msun$ ($51$ O1--O3, $53$
O4a, and $56$ O4b), using the measured O1--O4b selection. This is a diagnostic; Sec.~\ref{sec:results:split} gives the
joint inference with a field channel. Marginalizing $\fgc$ absorbs the absolute count, so $(s,r_h)$ is constrained only
by the detected high-mass $(\mc,\eta,\chieff,z)$ shape.

At zero natal spin, clusters predict a narrow, zero-centered $\chieff$ below $45\,\Msun$ and a sharp broadening above
it (standard deviation from $\lesssim0.03$ to $\gtrsim0.2$). Mass ratios become less equal, and the high-mass population
is almost entirely second generation or higher.
The metallicity ensemble produces first- and second-generation mass-spectrum breaks near $35$ and $70\,\Msun$,
qualitatively consistent with Ref.~\cite{2026arXiv260407456G}. Metal-rich clusters extend to a median near
$14\,\Msun$; metal-poor clusters down to $0.007\,Z_\odot$ supply the hierarchical tail to $\simeq80\,\Msun$.

\begin{figure*}
  \centering
  \includegraphics[width=0.5\textwidth]{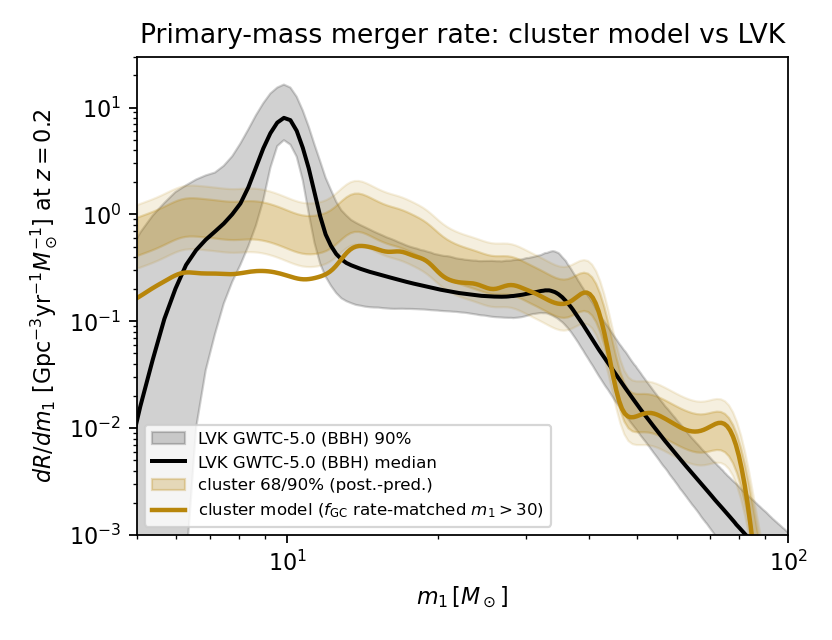}\hfill
  \includegraphics[width=0.49\textwidth]{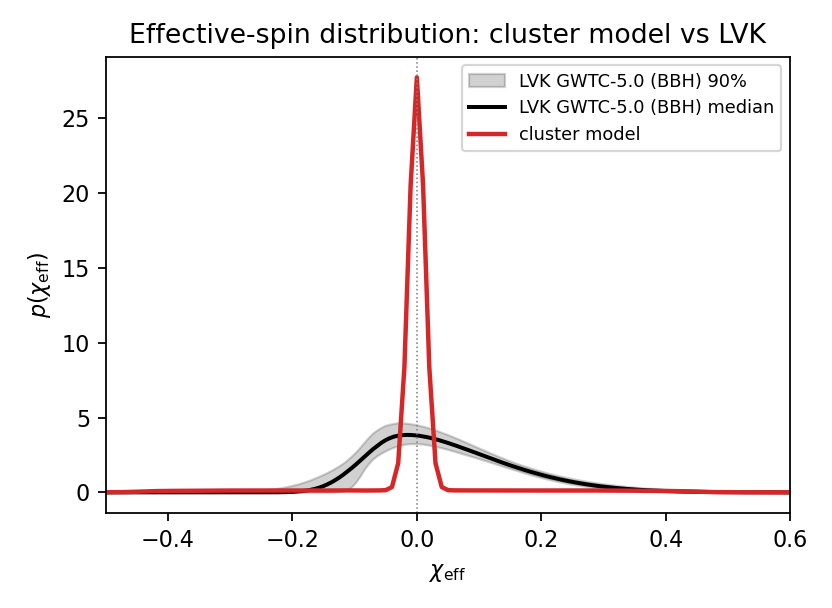}
  \caption{Best-fit cluster model versus the LVK GWTC-5.0 population data products
  (\cite{LIGO-GWTC5-populations-2026,data-GWTC5-pop}; skew-normal $(\chieff,\chi_p)$ analysis,
  Sec.~\ref{sec:methods:lvkproducts}), with $\fgc$ marginalized.
  \emph{Left:} intrinsic primary-mass merger-rate density $dR/dm_1$ at $z=0.2$
  (metallicity-ensemble cluster model, goldenrod median with posterior-predictive
  $68/90\%$ band, Sec.~\ref{sec:methods:rate}, vs LVK black median / gray $90\%$). The
  cluster is \emph{rate-matched} ($m_1>30$)---$\fgc$ set so its high-mass rate
  equals the LVK high-mass rate [Eq.~\eqref{eq:ratefac}]; goldenrod marks this
  normalization throughout, distinct from the fully posterior-marginalized results,
  because a cluster-\emph{only} count-anchored fit would be biased high by the
  low-mass peak the cluster cannot produce (Sec.~\ref{sec:results:forward}). It gives
  a physical $\hatfgc\simeq\result{HatFgc}$ and a local rate
  $\result{RateLocal}\,$Gpc$^{-3}$yr$^{-1}$ ($90\%$, posterior-predictive, dominated by the
  LVK high-mass-rate normalization; Sec.~\ref{sec:results}), and lies \emph{within
  the LVK $90\%$ band from $\simeq12\,\Msun$ through the high-mass tail to
  $\simeq80\,\Msun$}; it sits on the upper band edge at $25$--$30\,\Msun$ (a mild
  dynamical pile-up) and falls below the sharp $\simeq\result{FieldTunedMone}\,\Msun$ field-formation peak.
  \emph{Right:} effective-spin distribution $p(\chieff)$ (cluster model red vs LVK
  black/gray $90\%$): both center near zero, with the cluster channel narrower and
  symmetric---the isotropic spin--orbit tilts of dynamical assembly.}
  \label{fig:cmp_dRdm1}
  \label{fig:cmp_spin}
\end{figure*}

\emph{Overall rate and cluster mass efficiency.} Normalizing where clusters dominate gives
$\hatfgc\simeq\result{HatFgc}$ and
$R(z{=}0.2)=\result{RateLocal}\,$Gpc$^{-3}$yr$^{-1}$ ($90\%$ posterior predictive; Fig.~\ref{fig:cmp_dRdm1}), about
$\result{FracLVK}$ of the LVK total $R_0\simeq\result{LvkTotalRate}\,$Gpc$^{-3}$yr$^{-1}$. A cluster-only all-mass
count match would bias $\fgc$ high because clusters cannot supply the low-mass peak.

\begin{figure}
  \centering
  \includegraphics[width=\columnwidth]{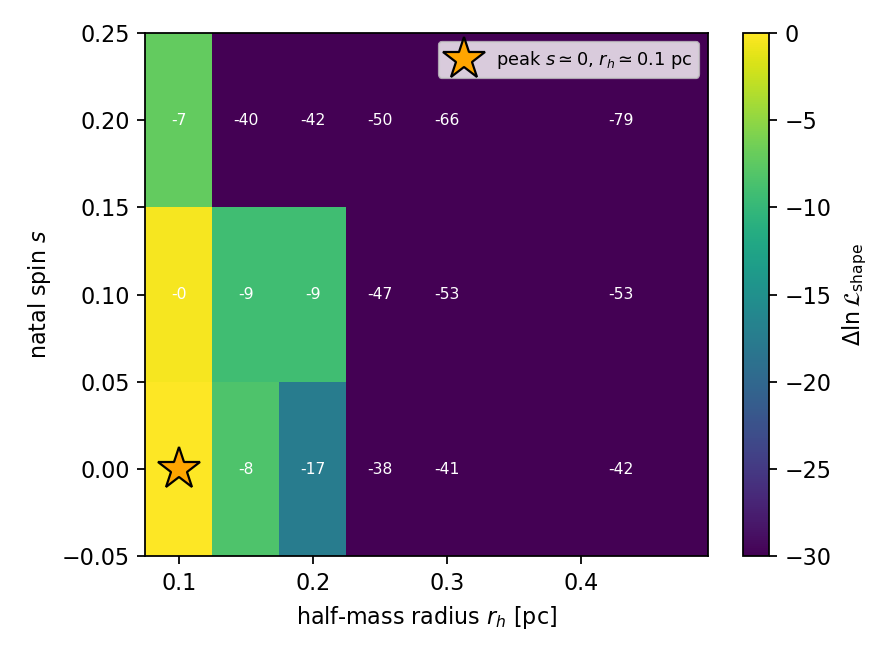}
  \caption{\emph{Fitting the cluster contribution to the high-mass population.}
  Shape likelihood $\Delta\ln\mathcal{L}(s,r_h)$ for the metallicity
  ensemble over natal spin and half-mass radius, scored on the high-mass
  $(\mc,\eta,\chieff,z)$ kernels against the combined O1--O4b high-mass catalog
  (160 events, measured O1--O4b selection, selection-free per-event placement of
  Sec.~\ref{sec:methods:likelihood}); $\fgc$ is marginalized, so this is a
  pure shape constraint. Cells show the raw $\Delta\ln\mathcal{L}$ ($8$ metallicity
  bins, $3$ seeds); the cell-to-cell irregularity ($\sim10\,\ln\mathcal{L}$, e.g.\ the
  dip at $s{=}0.1$, $r_h{=}0.15$) displays the Monte-Carlo scatter directly.
  The likelihood improves systematically toward smaller $r_h$ with no interior
  maximum---the maximum sits at the $r_h=\result{ShapeGridRhLo}\,$pc grid edge (star), a \emph{lower
  limit} on compactness---and strongly favors $s\simeq0$. More compact clusters
  retain more recoiling 2G remnants, building the high-mass population
  (Sec.~\ref{sec:compactness}).}
  \label{fig:posterior}
\end{figure}

\noindent \emph{Cluster properties.}
The cluster-only shape likelihood (Fig.~\ref{fig:posterior}) favors extreme parameters, but this changes once the field
is included.
Specifically, it favors $s\simeq0$ and the smallest $r_h$ allowed: across a grid from
$r_h=\result{ShapeGridRhHi}\,$pc down to $r_h=\result{ShapeGridRhLo}\,$pc the shape likelihood improves
\emph{monotonically} toward smaller $r_h$
($\Delta\ln\mathcal{L}\simeq-\result{ShapeDropAtLargest}$ at
$0.43\,$pc rising to the maximum at the $0.10\,$pc grid edge), with no interior
maximum and \emph{no sign of saturation}---each halving of $r_h$ still buys
$\sim\result{ShapeHalvingLo}$--$\result{ShapeHalvingHi}\,\ln\mathcal{L}$. This is a lower limit on compactness, not a
localized peak. Deeper potentials retain more recoiling remnants and build the broad high-mass $\chieff$ and
$>45\,\Msun$ tail, but the implied densities exceed typical young-globular-cluster values. This rail is only a
cluster-only diagnostic and disappears in the preferred two-channel model (Sec.~\ref{sec:results:split},
Fig.~\ref{fig:posterior_field}).

The left panel of Fig.~\ref{fig:cmp_dRdm1} compares $dR/dm_1$ with the LVK $90\%$ credible interval.

The right panel of Fig.~\ref{fig:cmp_dRdm1} shows the cluster-only $\chieff$ distribution, which is symmetric because
dynamical assembly isotropizes spin directions. Low natal spin, favored by the need to retain remnants, makes its population-averaged
distribution too narrow.
The agreement is mass dependent (Fig.~\ref{fig:cluster_chieff}). Above $45\,\Msun$, the model matches the
LVK width and symmetry (both broad about zero---the hierarchical-merger
signature), whereas at low and intermediate mass our distribution is too narrow and
the LVK one is broader with a slightly \emph{positive} mean (LVK $\langle\chieff
\rangle=+0.03$ at $5$--$15\,\Msun$ falling to $\simeq0$ above $45\,\Msun$).
The positive low-mass mean is the aligned-spin signature clusters cannot make: an isotropic dynamical subpopulation
sits atop an aligned field background \cite{dcc-Tong-Hierarchical-2025}. The high-mass match and low-mass deficit
together motivate two channels.

Two levers explain this result. Raising natal spin broadens low- and intermediate-mass $\chieff$ but also increases
recoil, suppressing retained remnants and the high-mass second generation. One parameter therefore controls both spin
width and high-mass amplitude \cite{2026arXiv260407456G}. Independently, isotropic clusters cannot make the narrow,
positive low-mass $\chieff$ peak, which requires an aligned field channel.

\subsection{Splitting out the field channel}
\label{sec:results:split}
The $\simeq\result{FieldTunedMone}\,\Msun$ peak and positive low-mass $\chieff$ motivate the parametric field channel
of Sec.~\ref{sec:methods:field}. At the best-fit cluster configuration, we maximize Eq.~\eqref{eq:twocomp} over
$\fgc$ and $R_{\rm field}$ using the $179$ all-mass O4 BBHs with measured $(m_1,\chieff)$ and $m_2\ge3\,\Msun$.

\begin{figure}
  \centering
            \includegraphics[width=\columnwidth]{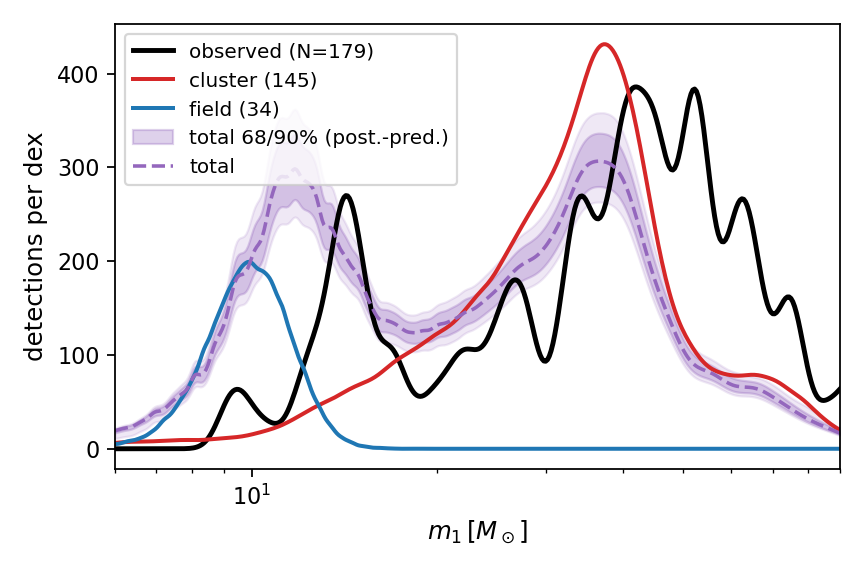}\\
  \includegraphics[width=\columnwidth]{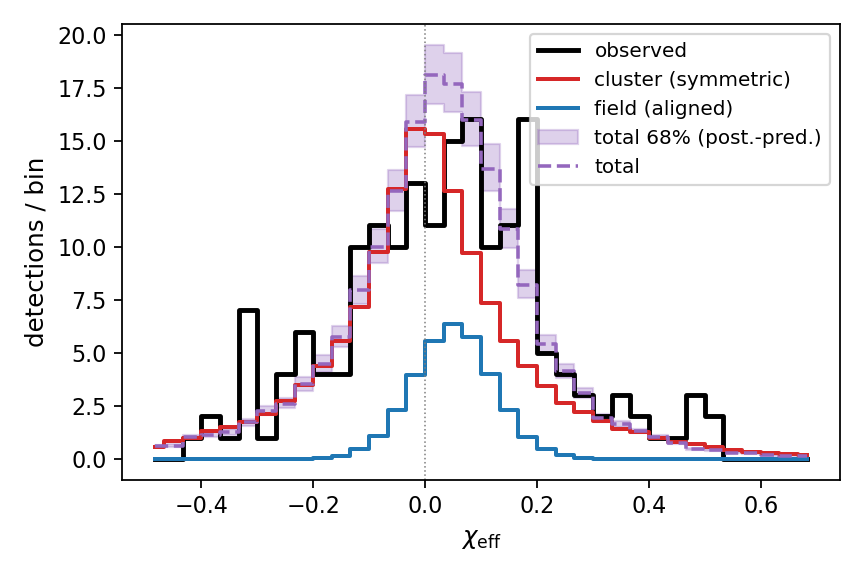}
  \caption{Two-component decomposition of the detected O4 population into the
  dynamical cluster model (red, amplitude $\fgc$) and a parametric field component
  (blue, fixed shape, free rate $R_{\rm field}$), with the total (purple, with
  posterior-predictive $68/90\%$ band, Sec.~\ref{sec:methods:rate}; components
  fiducial) vs observed (black). The observed curves are built from per-event
  point estimates (NAL centroids), so all model curves are measurement-smeared
  with the catalog's own per-event widths (App.~\ref{app:smear}).
  \emph{Top}, primary mass: the field carries the low-mass peak
  ($\simeq\result{FieldSplitFracLowMass}\%$
  of detections below $20\,\Msun$), the cluster the high-mass population ($\simeq100\%$
  above $20\,\Msun$). \emph{Bottom}, effective spin: the cluster's zero-spin
  first-generation spike, smeared, becomes the narrow central peak; the field
  supplies the mildly-\emph{aligned} component at
  $\chieff\simeq+\result{FieldTunedChi}$. The field improves the joint fit by
  $\Delta\ln\mathcal{L}\simeq+\result{FieldSplitDlnL}$ and accounts for a
  $\result{FieldSplitFrac}\pm\result{FieldSplitFracErr}$ detected fraction.}
  \label{fig:split}
\end{figure}

The field is strongly preferred. With its shape fixed \emph{a priori} at
Sec.~\ref{sec:methods:field}
[$\mu_m=\result{FieldRefMone}\,\Msun$, $\sigma_m=\result{FieldRefMoneSig}\,\Msun$,
$\mu_\chi=+\result{FieldRefChi}$, $\sigma_\chi=\result{FieldRefChiSig}$], adding it
improves the fit by $\Delta\ln\mathcal{L}\simeq+\result{FieldSplitDlnL}$ for one
additional free amplitude (Fig.~\ref{fig:split}). This is not a one-parameter model comparison: four fixed shape
parameters remain, Appendix~\ref{app:field} tunes them on the same data, and their uncertainty is not propagated.
Moreover, finite field samples produce several $\ln\mathcal{L}$ of seed scatter, so the last digit is not meaningful.
The fixed-shape field accounts for a detected fraction
$\result{FieldSplitFrac}\pm\result{FieldSplitFracErr}$, and---crucially---the
split is mass segregated: the field supplies
$\simeq\result{FieldSplitFracLowMass}\%$ of detections at
$m_1<20\,\Msun$ (the sharp peak the cluster channel cannot make) but
essentially none above $20\,\Msun$, where clusters dominate. The cluster efficiency
relaxes to $\hatfgc\simeq\result{FieldSplitFgc}$ once the field carries the
low-mass events. The result is a high-mass, symmetric-$\chieff$ dynamical population plus a low-mass, mildly aligned
isolated population. For comparison, tuning the field Gaussian to the same residual
(App.~\ref{app:field}) gives a primary-mass peak
$m_1\!\sim\!\mathcal{N}(\result{FieldTunedMone},\result{FieldTunedMoneSig})\,\Msun$
and $\chieff\!\sim\!\mathcal{N}(+\result{FieldTunedChi},\result{FieldTunedChiSig})$, improving on cluster-only by
$\Delta\ln\mathcal{L}\simeq+\result{FieldTunedDlnL}$. This in-sample tuning gain is not directly comparable to the
fixed-shape $+\result{FieldSplitDlnL}$. A joint inference over field shape and formation parameters is left for future work.

\paragraph*{Robustness to the compactness posterior.}
Because $r_h$ is weakly constrained within Monte-Carlo noise (Sec.~\ref{sec:compactness}), we refit and marginalize the
two-channel model at every $(r_h,s)$ node (App.~\ref{app:field}). Across the plausible low-spin range, the field
detected fraction is $\result{FieldDetFrac}$
($\result{FieldDetFracLo}$--$\result{FieldDetFracHi}$); the two-channel preference
$\Delta\ln\mathcal{L}\gtrsim\result{TwoChannelDlnLLo}$ (up to $\result{TwoChannelDlnL}$); the
intrinsic effective-spin width $\sigma_\chi\simeq\result{SigmaChiIntrLo}$--$\result{SigmaChiIntrHi}$
($\result{SigmaChiDetLo}$--$\result{SigmaChiDetHi}$ detected); the detected hierarchical ($2$G$+$)
fraction $\simeq\result{HierFracDetLo}$--$\result{HierFracDetHi}$ (intrinsically
$\simeq\result{HierFracIntrLo}$--$\result{HierFracIntrHi}$); and a \emph{field-dominated}
local-rate budget, with clusters supplying
$\result{ClusterFracTotalLo}$--$\result{ClusterFracTotalHi}\%$ of the total
($R_{\rm cl}\simeq\result{RclLo}$--$\result{RclHi}$ versus
$R_{\rm field}\simeq\result{RfieldLo}$--$\result{RfieldHi}\,$Gpc$^{-3}$yr$^{-1}$). The LVK high-mass anchor adds a
factor-$\sim2$ normalization uncertainty. Natal spin is better constrained: $s=0$ is favored over $s=0.2$ by
$\Delta\ln\mathcal{L}\simeq\result{NatalSpinDlnL}$, but this contrast uses the two deepest-sampled nodes. Each has the
same \result{NatalSpinSeeds} seeds; per-seed differences have standard deviation \result{NatalSpinSeedSd}, so only the
pooled estimate is informative. The sign is stable across all ensembles and follows independently from the structural
recoil--retention argument in Sec.~\ref{sec:compactness}.

The amplitude posteriors behind these numbers---$p(\fgc)$ and the joint
$(R_{\rm cl},R_{\rm field})$ budget---are collected in App.~\ref{app:amplitudes}
(Fig.~\ref{fig:fgc_rate_corner}).

\begin{figure*}
  \centering
  \includegraphics[width=0.49\textwidth]{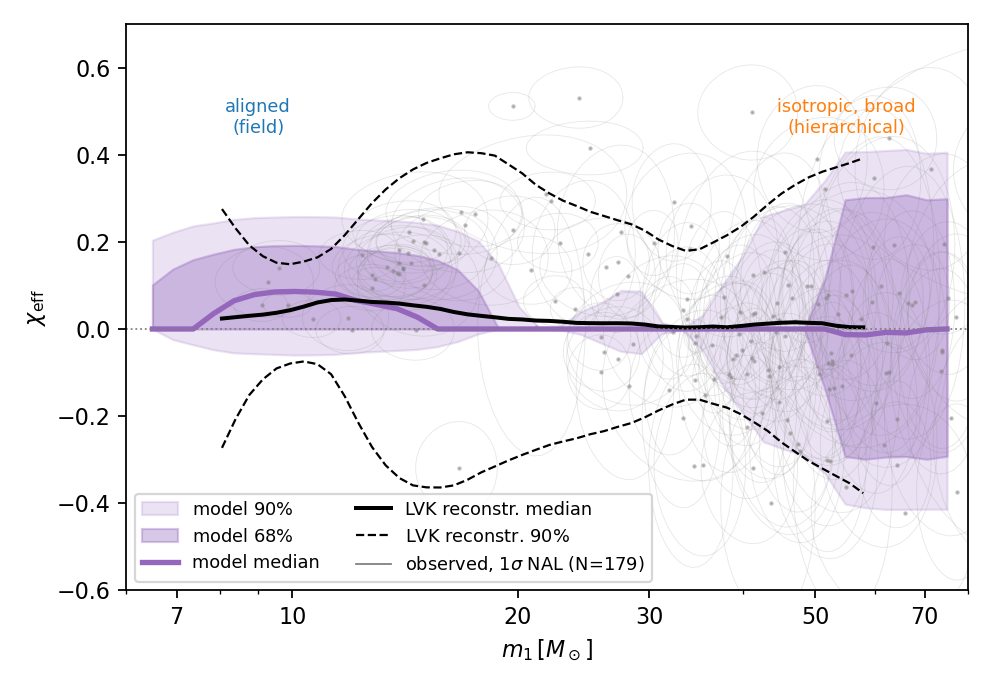}\hfill
  \includegraphics[width=0.49\textwidth]{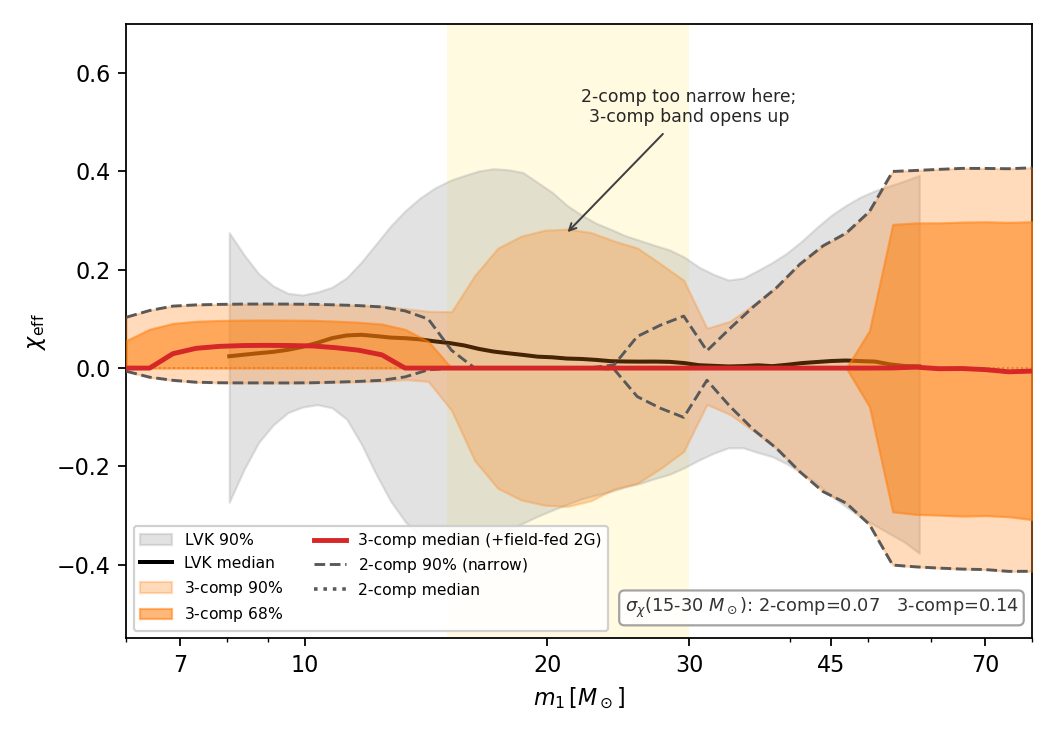}\\
  \includegraphics[width=0.49\textwidth]{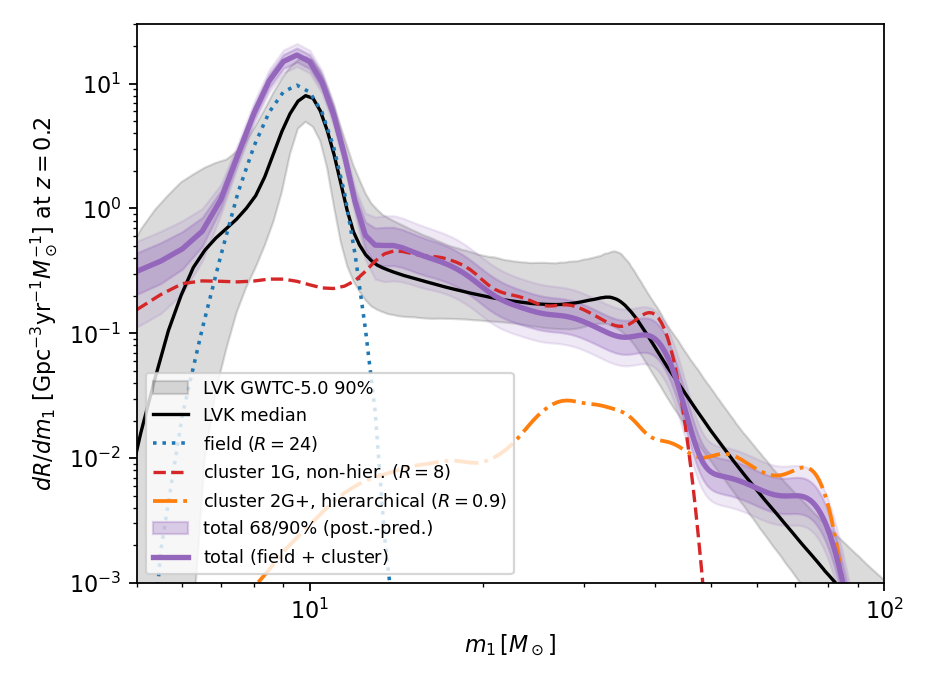}\hfill
  \includegraphics[width=0.49\textwidth]{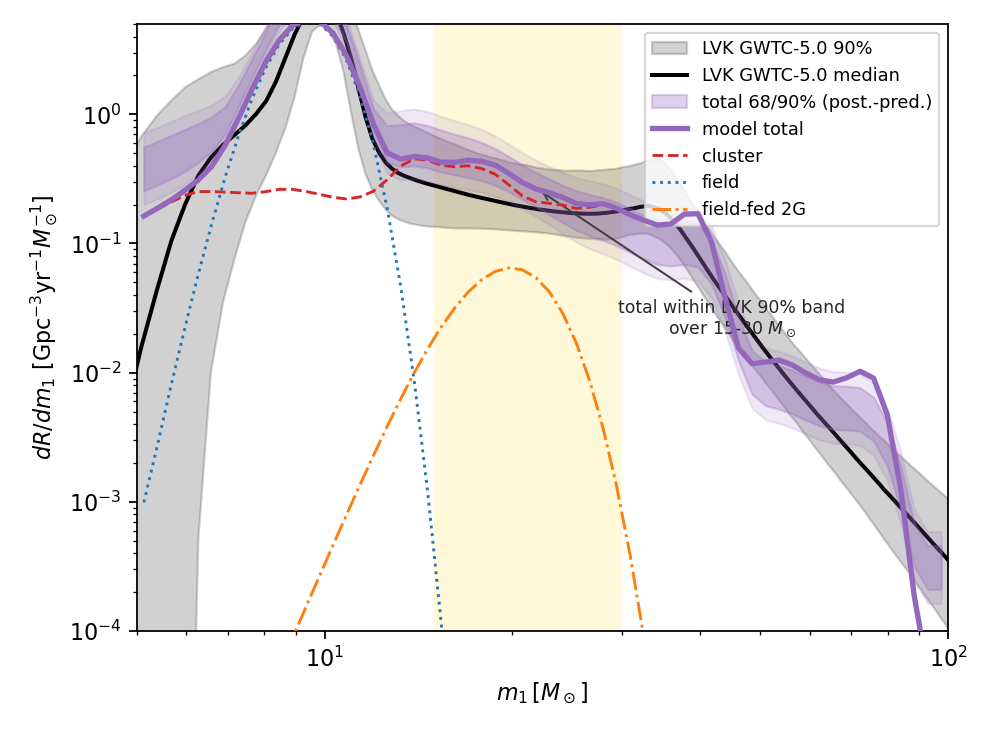}
  \caption{\textbf{Headline comparison of the dynamical-cluster$+$field model against
  the GWTC-5.0 population}, in effective spin (top row) and intrinsic rate (bottom
  row); left column the two-component (field$+$cluster) model, right column adding the
  field-fed second generation (Sec.~\ref{sec:results:fieldfed}).
  \emph{Top left:} predicted intrinsic $\chieff$--$m_1$ relation (purple median,
  $68/90\%$ population bands from sliding log-mass windows): $\chieff$ is narrow and
  \emph{aligned} at low mass (field), pinned near \emph{zero} at intermediate mass
  (non-hierarchical first generation, zero natal spin), and \emph{symmetric and broad}
  at high mass (hierarchical), against the LVK GWTC-5.0 reconstruction (black;
  \textsc{PixelPop} joint $\chieff$--$m_1$ product, Sec.~\ref{sec:methods:lvkproducts}) and
  observed O4 events (gray $1\sigma$ NAL marginal ellipses; App.~\ref{app:smear}).
  \emph{Top right:} the same observable with the field-fed second generation (color)
  versus the two-component model (gray dashed): the added isotropic-spin 2G lifts
  $\chieff$ at $15$--$30\,\Msun$ from near-zero toward the observed spread
  ($\sigma_\chi\simeq\result{SigmaChiTwoComp}\to\result{SigmaChiThreeComp}$, the value
  annotated in the panel), supplying the mid-mass spin scatter the minimal model
  under-produces. Both right-hand panels are drawn from the single \emph{reference}
  three-component fit defined in Sec.~\ref{sec:results:fieldfed}, and every count and
  width quoted for this component comes from that one fit.
  \emph{Bottom left:} intrinsic $dR/dm_1$ at $z{=}0.2$ decomposed by origin: the
  parametric field (blue dotted) supplies the $\simeq\result{FieldTunedMone}\,\Msun$ peak; the cluster is
  split by component generation into first-generation (red dashed) and hierarchical
  2G$+$ (orange dash-dot; only $\result{TwocompHierFrac}$ of cluster mergers, $R_{\rm 2G}^{\rm clu}\simeq\result{TwocompTwoGR}\,$Gpc$^{-3}$yr$^{-1}$),
  which takes over above the $\simeq45\,\Msun$ pair-instability cutoff and accounts for
  the entire high-mass tail and the bumps near $55$/$70\,\Msun$; the total (purple, with
  posterior-predictive $68/90\%$ band, Sec.~\ref{sec:methods:rate}; components fiducial)
  tracks the LVK band from $\simeq6$ to $\simeq80\,\Msun$ (computed from
  $\result{TwocompNmergers}$ cluster mergers, log-mass kernel).
  \emph{Bottom right:} $dR/dm_1$ for the three-component model: because a second
  generation near $\result{FieldFedTwoGPeakMone}$ is several times more detectable than the
  $\sim\result{FieldTunedMone}\,\Msun$ field, the handful of detections it supplies correspond to only
  $R_{\rm 2G}^{\rm ff}\simeq\result{FieldFedTwoGR}\,\mathrm{Gpc^{-3}\,yr^{-1}}$ intrinsic
  (the field-fed component's own rate, not the cluster's hierarchical
  $R_{\rm 2G}^{\rm clu}$ above), peaking at
  $\simeq\result{FieldFedTwoGPeak}\,\mathrm{Gpc^{-3}\,yr^{-1}\,\Msun^{-1}}$---an order of
  magnitude below the LVK $90\%$ ceiling there. The total (with the two-channel
  posterior-predictive band; the fiducial 2G added, its amplitude uncertainty negligible
  against the band width) lies within the LVK $90\%$ band across the whole window.}
  \label{fig:chieff_m1}
  \label{fig:twocomp_dRdm1}
  \label{fig:field_fed_2g}
  \label{fig:field_fed_dRdm1}
\end{figure*}

\paragraph*{The two channels reproduce the intrinsic mass spectrum.}
The summed intrinsic $dR/dm_1$ tracks the LVK median within its $90\%$ band from $\simeq12$ to
$\simeq80\,\Msun$ (Fig.~\ref{fig:twocomp_dRdm1}). The field supplies the sharp low-mass peak and clusters the smooth
high-mass tail. Their normalizations differ in reliability. The high-mass-anchored cluster gives
$R_{\rm cl}(z{=}0.2)\simeq\result{TwocompClusterR}\,$Gpc$^{-3}$yr$^{-1}$.
The field peak sits at $\result{FieldTunedMone}\,\Msun$, where detectability
falls steeply, so converting its fitted \emph{detected} count through the
parametric template's own selection function gives
$R_{\rm field}(z{=}0.2)\simeq\result{TwocompFieldR}\,$Gpc$^{-3}$yr$^{-1}$, but this raw template normalization is
uncertain by a factor $\sim2$--$3$ and overshoots the LVK total ($\simeq\result{LvkTotalRate}$). Anchoring instead to
the LVK low-mass rate gives
$R_{\rm field}\simeq\result{RfieldLo}$--$\result{RfieldHi}\,$Gpc$^{-3}$yr$^{-1}$
budget above. The generation split separates first-generation mergers ($g_1{=}g_2{=}1$; $91\%$ of
cluster mergers, $R\simeq\result{TwocompOneGR}\,$Gpc$^{-3}$yr$^{-1}$) from the
hierarchical second-and-higher-generation mergers ($\result{TwocompHierFrac}$,
$R_{\rm 2G}^{\rm clu}\simeq\result{TwocompTwoGR}\,$Gpc$^{-3}$yr$^{-1}$; this is the
cluster channel's \emph{own} hierarchical rate, distinct from the field-fed component's
$R_{\rm 2G}^{\rm ff}$ of Sec.~\ref{sec:fieldfed_rate}). First-generation mergers end sharply near the
$45\,\Msun$ pair-instability limit. Hierarchical mergers account for the entire higher-mass tail and its rate bumps
near $55$ and $70\,\Msun$, based on $\result{TwocompNmergers}$ simulated cluster mergers.

\paragraph*{The hierarchical signature in spin.}
The spin split gives the complementary signature. At near-zero natal spin, first-generation mergers have
$\chieff\simeq0$ (a near-delta spike, intrinsic width $\sigma\simeq0.001$), whereas
hierarchical (2G$+$) mergers inherit the $\simeq0.7$ spin of their merger-product
components and, with isotropic dynamical tilts, produce a \emph{broad, symmetric}
$\chieff$ ($\sigma\simeq0.26$). Within the cluster model, broad spin is exclusively hierarchical. Its fraction rises
from $\sim1\%$ near $10\,\Msun$ to unity above $45\,\Msun$, so the predicted $\chieff$ width grows with mass as in
Fig.~\ref{fig:chieff_m1}; the narrow, mildly aligned low-mass population is field formed.

\paragraph*{Predicted $\chieff$--mass relation.}
Figure~\ref{fig:chieff_m1} gives a falsifiable intrinsic spin--mass prediction. The field dominates at low mass, with
narrow aligned $\chieff$ (median $\simeq+0.10$). First-generation clusters dominate at intermediate mass and pin
$\chieff$ near zero. Hierarchical mergers dominate at high mass, producing a broad symmetric distribution ($\pm0.3$
at $68\%$). The model therefore reproduces the LVK low-mass alignment and high-mass broadening, but its
intermediate-mass distribution is too narrow. A direct test using prior-divided PE likelihoods for
$m_1\in[15,30]\,\Msun$ (App.~\ref{app:chieff_iso}) rejects a zero spike and favors a broad, two-sided intrinsic width
$\sigma_{\rm int}\simeq\result{ChiIsoSigmaInt}$ about a small positive mean
($\mu=\result{ChiIsoMu}\pm\result{ChiIsoMuErr}$), with
genuine large-$|\chieff|$ events of \emph{both} signs (GW241011 at
$\result{ChiIsoLoudChi}\pm\result{ChiIsoLoudSig}$ on the positive side, GW241110 at
$\result{ChiIsoChiMin}\pm\result{ChiIsoChiMinSig}$ on the negative; this
is independent of the NAL fit, where GW241011 is too narrow to score). The magnitudes require large spins, while
$\chieff<0$ requires misalignment. Raising cluster natal spin cannot repair the mismatch: recoil then suppresses the
high-mass tail by $\simeq\result{SpinTradeoffSuppression}\times$. The intermediate-mass scatter therefore points to a
separate spinning population, tested in Sec.~\ref{sec:results:fieldfed}, rather than to larger cluster natal spin.

\begin{figure}
  \centering
  \includegraphics[width=\columnwidth]{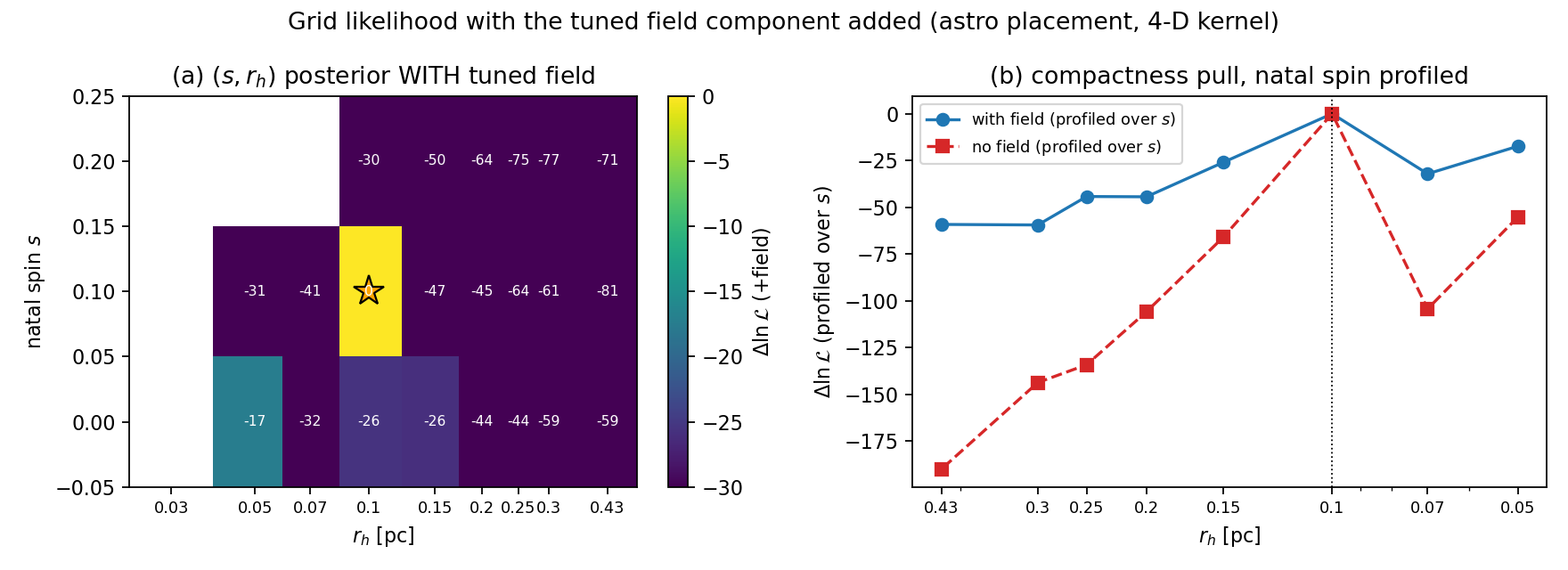}
  \caption{\emph{Jointly fitting a two-component universe to the whole population.}
  \emph{(a)} Shape likelihood over $(s,r_h)$ after adding the tuned field and profiling both amplitudes.
  \emph{(b)} At $s=0$, the field reduces the cluster-only compactness rail to a shallow variation comparable to
  Monte-Carlo scatter. The dense extensions do not resolve a turnover, so $r_h$ remains weakly constrained
  (Sec.~\ref{sec:results:split}).}
  \label{fig:posterior_field}
\end{figure}

\paragraph*{The field relaxes the compactness requirement.}
Adding the tuned field and profiling both amplitudes across the $(s,r_h)$ grid changes the compactness result
(Fig.~\ref{fig:posterior_field}). The cluster-only all-mass fit rails to maximum density because clusters are forced to
explain the low-mass peak. With the field carrying that peak, the likelihood varies by only
$\simeq\result{FieldGridRhSpan}\,\ln\mathcal{L}$ across the grid, comparable to Monte-Carlo scatter. The added dense
ensembles do not resolve either a turnover or a monotonic trend, so we infer only a weak lower limit on compactness,
not a preferred $r_h$. Low natal spin remains preferred.

This scan uses only $8$ metallicity bins and $3$ seeds, and its kernels omit redshift, a choice correlated with
compactness (App.~\ref{app:response}). These limitations do not restore the cluster-only rail, but they prevent a
quantitative residual $r_h$ measurement. A four-dimensional rerun remains future work.

\begin{figure*}
  \centering
  \includegraphics[width=0.49\textwidth]{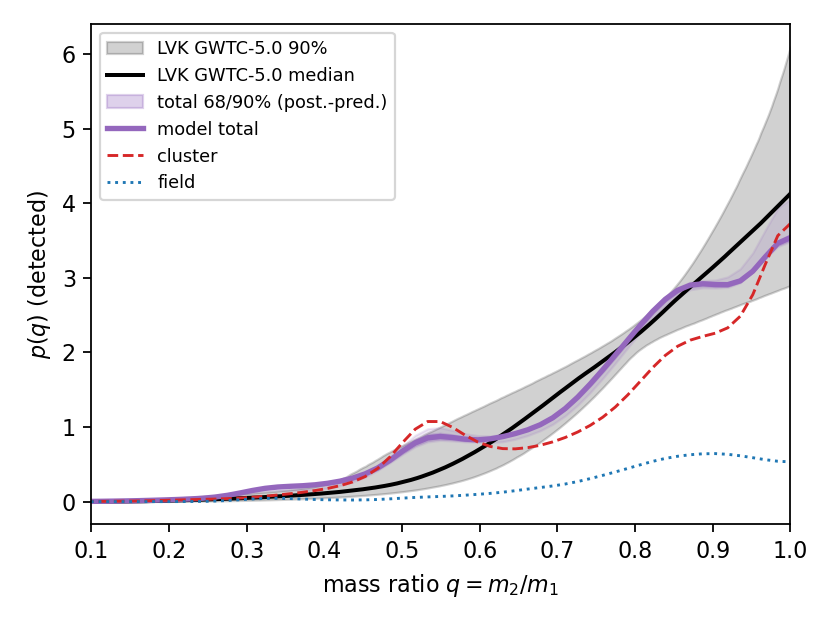}\hfill
  \includegraphics[width=0.49\textwidth]{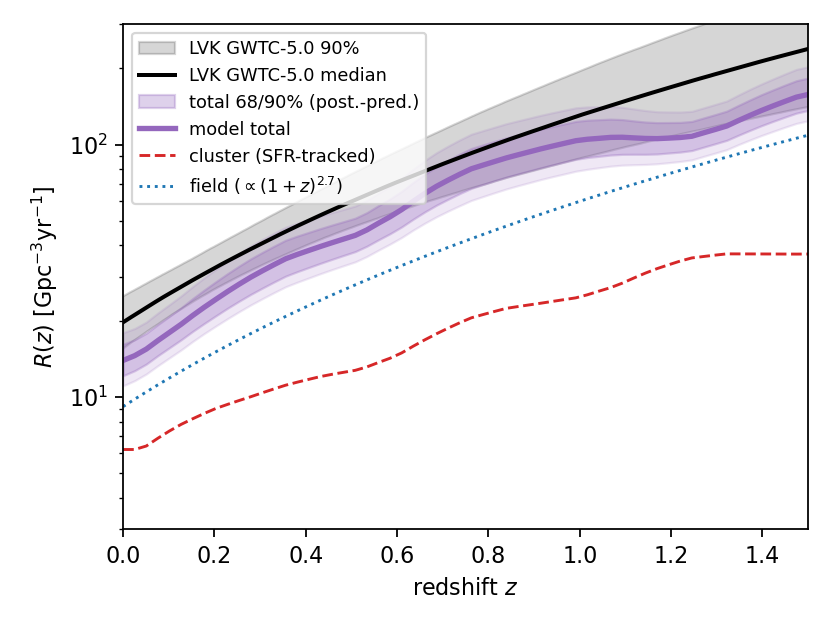}
  \caption{The two-component (field$+$cluster) model against two further GWTC-5.0
  products (updated default parametric model, Sec.~\ref{sec:methods:lvkproducts}).
  \emph{Left:} detected mass-ratio $p(q)$---model total (purple, with posterior-predictive
  $68/90\%$ band, Sec.~\ref{sec:methods:rate}; components fiducial) vs LVK (black/gray $90\%$);
  the cluster's $q\simeq0.5$ bump is the unequal-mass $1$G$+2$G hierarchical population.
  \emph{Right:} merger-rate evolution $R(z)$---cluster (SFR-tracked) plus field
  ($\propto(1+z)^{2.7}$, the assumed evolution) vs LVK; the banded total lies within the
  $90\%$ band to $z\simeq0.8$.}
  \label{fig:mixed_q}
  \label{fig:mixed_Rz}
\end{figure*}

\begin{figure}
  \centering
  \includegraphics[width=\columnwidth]{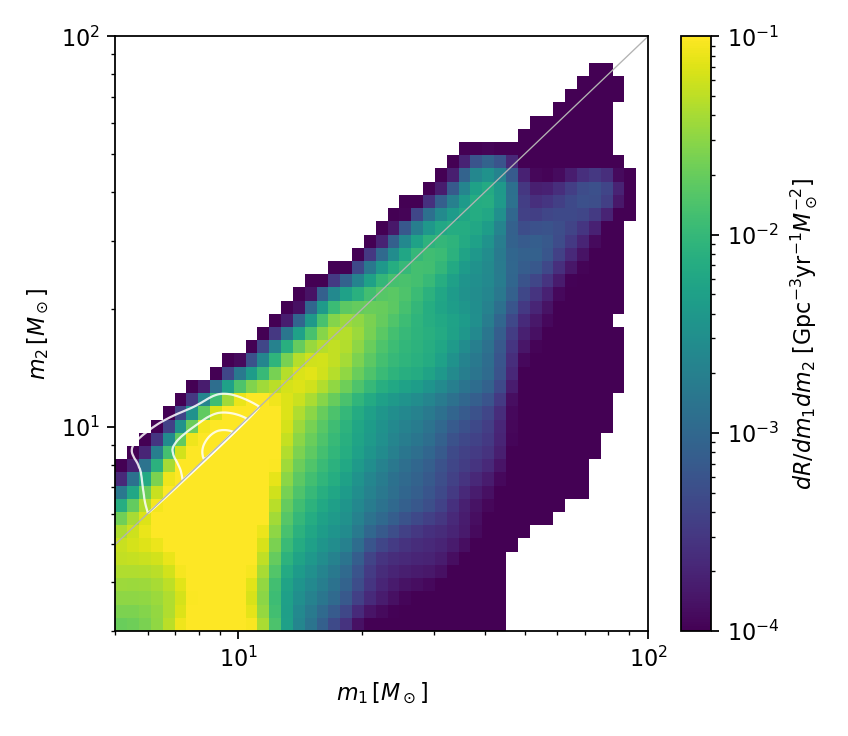}
  \caption{Joint intrinsic primary--secondary mass rate $dR/dm_1dm_2$ at $z=0.2$ for the
  two-component model: the field supplies the $\simeq\result{FieldTunedMone}\,\Msun$ near-equal-mass peak, the
  cluster the ridge extending to the high-mass tail (with the hierarchical $2$G feature near
  $m_1\simeq40\,\Msun$). White contours: the GWTC-5.0 \texttt{NotchFilter} binned-pairing
  joint-rate median (Sec.~\ref{sec:methods:lvkproducts}).}
  \label{fig:mixed_m1m2}
\end{figure}

\paragraph*{Mass ratio, redshift, and the joint mass distribution.}
The fit also reproduces three untuned GWTC-5.0 products. Detected $p(q)$ follows the LVK band toward equal mass
(Fig.~\ref{fig:mixed_q}), while clusters add a feature near
$q\simeq0.5$---the unequal-mass first-plus-second-generation ($1$G$+2$G) pairs whose
$\simeq2{:}1$ ratio is a hierarchical signature \cite{2023MNRAS.522..466A}. $R(z)$ stays within the LVK $90\%$ band
to $z\simeq0.8$ (Fig.~\ref{fig:mixed_Rz}); clusters follow delayed star formation, while the field evolution
$(1+z)^{2.7}$ is assumed rather than inferred here. Finally, $dR/dm_1dm_2$ shows the field's near-equal-mass peak at
$\simeq\result{FieldTunedMone}\,\Msun$ and the cluster ridge to the high-mass tail, including a hierarchical feature
near $40\,\Msun$ (Fig.~\ref{fig:mixed_m1m2}).

\begin{figure*}
  \centering
  \includegraphics[width=\textwidth]{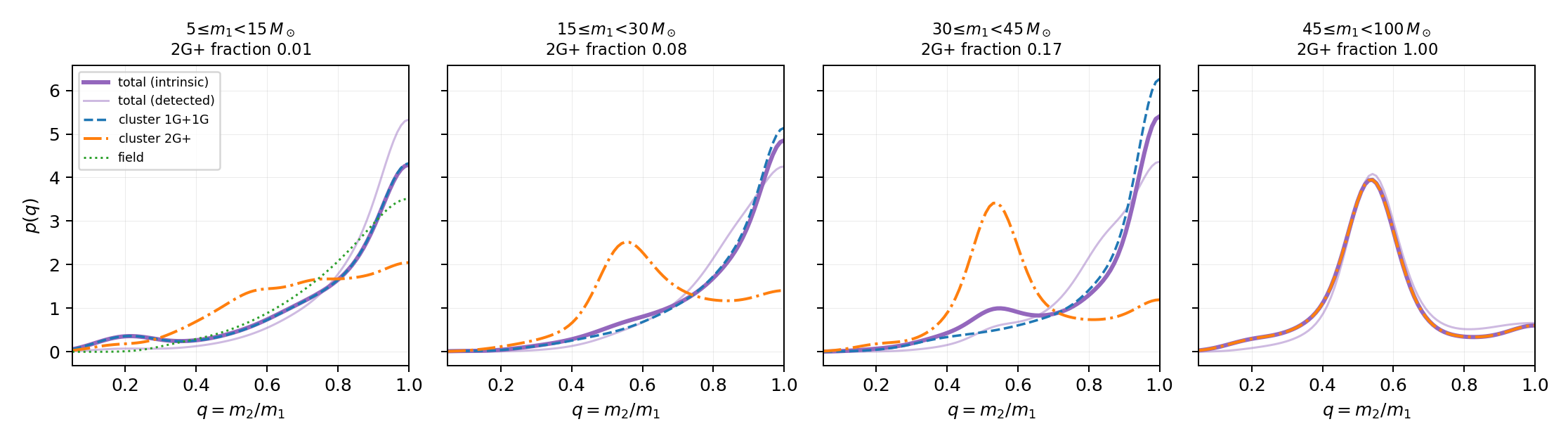}
  \caption{\textbf{The mass-ratio distribution evolves with primary mass---a direct
  prediction of the hierarchical interpretation.} $p(q)$ normalised within each
  primary-mass slice for the preferred cluster$+$field model (purple; faint purple is the
  detected-weighted version), decomposed into non-hierarchical cluster mergers (blue
  dashed), hierarchical $2$G$+$ cluster mergers (orange dash-dot) and the field (green
  dotted). Below the pair-instability edge the census is dominated by near-equal-mass
  pairings and the median $q$ stays near $0.82$. The hierarchical component instead peaks
  near $q\simeq0.55$---a merger remnant is roughly twice the mass of its parents, so
  pairing it with a first-generation companion is intrinsically asymmetric---with a
  secondary peak near $q\simeq1$ from $2$G$+2$G pairings. As the $2$G$+$ fraction grows
  with mass (quoted above each panel) this feature emerges as a shoulder, and above
  $45\,\Msun$, where every merger is hierarchical, it becomes the whole distribution and
  the median $q$ falls to $0.53$. A mass-averaged $p(q)$ such as Fig.~\ref{fig:mixed_q}
  cannot show this.}
  \label{fig:pq_slices}
\end{figure*}

\begin{figure*}
  \centering
  \includegraphics[width=\textwidth]{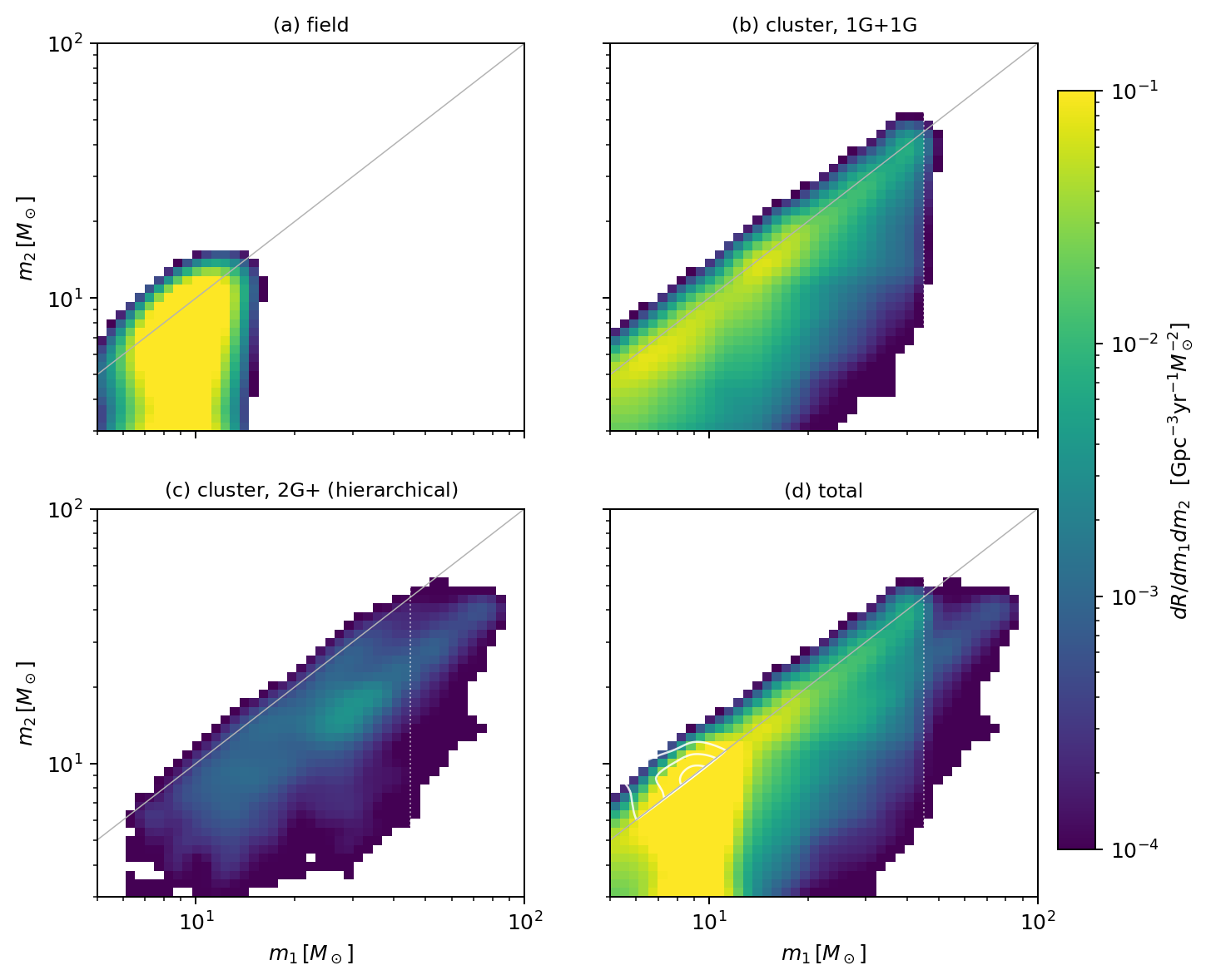}
  \caption{\textbf{The two-channel model decomposed in the mass plane.} Intrinsic
  $dR/dm_1dm_2$ at $z=0.2$ on a shared logarithmic colour scale.
  \emph{(a)} The parametric field component: a sharp, near-equal-mass peak at
  $\simeq\result{FieldTunedMone}\,\Msun$.
  \emph{(b)} Non-hierarchical cluster mergers ($g_1{=}g_2{=}1$), a broad ridge that
  terminates at the pair-instability edge (dotted line, $45\,\Msun$).
  \emph{(c)} Hierarchical ($2$G$+$) cluster mergers---the only component with support
  beyond that edge, where they carry $35\%$ of their rate against $0\%$ for the first
  generation.
  \emph{(d)} The total, with the GWTC-5.0 \texttt{NotchFilter} binned-pairing joint-rate
  median overlaid (white contours; levels set within the plotted window).
  The three components occupy visibly different regions, which is the two-channel
  interpretation of Sec.~\ref{sec:results:split} rendered directly.}
  \label{fig:m1m2_illustration}
\end{figure*}

\subsection{The pair-instability edge from the cluster high-mass population}
\label{sec:results:pisn}
The high-mass population gives only a conditional lower bound on the pair-instability edge, not a localized
measurement. The likelihood rises monotonically to the remnant ceiling of the adopted prescription, and changes by
only a few $\ln\mathcal{L}$, comparable to Monte-Carlo noise. We quote one-sided bounds from
$\Delta\ln\mathcal{L}=1.35$ and $0.50$ ($90\%$ and $68\%$) under a flat prior across the scan. They remain conditional
on the remnant prescription, ceiling, and pile-up treatment. This complements
parametric inferences of \cite{2025PhRvD.112f3040A,2026NatAs.tmp..111A}, who place the
edge at $\simeq44$--$46\,\Msun$ from the first-to-second-generation $\chieff$ transition,
and the second-generation-peak argument of \cite{2026arXiv260407456G}. Unlike those phenomenological fits, we vary the
edge inside the synthesis, allowing retention, hierarchical generations, and spin build-up to respond.

\begin{figure*}
  \centering
  \includegraphics[width=\textwidth]{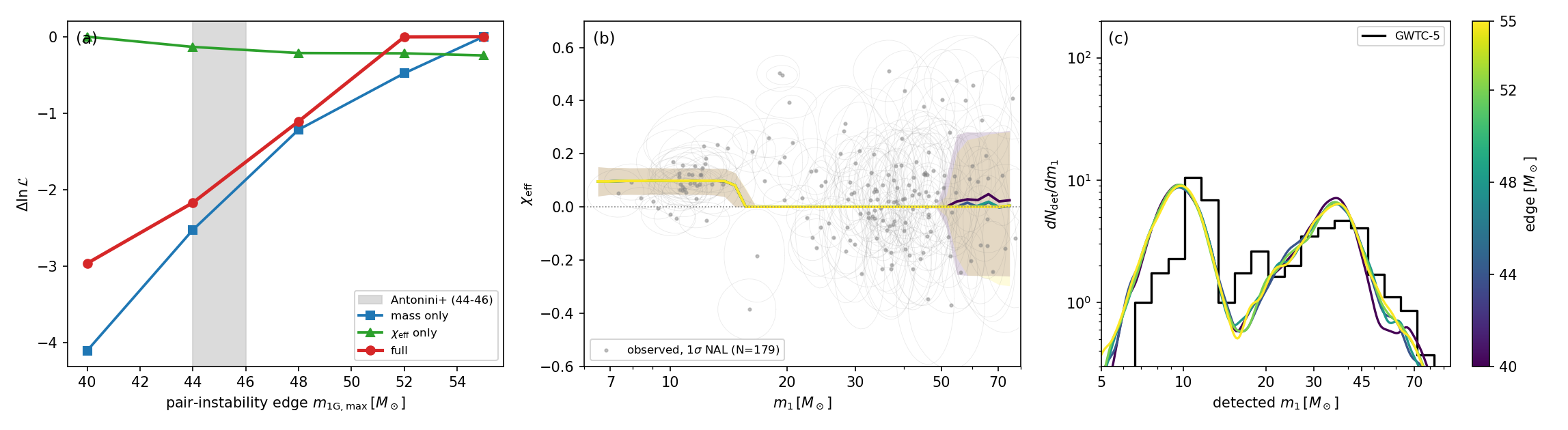}
  \caption{Forward-model constraint on the pair-instability edge (pulsational-PISN pile-up),
  the edge varied self-consistently inside the population synthesis (Spera--Mapelli
  \textsc{sevn} base, $s=\result{FiducialS}$, $r_h=\result{FiducialRh}\,$pc); grey band = the spin-transition measurement of
  \cite{2025PhRvD.112f3040A,2026NatAs.tmp..111A}. \emph{(a)} Two-channel log-likelihood
  versus edge, decomposed: the \emph{mass} term pulls strongly to a high edge (heavier
  second-generation mergers are needed for the $85$--$100\,\Msun$ events), while the
  \emph{$\chieff$} term prefers the low $\simeq44\,\Msun$ edge but far more weakly
  ($\Delta\ln\mathcal{L}\simeq0.2$ vs $\simeq4$), so the \emph{full} likelihood is
  mass-dominated. \emph{(b)} Why the spin cannot arbitrate: the detected-population
  $\chieff$--$m_1$ relation (medians for every edge, colored; $68\%$ bands shaded for the
  extreme edges) shifts its isotropic-broadening onset with the edge---a high edge fills
  $45$--$55\,\Msun$ with low-spin first-generation black holes---but the shift is small
  against the observed per-event uncertainties (gray $1\sigma$ NAL marginal ellipses;
  App.~\ref{app:smear}). \emph{(c)} Best-fit \emph{detected} primary-mass model per edge
  (color, measurement-smeared per App.~\ref{app:smear}) vs the observed point-estimate
  histogram (black). Smearing by the $\simeq9\,\Msun$ high-mass measurement errors
  visibly mutes the edge separation the intrinsic curves suggest---the honest visual
  counterpart of the few-$\ln\mathcal{L}$ discrimination in panel (a): the edge
  constraint rests on the likelihood's mass term, not on a feature visible by eye.}
  \label{fig:pisn_edge}
\end{figure*}

We scan the \textsc{sevn} edge from $40$ to $55\,\Msun$ at
$(s,r_h)=(\result{FiducialS},\result{FiducialRh}\,{\rm pc})$ with the two-channel O4 NAL likelihood. Pulsational
pair instability piles remnants at the edge, so the first-generation spectrum peaks there and the detected turnover
tracks it. The likelihood reaches a shallow plateau above $\simeq52\,\Msun$ (Fig.~\ref{fig:pisn_edge}): the edge is
bounded only from below, $\gtrsim\result{PisnEdgeLL}$ ($90\%$; $\gtrsim\result{PisnEdgeLLtight}$
at $68\%$, the lowest edge disfavored by $\Delta\ln\mathcal{L}\simeq\result{PisnLowEdgeDlnL}$),
railing to the \textsc{sevn} ceiling $\simeq\result{PisnPlateau}$, above the spin-transition value.

The mass term gains $\Delta\ln\mathcal{L}\simeq4$ from $40$ to $55\,\Msun$: a higher first-generation edge produces
the heavier second-generation mergers needed for the few $85$--$100\,\Msun$ events. The spin term instead prefers
$\simeq44\,\Msun$~\cite{2025PhRvD.112f3040A,2026NatAs.tmp..111A}, exactly as expected: a high
edge fills $45$--$55\,\Msun$ with low-spin first-generation black holes, suppressing the expected broad $\chieff$.
That preference is only $\Delta\ln\mathcal{L}\simeq0.2$: detections are sparse and their spin uncertainties are
$\sigma_{\chieff}\simeq0.2$, so mass dominates. Sharper inference requires better high-mass spins or transverse-spin
information. Delete-truncation erases even the mass signature and gives a flatter rail (App.~\ref{app:pisn_truncate});
we retain pile-up as the physical pulsational-PISN outcome.

\subsection{A field-fed second generation}
\label{sec:results:fieldfed}
We introduce the field-fed second generation as a diagnostic ansatz, not because a valid likelihood comparison demands
it. The limitation lies in our simulation resolution, not in the data.

\paragraph*{One reference fit.}
All quoted counts, rates, and widths come from one reference fit: the fiducial cluster ensemble, tuned field, and
reprocessed Gaussian, with three jointly maximized amplitudes and the all-mass four-dimensional O4 likelihood. The
reprocessed template uses $10^7$ samples; smaller samples fail on the narrow kernels that motivate it and shift both
its count and spin width. We label separately any range from lower-resolution refits across $(s,r_h)$.

\paragraph*{Why we do not quote an evidence for this component.}
We cannot robustly compare models with and without this component because the baseline cluster evidence is unresolved
for the decisive events. Finite importance-sampling sums cannot distinguish zero model probability from insufficient
simulation coverage, and their logarithms are biased low. Almost all apparent gain comes from unresolved events; where
the cluster is well sampled, the third component is mildly disfavored. Any quoted evidence would therefore measure
sampling depth rather than the census.

\begin{table}
  \centering
  \small
  \setlength{\tabcolsep}{3pt}
  \caption{\textbf{Intermediate-mass, strongly-aligned events, and which component the
  three-component fit assigns them to.} The $\result{GapN}$ events of
  $\result{GapNslice}$ in $\result{GapMoneLo}$--$\result{GapMoneHi}\,\Msun$ with
  $\chieff>\result{GapChiMin}$. Assigned fractions are $R_c c_j:R_f d_j:R_g e_j$ at the
  maximum-likelihood amplitudes (cluster fractions, $\simeq0.1$ or below, are omitted).
  The last column is the effective sample size behind the reprocessed evidence: it
  qualifies the assignment, and for GW241011 it shows the assignment cannot be
  believed.}
  \label{tab:gap}
  \begin{tabular}{lccccc}
\hline\hline
Event & $m_1/\Msun$ & $\chieff$ & \multicolumn{2}{c}{assignment} & reproc. \\
 & & & field & reproc. & ESS \\
\hline
GW241113\_16 & 24.2 & $+0.53$ & 0.00 & 0.94 & 6954 \\
GW241011\_23 & 19.7 & $+0.51$ & 0.00 & 1.00 & \textbf{1.0} \\
GW231118\_00 & 25.0 & $+0.42$ & 0.00 & 0.78 & 18309 \\
GW231118\_09 & 21.8 & $+0.31$ & 0.97 & 0.00 & 10781 \\
\hline
\end{tabular}

\end{table}

\paragraph*{Which events actually require it.}
Among the
$\result{GapNslice}$ events in $\result{GapMoneLo}$--$\result{GapMoneHi}\,\Msun$,
$\result{GapN}$ have $\chieff>\result{GapChiMin}$ (Table~\ref{tab:gap}). All lie
$\gtrsim\result{GapFieldSigMone}\sigma$ above the field component's mass peak and
$\gtrsim\result{GapFieldSigChi}\sigma$ above its spin peak; the field template has no samples in their mass--spin region.

Description and mixture assignment differ. On shape alone, the field density is
$\result{GapShapeDecadesMin}$--$\result{GapShapeDecadesMax}$ orders of magnitude below the reprocessed component for
these events. Yet assignment weights density by rate, so the much more abundant field still absorbs two of four. That
reflects abundance, not a good event-level description.

GW241113 is assigned mainly to the reprocessed component with resolved evidence. GW241011 is assigned entirely to it,
but every component is unresolved for that event, so the assignment is not interpretable.

The more robust result is collective: the component raises
$\sigma_{\chieff}(\result{GapMoneLo}$--$\result{GapMoneHi}\,\Msun)$ from
$\result{FitSigchiTwo}$ to $\result{FitSigchiThree}$, against an observed
$\simeq\result{ChiIsoSigmaInt}$, although its rate is fit to the full catalog rather than this width. Its reference-fit
detected budget is
$\result{FitNreproc}$ of
$\result{NevScored}$ events, alongside $\result{FitNcluster}$ from clusters and
$\result{FitNfield}$ from the field. These are summed assignment probabilities, not identified members. The observed
$\simeq\result{ChiIsoSigmaInt}$ is a single-Gaussian width across $15$--$30\,\Msun$ (App.~\ref{app:chieff_iso}),
compressing a narrow bulk and large-$|\chieff|$ outliers into one value. It is therefore a reference, not a tuning target.

The ansatz supplies large spins with isotropic orientations at intermediate mass, matching events of both $\chieff$
signs better than an aligned field. We fit its rate but attach no evidence and do not claim that the census requires
this realization over another component occupying the same region.

The relevant joint signature is $q\simeq0.4$--$0.6$ and $\chieff\simeq+0.2$--$+0.5$. Pairing a rapidly spinning merger
remnant with a first-generation black hole naturally produces both asymmetry and large spin. Neither feature is unique,
but their conjunction is informative and requires the mass-ratio dimension of the event kernels. The component chiefly
supplies intermediate-mass $\chieff$ width, not any single event.

\begin{figure}
  \centering
  \includegraphics[width=\columnwidth]{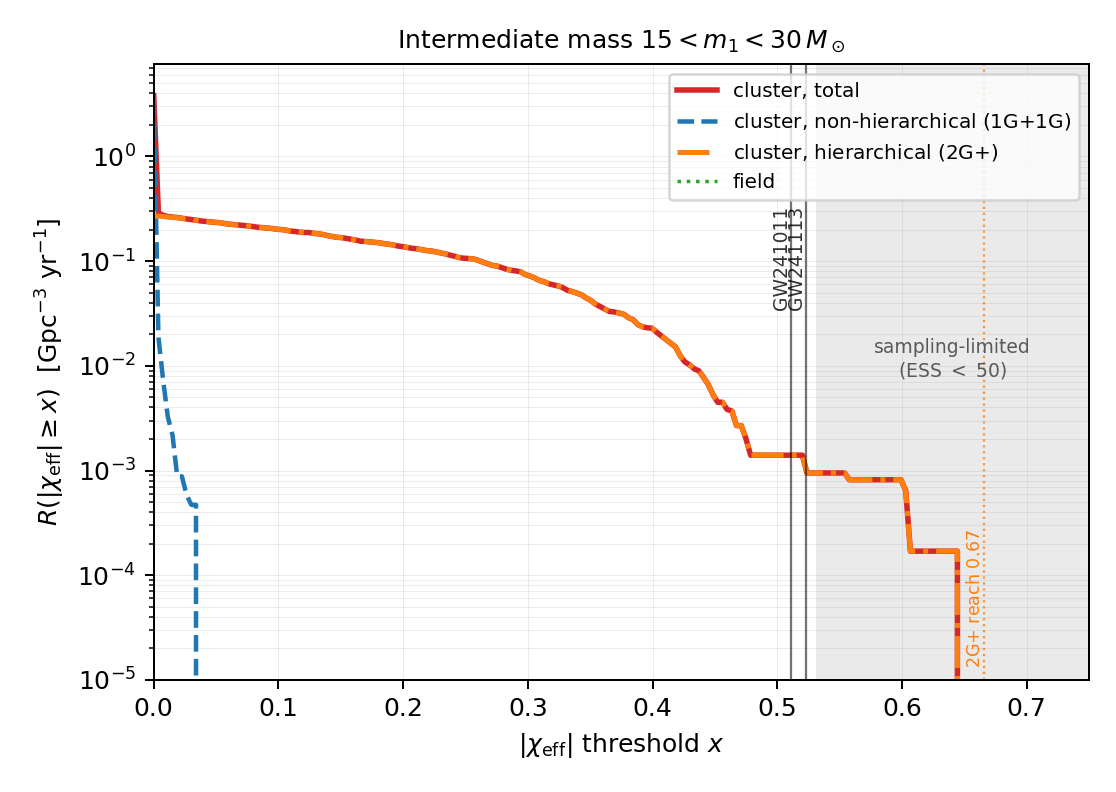}
  \caption{\textbf{The intermediate-mass effective-spin tail, by generation.} Intrinsic
  merger rate with $|\chieff|$ above a threshold, for cluster mergers with
  $15<m_1<30\,\Msun$. The non-hierarchical ($g_1{=}g_2{=}1$) population (blue dashed)
  holds $93\%$ of the rate at $x=0$ but terminates near $|\chieff|\simeq0.05$, as zero
  natal spin requires; the hierarchical $2$G$+$ population (orange dash-dot) is only
  $7\%$ of the rate yet supplies the entire tail beyond that, reaching
  $|\chieff|\simeq\result{ClusterReachChi}$, so the cluster total (red) coincides with it
  above $\simeq0.05$. The tuned field component (green dotted) is confined to small
  aligned spins. Vertical lines mark the two strongly-aligned intermediate-mass events,
  GW241113 and GW241011; both lie within the hierarchical tail. GW241113 carries the
  evidence for the field-fed component (Sec.~\ref{sec:results:fieldfed}); GW241011 is
  drawn for reference only, since it is excluded from that comparison---too few simulated
  binaries fall under its very narrow measurement kernel for its contribution to be
  estimated. The shaded
  region is where the Kish effective sample size of the simulated tail falls below $50$
  ($\simeq14$ binaries above $0.6$): there the staircase is finite Monte-Carlo sampling of
  a rare population, not structure, and the rate is uncertain by a factor of a few.}
  \label{fig:chieff_tail}
\end{figure}

Figure~\ref{fig:chieff_tail} exposes the tail hidden beneath the first-generation bulk in Fig.~\ref{fig:chieff_m1}.
Hierarchical mergers are only $7\%$ of the cluster rate in this band but supply all support above
$|\chieff|\simeq0.05$, extending three decades to $\result{ClusterReachChi}$. Both aligned events lie in this rare
$\sim10^{-3}\,$Gpc$^{-3}$yr$^{-1}$ tail, so clusters do not forbid them and a dedicated component yields little robust
evidence gain.

Monte-Carlo resolution fails near the events: the tail's Kish effective sample size falls below $50$ at
$|\chieff|\simeq0.53$, and only $\simeq14$ simulated binaries exceed $0.6$. The tail exists robustly below
$|\chieff|\lesssim0.45$, but event-level rates and the apparent gain may shift by a factor of a few. The field-fed
component is therefore motivated but not sharply constrained.

The intermediate-mass ($15$--$30\,\Msun$) effective-spin width rises from
$\sigma_\chi\simeq\result{SigmaChiTwoComp}$ to $\simeq\result{SigmaChiThreeComp}$ in the
reference fit. Lower-resolution refits across $(s,r_h)$ vary by
$\lesssim\result{SigmaChiThreeCompSpread}$. This approaches but does not reach the intrinsic
$\simeq\result{ChiIsoSigmaInt}$ inferred from event likelihoods (App.~\ref{app:chieff_iso}); because that value fits one
Gaussian to a mixture, it is an upper reference rather than a target. Figure~\ref{fig:field_fed_2g} shows the symmetric
broadening. The component remains a phenomenological stand-in: unlike the self-consistent calculation of
Ref.~\cite{2026arXiv260614846O}, \Rapster{} does not reprocess primordial-binary remnants here, and the component's
mass, spin, and rate are free. Its success nevertheless motivates that physical extension.

The broadening does not overproduce intermediate-mass mergers. A second generation near
$\result{FieldFedTwoGPeakMone}$ is several times more detectable than the $\sim\result{FieldTunedMone}\,\Msun$ field,
so its $\simeq\result{FitNreproc}$ assigned detections imply only
$R_{\rm 2G}^{\rm ff}\simeq\result{FieldFedTwoGR}\,\mathrm{Gpc^{-3}\,yr^{-1}}$
versus $\simeq\result{FieldFedFieldR}$ for the field. Its $dR/dm_1$ peaks at
$\simeq\result{FieldFedTwoGPeak}\,\mathrm{Gpc^{-3}\,yr^{-1}\,\Msun^{-1}}$, an order of magnitude below the LVK
$90\%$ ceiling; the total remains inside the LVK band (Fig.~\ref{fig:field_fed_dRdm1}).

\section{Discussion}
\label{sec:discussion}

\subsection{Multi-channel model required}
Our \Rapster{} cluster population cannot reproduce the full GW census. It lacks
the $\simeq\result{FieldTunedMone}\,\Msun$ primary-mass peak without
overproducing $25$--$30\,\Msun$ binaries, and isotropic spins cannot produce the
positive low-mass $\chieff$ mean. Forced to fit alone, it also rails toward
$r_h=0.1\,$pc and half-mass densities $\gtrsim10^7\,\Msun\,$pc$^{-3}$. A
single-code account of the full population
\cite{2026arXiv260614472F,2025PhRvL.134a1401A} must therefore confront both the
extreme compactness and the missing aligned-spin component.

Adding a field (isolated-binary) population resolves both failures and improves
the fit by $\Delta\ln\mathcal{L}\simeq\result{FieldSplitDlnL}$ for one
parameter. It supplies the low-mass aligned peak and
$\simeq\result{FieldSplitFrac}$ of detections; clusters supply the high-mass,
symmetric-$\chieff$ population. The resulting cluster parameters---near-zero
natal spin, $r_h\simeq\result{ShapeGridRhLo}$--$\result{FiducialRh}\,$pc, and
$\hatfgc\simeq\result{HatFgc}$---are compatible with dense-cluster initial
conditions and galaxy-based GC-formation models
\cite{2026A&A...708A.364A,2019MNRAS.482.4528E}.

\subsection{Comparison with cluster simulations}
The inferred local cluster rate,
$R(z{=}0.2)\simeq\result{RateMatched}\,$Gpc$^{-3}$yr$^{-1}$, agrees with
semi-analytic and direct predictions: $7.2$~\cite{2020PhRvD.102l3016A},
$4$--$8$~\cite{2022MNRAS.511.5797M}, and $\simeq12$ from Dragon-II
\cite{2024MNRAS.528.5140A}. All are subdominant to the total rate. Direct
simulations likewise require dense clusters
($\rho_{h,0}\gtrsim10^4\,\Msun\,$pc$^{-3}$) for the hierarchical tail
\cite{2023MNRAS.522..466A}, and predict that its fraction falls from $>10\%$ at
$\chi_{\rm birth}\simeq0$ to $\sim1\%$ at $0.5$
\cite{2019PhRvD.100d3027R} because recoil expels remnants from shallow potentials
\cite{2021NatAs...5..749G}. Our intrinsic
$\simeq\result{HierFracIntrLo}$--$\result{HierFracIntrHi}$ and detected
$\simeq\result{HierFracDetLo}$--$\result{HierFracDetHi}$ hierarchical fractions
are consistent with these calculations
\cite{2019PhRvD.100d3027R,2022MNRAS.511.5797M}.

Branching analyses have inferred a $\sim50\%$ dynamical share of detections
\cite{2021ApJ...910..152Z}, or more recently $\simeq79\%$ isolated,
$\simeq15\%$ globular-cluster, and $\simeq2.5\%$ higher-generation
\cite{2026ApJ..1005L..55R}. Both support a field-dominated rate with a smaller
hierarchical component. Clusters also struggle to form the
$\simeq\result{FieldTunedMone}\,\Msun$ peak
\cite{2023MNRAS.522..466A}. Full-lifetime simulations with realistic primordial
binaries now recover the GWTC-5.0 mass and spin distributions with coexisting
primordial and dynamically assembled populations
\cite{2026arXiv260614846O}.

Our one-sided pair-instability bound ($\gtrsim\result{PisnEdgeLL}$ at $90\%$) is consistent with the mass-based
$\gtrsim57\,\Msun$ bound~\cite{2025arXiv251018867R}. The scan spans only
$\simeq3\,\ln\mathcal{L}$, however, and does not exclude the $44$--$46\,\Msun$ spin-transition estimate
\cite{2025PhRvD.112f3040A}; other mass fits also favor a lower edge~\cite{2025arXiv250904151T}.
Masses pull the edge high, whereas aligned-spin information pulls it low.

\subsection{Consistency with \emph{Gaia}'s dormant black holes}
The field channel makes an independent prediction: a population of
$\simeq8$--$10\,\Msun$ black holes formed from isolated binaries. This is precisely
the mass scale of the dormant black holes found astrometrically by \emph{Gaia}---Gaia
BH1 ($\simeq9.6\,\Msun$;~\cite{2023MNRAS.518.1057E}) and Gaia BH2
($\simeq8.9\,\Msun$;~\cite{2023MNRAS.521.4323E}), both orbiting ordinary stars and
naturally attributed to isolated binary evolution. Their black-hole masses coincide
with our tuned field peak ($m_1\simeq\result{FieldTunedMone}\,\Msun$), an encouraging cross-check from a
wholly independent observational channel. The metal-poor, higher-mass Gaia
BH3~\cite{2024A&A...686L...2G} ($\simeq33\,\Msun$) instead sits in the regime our
low-metallicity cluster component populates.

\subsection{Formation efficiency per unit star-forming mass}
The two normalizations imply BBH-merger formation efficiencies that differ sharply
between channels. The cluster yield is computed directly from the simulated clusters,
giving a time-integrated efficiency $\epsilon_{\rm cl}\simeq10^{-4}\,\Msun^{-1}$
(roughly one merger per $10^{4}\,\Msun$ of cluster stars, consistent with semi-analytic
cluster suites~\cite{2020PhRvD.102l3016A}). The field efficiency follows from its local
rate by deconvolving the cosmic star-formation history with a $\propto1/\tau$
delay-time distribution---necessary because the rate is redshift-dependent and
present-day mergers trace earlier, higher star formation---giving
$\epsilon_{\rm field}\simeq8\times10^{-7}\,\Msun^{-1}$ per unit field star-forming
mass. Dynamical assembly is thus $\sim10^{2}\times$ more efficient \emph{per unit
cluster mass}; but because only $\hatfgc\simeq\result{FieldSplitFgc}$ of star formation occurs in dense
clusters, the two channels contribute comparably \emph{per unit total star-forming
mass} ($\sim4$ vs $\sim8\times10^{-7}\,\Msun^{-1}$), with the field dominating the total
local rate ($R_{\rm field}\simeq\result{Rfield}$ vs
$R_{\rm cl}\simeq\result{Rcl}\,$Gpc$^{-3}$yr$^{-1}$). The
field value is an order-of-magnitude estimate, sensitive to the assumed delay-time
distribution and to attributing all field star formation to the channel.

\subsection{Cluster-model simplifications}
Our synthetic universe is population-averaged, and four simplifications are worth
flagging against galaxy-simulation cluster suites~\cite{2026A&A...708A.364A,
2019MNRAS.482.4528E}, in rough order of importance. (i)~\emph{Metallicity floor.}
We impose a sharp metallicity floor at $0.007\,Z_\odot$, already below the
$\sim0.01\,Z_\odot$ reached by CMC and the GAMESH-coupled model; since the lowest-$Z$
bin sets the maximum 1G black-hole mass, this floor is low enough to populate the
$>45\,\Msun$ tail (Sec.~\ref{sec:results}), and pushing it lower has diminishing returns
because metal-poor star formation is rare in the MDF. The residual idealization is the
\emph{sharp} cutoff and the discrete $Z$ grid rather than the floor value. (ii)~\emph{Separable MDF.} Our weight
[Eq.~\eqref{eq:wk}] integrates a global MDF over all formation redshifts, decoupling
$Z$ from $z_f$ within each bin; physically, metal-poor clusters form preferentially
early, so the joint $p(z_f,Z)$ should be retained (assigning each bin the
\emph{conditional} formation history $\propto\psi_{\rm SFR}(z)|dt/dz|\,
\mathcal{N}(\log Z_k;\mu_Z(z),\sigma_Z)$). Because $p_{\rm det}$ falls steeply with
$z$, this can bias the detected high-mass yield. (iii)~\emph{Single half-mass
radius.} We infer one characteristic $r_h$, whereas the hierarchical (2G$+$) yield
depends steeply on the escape velocity $\propto(M_{\rm cl}/r_h)^{1/2}$ and hence on
the low-$r_h$ tail of the $r_h$ distribution~\cite{2026A&A...708A.364A,
2020PhRvD.102l3016A}; sampling $r_h$ (rather than fixing it) would broaden the
high-mass $\chieff$ and lengthen the 2G tail. (iv)~\emph{Constant $\fgc$.} A single
amplitude assumes cluster-formation efficiency is independent of host galaxy and
redshift, unlike the environment-dependent $\Gamma(\Sigma_{\rm SFR})$ of
\cite{2026A&A...708A.364A,2019MNRAS.482.4528E}; this is correct in total mass but
not in its distribution over $(z,Z)$, which enters the shape. None of these is fatal
to the high-mass cluster interpretation, and each is tractable.

\subsection{Robustness to the remnant and pair-instability prescription}
\label{sec:disc:remnant}
The first-generation black-hole spectrum, and with it the pair-instability edge,
depends on the adopted remnant-mass model---precisely the modeling freedom most debated
in the pair-instability literature. \Rapster\ exposes this as a selectable,
metallicity-dependent prescription, so we re-ran the representative cluster ensemble
($s=\result{FiducialS}$, $r_h=\result{FiducialRh}\,$pc) under three: the fiducial \cite{2012ApJ...749...91F} delayed
model (lower pair-instability edge $\simeq45\,\Msun$), its rapid variant, and the
Spera--Mapelli \textsc{sevn} model \cite{2015MNRAS.451.4086S} (edge $\simeq55\,\Msun$).
The first-generation edge moves as expected ($43\to55\,\Msun$), but the \emph{observable}
cluster predictions are stable: the detected primary-mass median ($36$--$38\,\Msun$) and
the high-mass detected fraction ($f_{m_1>45}=0.17$--$0.19$) change by $\lesssim7\%$, and
the rate-matched ($m_1>30$) $\fgc$ by $\lesssim16\%$---all within the Monte-Carlo and
rate-normalization uncertainties. The most model-sensitive quantity is the
intermediate-mass $\chieff$ width ($\sigma_{\rm int}\simeq0.03$ for the $55\,\Msun$ edge
versus $\simeq0.08$ for the fiducial, since a higher edge leaves more of the
intermediate-mass population first-generation rather than hierarchical); but it remains
far below the observed $\simeq\result{ChiIsoSigmaInt}$ (App.~\ref{app:chieff_iso}) in every case,
so the conclusion that the dynamical
channel cannot by itself supply the intermediate-mass spin width---and the consequent
need for an aligned field component---is unchanged. The stellar IMF enters more weakly
still: the high-mass spectrum is pair-instability--truncated below the $150\,\Msun$ upper
IMF limit, so it is insensitive to the maximum stellar mass, while the high-mass slope
rescales the number of massive stars (and hence $\fgc$) without altering the spectral
shape. The headline result---a hierarchical cluster population supplying the
$>45\,\Msun$ events---is therefore robust to these modeling choices.

\subsection{The inferred compactness in the cluster context}
\label{sec:compactness}
In the cluster-only fit, the likelihood increases down to our $r_h=0.10\,$pc
grid edge without an interior maximum. Here $r_h$ is the \emph{natal} half-mass
radius, when density most strongly controls binary formation and hardening
\cite{2020PhRvD.102l3016A,2020MNRAS.492.2936A}, not a present-day radius. Values
$r_h\simeq0.2$--$0.25\,$pc are compact but close to ranges used in CMC
\cite{2020ApJS..247...48K}, GAMESH~\cite{2026A&A...708A.364A}, and MOCCA
\cite{2017MNRAS.464L..36A}, and to embedded-cluster relations
\cite{2012A&A...543A...8M} and dense young clusters such as R136 and the Arches
\cite{2010ARA&A..48..431P}. The field-inclusive grid extends to
$r_h=\result{FieldGridRhLo}\,$pc, but Monte-Carlo scatter prevents localization of an interior optimum.

Present Galactic globular clusters are larger ($\simeq2$--$4\,$pc)
\cite{1996AJ....112.1487H,2018MNRAS.478.1520B} because relaxation, early mass
loss and gas expulsion~\cite{2007MNRAS.380.1589B}, tidal stripping
\cite{2003MNRAS.340..227B}, and black-hole-subsystem heating
\cite{2013MNRAS.432.2779B,2008MNRAS.386...65M,2020MNRAS.492.2936A} erase the
dense birth state.

The inferred compactness implies a half-mass density
$\rho_h=3M_{\rm cl}/(8\pi r_h^3)\sim10^{6}$--$10^{7}\,\Msun\,$pc$^{-3}$ for
$M_{\rm cl}\sim10^{5}$--$10^{6}\,\Msun$---two to four orders of magnitude above the
fiducial initial GC density $\sim10^{4}\,\Msun\,$pc$^{-3}$~\cite{2020PhRvD.102l3016A}.
The associated escape velocity,
\begin{equation}
  v_{\rm esc}\approx40\,{\rm km\,s^{-1}}
  \left(\frac{M_{\rm cl}}{10^{5}\Msun}\right)^{1/3}
  \left(\frac{\rho_h}{10^{5}\Msun{\rm pc^{-3}}}\right)^{1/6}
  \propto\left(\frac{M_{\rm cl}}{r_h}\right)^{1/2}
  \label{eq:vesc}
\end{equation}
\cite{2016ApJ...831..187A}, rises to $\sim60$--$200\,{\rm km\,s^{-1}}$ at
$r_h\simeq0.2$--$0.25\,$pc, versus only $\sim15$--$50\,{\rm km\,s^{-1}}$ for
present-day-sized clusters. This is why the data favor compact clusters:
gravitational-wave recoil kicks are tens to a few hundred $\,{\rm km\,s^{-1}}$ for the
low-spin, near-equal-mass mergers we infer~\cite{2008PhRvD..77d4028L}, and retaining a
second-generation (2G) remnant requires $v_{\rm esc}\gtrsim50\,{\rm km\,s^{-1}}$, with
an \emph{efficient} hierarchical channel needing
$\gtrsim100\,{\rm km\,s^{-1}}$~\cite{2019PhRvD.100d1301G,2021NatAs...5..749G,
2019PhRvD.100d3027R}. Compact clusters clear these thresholds and so retain the
recoiling 2G remnants that build the high-mass population the data require; diffuse
clusters do not.

If this edge preference survived a multi-channel fit, it would point toward
nuclear star clusters (NSCs) or extreme young massive clusters. NSCs reach
$10^6$--$10^8\,\Msun\,$pc$^{-3}$~\cite{2020A&ARv..28....4N} and support longer
hierarchical chains than GCs, where black-hole heating commonly halts growth by
the third generation~\cite{2024A&A...688A.148T}. Simulations confirm enhanced
remnant growth in NSCs~\cite{2016ApJ...831..187A,2019MNRAS.486.5008A,
2021MNRAS.505..339M,2026ApJ...998..138M}, approaching gas-assisted AGN-disk
formation in the deepest nuclei~\cite{2018ApJ...866...66M,2019PhRvL.123r1101Y}.
Adding the field flattens the $r_h$ likelihood, and the denser extensions do not resolve a turnover within
Monte-Carlo scatter (Sec.~\ref{sec:results:split}). The result is a weak compactness constraint consistent with
ordinary globular-cluster birth radii, not a requirement for NSCs. Moreover, $r_h$ is an effective population
parameter, not the radius of any observed cluster.

\subsection{The second-generation rate as a reprocessing efficiency}
\label{sec:fieldfed_rate}
The fitted field-fed rate is
$R_{\rm 2G}^{\rm ff}\simeq\result{FieldFedTwoGR}\,\mathrm{Gpc^{-3}\,yr^{-1}}$,
distinct from the cluster channel's
$R_{\rm 2G}^{\rm clu}\simeq\result{TwocompTwoGR}\,\mathrm{Gpc^{-3}\,yr^{-1}}$.
It is $\simeq\result{FieldFedRateFrac}$ of the field rate
$R_{\rm field}\simeq\result{FieldFedFieldR}\,\mathrm{Gpc^{-3}\,yr^{-1}}$,
consistent with the fitted $\result{Rfield}\,\mathrm{Gpc^{-3}\,yr^{-1}}$.
Because each second-generation binary uses two remnants, the implied per-remnant
efficiency is
$\varepsilon=2R_{\rm 2G}^{\rm ff}/R_{\rm field}\simeq\result{FieldFedEpsilon}$.
This is a lower bound because only field binaries in clusters can be reprocessed.

Writing $\varepsilon\simeq f_{\rm ret}f_{\rm pair}$, low-spin, near-equal-mass
mergers receive tens-to-hundreds of $\mathrm{km\,s^{-1}}$ recoil rather than
spin-driven superkicks~\cite{2008PhRvD..77d4028L}. For our
$v_{\rm esc}\sim60$--$200\,\mathrm{km\,s^{-1}}$ clusters, an estimated
$f_{\rm ret}\sim0.3$--$0.7$ combined with a re-pairing probability of a few
tenths naturally gives a few-to-ten percent, consistent with low-natal-spin
cluster simulations~\cite{2021NatAs...5..749G,2019PhRvD.100d3027R}.

Larger efficiencies overproduce $15$--$30\,\Msun$ mergers; smaller ones fail to
broadly reproduce their $\chieff$ distribution. This is nevertheless an
order-of-magnitude check: $f_{\rm ret}$ and $f_{\rm pair}$ are estimated, not
computed. A cluster calculation that evolves primordial binaries and retained
remnants together would predict the ratio directly
\cite{2026arXiv260614846O}.

\subsection{The very-low-spin first generation and its observational status}
\label{sec:lowspin}
A direct prediction is a large very-low-spin population. The preferred $s=0$
pins first-generation mergers near $\chieff=0$ and suppresses recoil; substantial
spin appears mainly in high-mass hierarchical products. Detecting a distinct
zero-spin component is difficult, however, because individual spins are weakly
measured and mixture inferences depend on priors and likelihood normalization
\cite{2024ApJ...964L...6A,gwastro-mergers-hierarchical-genealogy-Kimball2020}.

Formation-informed emulation instead finds a common natal spin
$\chi_b\simeq0.04$, with support from $0$ to $0.1$
\cite{2025ApJ...988..189C}, consistent with our $s\lesssim0.1$ result. Clusters
need not be exactly spinless: black-hole--star collisions and accretion may give
up to $\sim40\%$ of merging binaries $\chi\gtrsim0.2$
\cite{2025ApJ...979..237K}. That omitted channel could soften the $s=0$
preference without erasing the high-mass hierarchical signal.

\subsection{What does an ``isotropic fraction'' measure?}
\label{sec:disc:isotropic}
Phenomenological fits often split spins into aligned and isotropic components
\cite{LIGO-GWTC5-populations-2026,dcc-Tong-Hierarchical-2025,
2026arXiv260614472F,gwastro-pop-Zeeshan-O4aMixture}, sometimes finding a
low-mass spinning, isotropic population
\cite{dcc-Tong-Hierarchical-2025,2026arXiv260700565F}. Their ``isotropic
fraction'' is not a dynamical-channel fraction. When $s\simeq0$, tilt is
unobservable: aligned and isotropic binaries both give $\chieff\simeq0$, and a
mixture assigns them according to its priors. This fragility appears in both
zero-spin mixtures~\cite{2024ApJ...964L...6A} and nonparametric tilt peaks
\cite{2026arXiv260505300W}. The valid comparison is therefore the mass-resolved
$\chieff$ distribution, not the nominal component fraction.

Low-mass \emph{spinning} isotropy is a genuine test: our $s\simeq0$ first
generation cannot produce systems such as GW241011
\cite{LIGO-O4-HierarchicalPair-2025}. The field-fed component instead predicts
$\chi\simeq0.7$ isotropic binaries at $15$--$30\,\Msun$, with rate
$R_{\rm 2G}^{\rm ff}\simeq\result{FieldFedTwoGR}\,$Gpc$^{-3}$yr$^{-1}$
($\simeq\result{FieldFedRateFrac}$ of the field rate), and could explain the
reported low-mass isotropic population
\cite{dcc-Tong-Hierarchical-2025,2026arXiv260700565F}. Measurable isotropy should
then rise from percent-level near $10\,\Msun$ to order unity above
$45\,\Msun$. A larger intermediate-mass isotropic share, or strong high-mass
alignment~\cite{2026PhRvL.137b1407L,2026arXiv260612205A,
2026arXiv260623305R}, would falsify the model. Mild alignment is less decisive
because repeated encounters randomize tilts gradually
\cite{2026arXiv260621691M}.

\section{Conclusions}
\label{sec:conclusions}
We confronted the GW census with physically normalized globular-cluster
populations from \Rapster~\cite{2024PhRvD.110d3023K}, a parametric field
(isolated-binary) population, and a diagnostic population of reprocessed field
remnants. The census requires at least two channels. The field produces the
low-mass, preferentially aligned peak; clusters produce the high-mass,
symmetric-$\chieff$ population through hierarchical growth. This division agrees
with direct cluster calculations
\cite{2026arXiv260614846O,2020PhRvD.102l3016A,2022MNRAS.511.5797M,
2023MNRAS.522..466A,2024MNRAS.528.5140A,2025PhRvL.134a1401A,
2026arXiv260614472F}.

The high-mass data favor low black-hole natal spin. Larger natal spins broaden
the first-generation $\chieff$ distribution and increase recoil, removing the
remnants needed to build the high-mass tail. This conclusion is consistent with
cluster models~\cite{2019PhRvD.100d3027R,2021NatAs...5..749G,
2026arXiv260614846O}, isolated-binary calculations
\cite{2019PhRvD.100d3012W,popsyn-KB-LowNatalBHSpin-2017,
popsyn-gwastro-STInterpFinal-Vera2023}, formation-model emulation
\cite{2025ApJ...988..189C}, and efficient angular-momentum transport in massive
stars~\cite{starev-Fuller-LowNatalSpin}. It is independent of the low-mass
argument for a field channel.

Compactness requires a two-channel interpretation. A cluster-only fit rails to
the densest grid edge, improving by
$\simeq\result{ShapeDropAtLargest}\,\ln\mathcal{L}$ from the diffuse end. With
the field included, the variation shrinks to
$\simeq\result{FieldGridRhSpan}\,\ln\mathcal{L}$, comparable to Monte-Carlo
scatter; denser extensions do not resolve a turnover. We therefore obtain only a
weak lower limit, not a preferred natal radius. A free compactness mixture assigns
$\simeq\result{RhMixDiffuseFrac}$ to ordinary diffuse clusters and
$\simeq\result{RhMixDenseFrac}$ to dense cores (App.~\ref{app:chieff_iso}). The
multi-channel result permits ordinary globular-cluster natal compactness and does
not require nuclear-star-cluster conditions.

The rate budget is field dominated. The cluster-only normalization gives
$\hatfgc\simeq\result{HatFgc}$ ($90\%$ interval
$\result{HatFgcLo}$--$\result{HatFgcHi}\%$), while the two-channel fit gives
$\simeq\result{FieldSplitFgc}$. At $z\simeq0.2$, we infer
$R_{\rm field}\simeq\result{Rfield}$
($\result{RfieldLo}$--$\result{RfieldHi}$) and
$R_{\rm cl}\simeq\result{Rcl}$
($\result{RclLo}$--$\result{RclHi}$)~Gpc$^{-3}$yr$^{-1}$: clusters supply
$\result{ClusterFracTotalLo}$--$\result{ClusterFracTotalHi}\%$ of the total,
subject to a common factor-$\sim2$ rate-calibration uncertainty.

Two predictions distinguish this picture. First, $dR/dm_1$ should break near
twice the first-generation maximum mass
\cite{2026arXiv260407456G,2026arXiv260701121L}. Second, measurable isotropy
should rise with mass as spinning remnants become common. Flexible
phenomenological results are broadly compatible
\cite{LIGO-GWTC5-populations-2026,dcc-Tong-Hierarchical-2025,
2026arXiv260614472F,2026PhRvL.137b1404P,2026arXiv260612205A}, but their isotropic
fractions are not channel fractions: near-zero-spin orientations are
unobservable. Our model also omits the multiple low-mass isotropic populations
allowed in some fits~\cite{LIGO-GWTC5-populations-2026,
dcc-Tong-Hierarchical-2025,2026arXiv260614472F,
gwastro-pop-Zeeshan-O4aMixture}; a companion study explores a wider set of natal
populations and interactions
\cite{gwastro-PopulationReconstruct-Hierarchical-PhysicalParam2-Coagulation}.

The optional third component identifies low-mass spinning, misaligned systems
with field remnants later reprocessed in clusters. It could accommodate
GW241011~\cite{LIGO-O4-HierarchicalPair-2025}, otherwise difficult for our
low-natal-spin clusters and variously interpreted as hierarchical
\cite{2026arXiv260704663H} or as a mass-ratio-reversed isolated binary
\cite{2026arXiv260627852H}. Current data do not require this component, so it is
a testable hypothesis rather than a headline inference.

The principal limitations are explicit. The field shape is fixed rather than
inferred; the cluster ensemble uses a sharp metallicity floor, separable
metallicity and formation time, one effective $r_h$, and constant $\fgc$; event
likelihoods approximate full posterior samples; and selection is
spin-marginalized and semianalytic for O1/O2. The measured-mass catalog cut is
scored against a true-mass-truncated model, allowing migration across the
$30\,\Msun$ boundary (median high-mass uncertainty
$\simeq\result{SmearSigmaMoneHigh}\,\Msun$). Redshift is factorized from the
mass--spin kernel, and finite simulation libraries make narrow-kernel evidence
sums noisy and biased low; some such events are excluded. Thus evidence
differences are lower bounds where stated, not precision measurements.

Next steps are a joint $(s,r_h,\fgc)$ and field-shape fit, spin-conditional
selection, full event likelihoods with threshold migration and mass--redshift
covariance, deeper simulation libraries, and a unified calculation that evolves
primordial binaries and retained remnants self-consistently
\cite{2026arXiv260614846O}. These tests can turn the present channel assignment
into a sharper, falsifiable account of how the observed black-hole population
was assembled.

\begin{acknowledgments}
    We thank Vera Delfavero, Christopher Berry, and Suvodip Mukerjee for suggestions on the manuscript.
  This work made use of the \texttt{popsynth\_hyperpipe} framework, including
  its \texttt{response} package for the archived simulation-to-observation
  likelihood of App.~\ref{app:response}; of
  \gwkokab~\cite{gwastro-mergers-zeeshan-gwkokab,code-Zeeshan-gwkokab}; and of
  RIFT's \texttt{simulation\_manager}~\cite{code-RIFT-research-projects-RIT}.
    ROS acknowledges support from  NSF PHY 2012057, 2309172 and 2206321.  ROS also acknowledges assistance from
    Claude/Codex during the implementation, operation, and drafting of the calculations described in this manuscript. The authors are
  grateful for computational
  resources provided by the LIGO Laboratory and supported by National Science
  Foundation Grants PHY-0757058 and PHY-0823459. This material is based upon
  work supported by NSF's LIGO Laboratory which is a major facility fully funded
  by the National Science Foundation. This research has made use of data or
  software obtained from the Gravitational Wave Open Science Center (gwosc.org),
  a service of the LIGO Scientific Collaboration, the Virgo Collaboration, and
  KAGRA. This material is based upon work supported by NSF's LIGO Laboratory
  which is a major facility fully funded by the National Science Foundation, as
  well as the Science and Technology Facilities Council (STFC) of the United
  Kingdom, the Max-Planck-Society (MPS), and the State of Niedersachsen/Germany
  for support of the construction of Advanced LIGO and construction and
  operation of the GEO600 detector. Additional support for Advanced LIGO was
  provided by the Australian Research Council. Virgo is funded, through the
  European Gravitational Observatory (EGO), by the French Centre National de
  Recherche Scientifique (CNRS), the Italian Istituto Nazionale di Fisica
  Nucleare (INFN) and the Dutch Nikhef, with contributions by institutions from
  Belgium, Germany, Greece, Hungary, Ireland, Japan, Monaco, Poland, Portugal,
  Spain. KAGRA is supported by Ministry of Education, Culture, Sports, Science
  and Technology (MEXT), Japan Society for the Promotion of Science (JSPS) in
  Japan; National Research Foundation (NRF) and Ministry of Science and ICT
  (MSIT) in Korea; Academia Sinica (AS) and National Science and Technology
  Council (NSTC) in Taiwan.
\end{acknowledgments}

\appendix

\section{The simulation-to-observation response}
\label{app:response}
All likelihoods reduce to one response function: the simulation's differential
merger yield per unit formed stellar mass,
$r(\svec,t_d\,|\,\theta)=M_\star^{-1}\,d^2N/(d\svec\,dt_d)$, in
source-frame properties $\svec$ and delay time $t_d$, conditioned on
$\theta=(r_h,s,Z)$. Its astrophysical prediction is the Poisson intensity over
observable space $\svecz=(\svec,z)$ on our past light cone,
\begin{multline}
  n_{\rm ast}(\svecz\,|\,\pvec;\theta)=T_{\rm obs}\!\int\!dz_f\;\varphi(z_f;\pvec)\,
  f_{\rm form}(z_f,Z;\pvec)\\
  \times\!\int\!dt_d\;r(\svec,t_d\,|\,\theta)\,V(z)\,
  \delta\big(z-z_m(z_f,t_d)\big),
  \label{eq:response}
\end{multline}
where $\varphi(z_f;\pvec)\equiv\fgc\,\psi_{\rm SFR}(z_f)\,|dt/dz|_{z_f}$
is the natal function---cluster stellar mass formed per comoving volume
per unit redshift [Eqs.~\eqref{eq:pzf} and~\eqref{eq:sigmastar}]---$f_{\rm form}$ an optional metallicity-resolved formation reweighting
(unity for the fiducial history),
$V(z)\equiv[(dV_c/dz)/((1+z)|dt/dz|)]_z$ the
light-cone volume factor, and $z_m(z_f,t_d)$ the
merger redshift at lookback time
$t_{\rm lookback}(z_f)-t_d$, with no support once
$t_d>t_{\rm lookback}(z_f)$ (the merger postdates today). No selection
appears in Eq.~\eqref{eq:response}. The data then
require exactly two contractions of this intensity, against the
detection probability $p_{\rm det}$ and against the per-event
likelihoods $\ell_n$ (Sec.~\ref{sec:methods:likelihood}),
\begin{align}
  \mu(\pvec) &= \int d\svecz\; p_{\rm det}(\svec,z)\,
  n_{\rm ast}(\svecz\,|\,\pvec;\theta),
  \label{eq:mu_response}\\
  I_n(\pvec) &= \int d\svecz\; \ell_n(\svecz)\,n_{\rm ast}(\svecz\,|\,\pvec;\theta),
  \label{eq:In_response}
\end{align}
the first the expected detected count $\Nexp$ of Eq.~\eqref{eq:Nexp},
with $z$ inside $p_{\rm det}$ the merger redshift pinned by the delta
function of Eq.~\eqref{eq:response}. Selection enters $\mu$ and only
$\mu$: for a catalog defined by a threshold on the data, each per-event
likelihood $\ell_n$ already conditions on the event's data having
crossed that threshold, so a $p_{\rm det}$ factor inside $I_n$ would
count the selection twice \cite{2004AIPC..735..195L,LIGO-P1600187-Farr-SelectionBiasaAndMonteCarlo}.
The
Poisson likelihood of Eq.~\eqref{eq:poisson_like} reduces to
$\ln\mathcal{L}=\sum_n\ln I_n-\mu$ up to a $\pvec$-independent constant.
A library of simulations enters as the mixture
$n_{\rm ast}=\sum_\alpha\lambda_\alpha\,n_{\rm ast}(\svecz\,|\,\pvec;\theta_\alpha)$, with
$\lambda_\alpha$ the mixture weights of
Sec.~\ref{sec:methods:hyperpipe} (the channel amplitudes $\fgc$ and
$R_{\rm field}$ occupy the same slot). Every quantity the analysis needs
is thus an integral against $r$; the hyperparameters act only on the
factors---$\varphi$, $f_{\rm form}$, the projections $V$ and
$p_{\rm det}$, the mixture weights---never on
$r$ itself, which is what makes archived post-processing possible.

A Monte Carlo simulation delivers $r$ as a weighted empirical measure
over its catalog rows $\svecz_i$ (masses, spins, merger and formation
redshift) and simulated cluster stellar mass $M_{{\rm sim},\alpha}$,
\begin{equation}
  r_\alpha(\svec,t_d)=\frac{1}{M_{{\rm sim},\alpha}}
  \sum_{i\in\alpha}\delta(\svec-\svec_i)\,\delta(t_d-t_{d,i}),
  \label{eq:remp}
\end{equation}
with $t_{d,i}=t_{\rm lookback}(z_{f,i})-t_{\rm lookback}(z_{m,i})$
(App.~\ref{app:delay}). Substituting into Eq.~\eqref{eq:response}
collapses every integral to a sum: the $t_d$ integral pins each sample's
merger redshift along the replayed formation history, $z_{m,i}(z_f)$,
and each merger acquires the ($z$-marginal) detected weight
\begin{multline}
  w_i(\pvec)=\kappa_i\,T_{\rm obs}\!\int\!dz_f\;p(z_f)\,
  f_{{\rm form},i}(z_f;\pvec)\\
  \times\,p_{\rm det}\big(\svec_i,z_{m,i}(z_f)\big)\,
  V\big(z_{m,i}(z_f)\big),
  \label{eq:wdet_time}
\end{multline}
together with its selection-free counterpart $w^{\rm ast}_i$, the same
expression with the $p_{\rm det}$ factor omitted,
where the comoving density $\kappa_i$ of clusters like merger $i$'s host
[Eq.~\eqref{eq:kappak}] and the formation distribution $p(z_f)$
[Eq.~\eqref{eq:pzf}] tabulate the natal function per unit simulated mass,
apportioned over the metallicity grid:
$\kappa_i\,p(z_f)=w_k\,\varphi(z_f;\pvec)/M_{{\rm sim},k}$
[Eq.~\eqref{eq:wk}]. For this paper's \Rapster\ catalogs the realized
pair $(z_{f,i},z_{m,i})$ is one Monte Carlo draw from the formation
history, and scoring each merger only at that
epoch---the one-draw estimate of Eq.~\eqref{eq:wdet_time}, a response
that is a delta function in formation time---gives the per-merger
detected weight used throughout this paper,
\begin{equation}
  w_i(\pvec)=\kappa_i\,f_{{\rm form},i}(\pvec)\,p_{{\rm det},i}\,
  V(z_{m,i})\, T_{\rm obs},
  \label{eq:wdet}
\end{equation}
the summand of Eq.~\eqref{eq:Nexp}, again with a selection-free
counterpart $w^{\rm ast}_i$ lacking the $p_{{\rm det},i}$ factor. The
contractions
[Eqs.~\eqref{eq:mu_response} and~\eqref{eq:In_response}] become the Monte
Carlo estimators
$\mu=\sum_\alpha\lambda_\alpha\sum_{i\in\alpha}w_i(\pvec)$ and
$I_n=\sum_\alpha\lambda_\alpha\sum_{i\in\alpha}w^{\rm ast}_i(\pvec)\,
\ell_n(\svecz_i)$, the empirical image of the selection split above. The
per-merger selection and volume factors are archived as separate columns
alongside the kernel values $\{\ell_n(\svecz_i)\}$, once per reference
simulation---both weight sets derive from the same frozen arrays, and
scoring a single simulation at fixed $\pvec$ is one contraction of
them---and mixture
inference over the archived library is closed-form in the weights,
\begin{equation}
  \ln\mathcal{L}(\{\lambda_\alpha\})=\sum_n\ln\sum_\alpha\lambda_\alpha
  e^{\ln I_{n,\alpha}}-\sum_\alpha\lambda_\alpha\mu_\alpha,
  \label{eq:mixlnL}
\end{equation}
with $\ln I_{n,\alpha}$ and $\mu_\alpha$ archived per simulation, so the
amplitude and simplex fits of Sec.~\ref{sec:methods:fit} reduce to cheap,
exact-gradient optimizations over a few hundred archived simulations,
with no re-simulation. Retaining the $z_f$ integral of
Eq.~\eqref{eq:wdet_time} is one additional contracted axis---frozen
per-(merger, epoch) selection and volume factors contracted against
formation-history weights---while the kernel values $\ell_n(\svecz_i)$
depend only on $\svec_i$ and are untouched, so the estimators stay
closed-form and differentiable. Analytic ingredients---a delay-time
distribution $p(t_d\,|\,\svec;\pvec)$, or an analytic $r$ from a
parametric channel---enter by replacing the empirical measure inside the
same integrals: the factorization, not the sampling, is the contract.

One bookkeeping choice in this construction must be recorded, because
it differs from the convention widely used previously: that
convention contracted the per-event kernels against the detected weights
$w_i$ themselves---$p_{\rm det}$ inside $I_n$ as well as $\mu$. Were
every kernel $\ell_n$ narrow across $p_{\rm det}$, that convention would
differ from Eq.~\eqref{eq:In_response} only by the $\pvec$-independent
constant $\sum_n\ln p_{\rm det}(\svecz_n)$, leaving every fit unchanged.
With redshift-free $(\mc,\eta,\chieff)$ kernels it is not in that
regime: each kernel is broad in $z$, the coordinate along which
$p_{\rm det}$ falls by orders of magnitude (Fig.~\ref{fig:pdet}). A
dedicated grid comparison of the two conventions---identical detected
$\mu$ in both, so the Poisson rate term cancels exactly---quantified the
consequence: on the production $18$-point $(s,r_h)$ grid the offset
$\ln\mathcal{L}_{\rm det}-\ln\mathcal{L}_{\rm ast}$ is not constant but
correlates with compactness, rising by $\approx26$--$28\,\ln\mathcal{L}$
from $r_h=0.1$ to $0.43\,$pc along the low-spin rows (the correct
placement favors compact clusters), with a ranking-relevant spread of up
to $\result{SelPlaceDistortion}\,\ln\mathcal{L}$ and a
joint-$\ln\mathcal{L}$ rank correlation between conventions of only
$\result{SelPlaceSpearman}$ ($\result{SelPlaceSpearmanOFour}$, O4-only);
only the \emph{shape-only} ranking was convention-robust (Spearman
$0.99$ in both eras). This analysis therefore adopts the selection-free
placement of Eq.~\eqref{eq:In_response} throughout, and adds each
event's redshift (GWOSC median and $90\%$ interval) as a fourth kernel
dimension, which restores the narrow-kernel regime in $z$: with the
$(\mc,\eta,\chieff,z)$ kernels the det--ast offset varies by only
$\result{ZKernOffsetVar}\,\ln\mathcal{L}$
($\result{ZKernOffsetVarOFour}$, O4-only) across the grid---below the
per-point seed scatter---so the placement choice no longer moves any
fit. Under the adopted conventions---this placement together with the
per-merger natal-time MDF weighting of Eq.~\eqref{eq:ri}---the
combined-era joint best fit sits
at $(s,r_h)=(\result{JointBestS},\result{JointBestRh})$, and the shape
surface of Fig.~\ref{fig:posterior} keeps its maximum at the $s=0$,
$r_h=\result{ShapeGridRhLo}\,$pc grid edge.

Two features of this bookkeeping deserve emphasis. The first is the
merger-\emph{redshift} placement in Eq.~\eqref{eq:wdet}, which materially
affects the effective sample size: the redshift range accessible to
observations is narrow (Fig.~\ref{fig:pdet}), so weighting each merger at
its exact $z_{m,i}$ can starve the archive of usable samples, and it
sometimes helps to treat mergers as prompt---placed at their formation
redshift, with no merger-time tracking; App.~\ref{app:delay} quantifies
exactly this choice (detected shapes unchanged at the $10^{-3}$-bit level,
$\Nexp$ and $R_0$ shifted by tens of percent). Retaining the replay
integral of Eq.~\eqref{eq:wdet_time} instead lets each merger contribute
at every formation epoch that lands it inside the narrow accessible
window, rather than once, directly relieving that starvation. The replay
does hold the merger's simulated metallicity fixed while the cosmic MDF
[Eq.~\eqref{eq:pZz}] shifts with $z_f$, so it requires the
metallicity-resolved reweighting factor $f_{{\rm form},i}$ retained in
Eq.~\eqref{eq:wdet_time}---or a $z_f$ window restricted to MDF-consistent
epochs.

The second feature is the amplitude direction: the continuous amplitude
$\fgc$ enters
the weights linearly through $\kappa_i$ [Eq.~\eqref{eq:wdet}], so
$\ln\mathcal{L}(\fgc)$ is analytic along that direction---scans and
gradients are free, with no re-simulation---which is what the no-free-scale
normalization of Sec.~\ref{sec:results:forward} exploits.

The construct is implemented as the \texttt{response} package of the
\texttt{popsynth\_hyperpipe} framework underlying this analysis---including
the formation-history replay of Eq.~\eqref{eq:wdet_time}.

\section{The prompt-merger approximation}
\label{app:delay}
Our rate calculation assigns each merger its host's cosmic-MDF formation weight
[Eq.~\eqref{eq:kappak}] and its \Rapster{} merger redshift, without explicitly
convolving over delay time $t_d$. We test the shape and rate consequences here.

For the fiducial 29-metallicity ensemble we compute
$t_d=t_{\rm lookback}(z_f)-t_{\rm lookback}(z_{\rm merge})$. Its
formation-weighted median is $\simeq\result{DelayMedianMyr}$, and
$\simeq\result{PromptFracInt}$ of mergers
($\simeq\result{PromptFracDet}$ of detections) occur within $1\,$Gyr. Short-
and long-delay halves differ by Jensen--Shannon divergence
$<\result{JsDelayMax}$ in both $m_1$ and $\chieff$, intrinsically and after
selection (Table~\ref{tab:delay}). Early mergers are slightly heavier and more
often hierarchical, but the fitted shapes are insensitive to delay treatment.

Delay treatment primarily affects $R(z)$, not the shape inference. The placement
test below bounds that effect; Eq.~\eqref{eq:wdet_time} gives the full
formation-history replay needed for a self-consistent treatment.

\begin{table}
  \centering
  \caption{\label{tab:delay}\textbf{Merger properties versus formation-to-merger delay time $t_d$.} Intrinsic (MDF-weighted) and, in parentheses, detected weight fraction per delay bin, with the (kap-weighted) median primary mass, effective-spin dispersion, and detected hierarchical ($2$G$+$) fraction. The \emph{conditional} properties do vary appreciably across $t_d$: from the shortest to the longest delay bin the effective-spin dispersion falls by a factor $\simeq2.2$ and the detected hierarchical fraction by a factor $\simeq3.8$, while the median primary mass changes only mildly. What is weak is the \emph{prompt-weighted integral}: because $\simeq82\%$ of mergers occur within $1\,$Gyr, the delay-integrated $dR/dm_1$ and $p(\chieff)$ that we actually fit are insensitive to the trend (Jensen--Shannon divergence between the short- and long-delay halves $<0.04$ bits; App.~\ref{app:delay}). Produced by \texttt{analyses/delay\_time\_check.py}.}
  \begin{tabular}{lcccc}
    \hline\hline
    $t_d$ [Gyr] & frac. (det.) & $m_1^{\rm med}$ [$\Msun$] & $\sigma_{\chieff}$ & hier. frac. \\
    \hline
    $0$--$1$ & 0.82 (0.73) & 19.5 & 0.09 & 0.30 \\
    $1$--$3$ & 0.09 (0.10) & 18.0 & 0.05 & 0.14 \\
    $3$--$6$ & 0.06 (0.10) & 17.3 & 0.04 & 0.10 \\
    $6$--$14$ & 0.03 (0.07) & 17.2 & 0.04 & 0.08 \\
    \hline\hline
  \end{tabular}
\end{table}

\subsection*{A prompt-lite cross-validation of the response machinery}
We use redshift placement as an end-to-end response test. The \emph{prompt-lite}
variant sets $z_{m,i}=z_{f,i}$ in the detected weights and $R(z)$, the zero-delay
limit. The mass cut and $(\mc,\eta,\chieff)$ kernels are unchanged, so this
perturbation isolates the redshift-dependent weighting and contraction.

We score both placements, for the combined O1--O4b high-mass catalog and its
O4-only subset (Sec.~\ref{sec:methods:likelihood}), across the production
$N_Z{=}29$ ensemble posterior grid ($18$ $(s,r_h)$ points), the wider natal-spin
scan grid ($N_Z{=}4$, $25$ points, $s\le0.3$), and $11$ follow-up single universes
scored one at a time through the one-off single-simulation path
(App.~\ref{app:response}); detected shapes are compared on the fiducial
($s{=}0$, $r_h{=}0.25\,$pc) $N_Z{=}29$ ensemble pooled over $13$ seeds ($167$
universes), with the binning conventions of Table~\ref{tab:delay}.
Table~\ref{tab:prompt_lite} collects the outcome. The detected shapes are
essentially untouched: the $dR/dm_1$ and $p(\chieff)$ distributions agree to a
Jensen--Shannon divergence $\le0.0024$ bits---an order of magnitude below even the
weak short-versus-long-delay contrast of Table~\ref{tab:delay}---the
$m_1$--$\chieff$ trend shifts by $|\Delta\langle\chieff\rangle|\le0.002$ per mass
bin, the shape-term ranking over $(s,r_h)$ is preserved (Spearman rank correlation
$0.95$--$1.00$) with the shape best-fit point unchanged (the one exception a
$0.09$-$\ln\mathcal{L}$ tie), and the ranking-relevant shape-term distortion is
$\lesssim5$ in $\ln\mathcal{L}$ on the ensemble grids---well below the
tens-of-$\ln\mathcal{L}$ contrasts that drive our conclusions
(Sec.~\ref{sec:methods:fit}). The quantities that do move are exactly those the
prompt approximation is documented to affect, the rate versus redshift: $\Nexp$
shifts by $10$--$12\%$ (median; $\le19\%$) on the ensemble grids ($24$--$29\%$ on
the seed-limited single universes) and the local rate $R_0$ by $40$--$70\%$
(medians), since relocating mergers to their higher formation redshift pushes
detection-weighted events toward the horizon of Fig.~\ref{fig:pdet}. The lower
\emph{joint}-likelihood rank correlations of the single-universe rows
($0.84$--$0.86$; $1$--$6$ seeds per point) decompose entirely into the Poisson
rate term acting on per-universe $\Nexp$ seed jitter: with the rate term removed,
the same points rank at $0.95$--$0.96$ with the shape best fit unchanged.

Thus both the one-off and pooled response paths pass an end-to-end shape test at
the $10^{-3}$-bit level. Delay treatment is subdominant to Monte-Carlo scatter
for natal spin, compactness, and the $m_1$--$\chieff$ trend, but changes $\Nexp$
and $R_0$ by tens of percent. Prompt-lite is a sensitivity bound, not a physical
model, because it forbids mergers below the formation-redshift floor.

\begin{table}
  \centering
  \caption{\label{tab:prompt_lite}\textbf{Prompt-lite placement ($z_m\to z_f$) versus the
  standard response.} \emph{Top:} per simulation grid and observed catalog, the median
  (maximum) relative shift $\delta x\equiv|\Delta x|/x$ (in percent) of the expected detected count
  $\Nexp$ and the local rate $R_0$, and the Spearman rank correlation across grid points of the joint log-likelihood
  [$r_s(\ln\mathcal{L})$] and of its shape term [$r_s({\rm sh})$; the joint
  $\ln\mathcal{L}$ minus the Poisson rate term]. ``$29$-$Z$'' is the production posterior
  grid ($18$ points), ``$4$-$Z$'' the natal-spin scan ($25$ points), ``single'' the
  follow-up single universes ($11$ points, $1$--$6$ seeds each). \emph{Bottom:} fiducial-ensemble ($s{=}0$,
  $r_h{=}0.25\,$pc) detected-shape comparison, with the binning conventions of
  Table~\ref{tab:delay} and per-$m_1$-bin extrema restricted to bins holding $>1\%$ of the
  detected weight. Produced by \texttt{analyses/prompt\_lite\_validation.py}
  (\texttt{data/prompt\_lite\_validation.json}).}
  \setlength{\tabcolsep}{3.5pt}
  \begin{tabular}{llcccc}
    \hline\hline
    grid & catalog & $\delta\Nexp$ & $\delta R_0$ &
    $r_s(\ln\mathcal{L})$ & $r_s({\rm sh})$ \\
    \hline
    $29$-$Z$ & O1--O4b & $10\,(18)$ & $40\,(53)$ & $0.979$ & $1.000$ \\
                      & O4      & $12\,(19)$ & $40\,(53)$ & $0.948$ & $0.996$ \\
    $4$-$Z$  & O1--O4b & $11\,(14)$ & $70\,(78)$ & $0.965$ & $0.998$ \\
                      & O4      & $12\,(16)$ & $70\,(78)$ & $0.982$ & $0.987$ \\
    single   & O1--O4b & $24\,(28)$ & $58\,(93)$ & $0.855$ & $0.964$ \\
                      & O4      & $29\,(33)$ & $58\,(93)$ & $0.836$ & $0.955$ \\
    \hline
    \multicolumn{2}{l}{fiducial detected shapes} & \multicolumn{2}{c}{O1--O4b} & \multicolumn{2}{c}{O4} \\
    \hline
    \multicolumn{2}{l}{JS$(dR/dm_1)$ [bits]} & \multicolumn{2}{c}{$0.0024$} & \multicolumn{2}{c}{$0.0011$} \\
    \multicolumn{2}{l}{JS$[p(\chieff)]$ [bits]} & \multicolumn{2}{c}{$0.0008$} & \multicolumn{2}{c}{$0.0009$} \\
    \multicolumn{2}{l}{$\max_j|\Delta\langle\chieff\rangle_j|$} & \multicolumn{2}{c}{$0.0021$} & \multicolumn{2}{c}{$0.0017$} \\
    \multicolumn{2}{l}{$\max_j|\Delta\sigma_{\chieff,j}|$} & \multicolumn{2}{c}{$0.0074$} & \multicolumn{2}{c}{$0.0075$} \\
    \hline\hline
  \end{tabular}
\end{table}

\section{Tuning the parametric field component}
\label{app:field}
We forward-model the parametric field population [Eq.~\eqref{eq:field}] through
the same selection function as the clusters, then tune
$(\mu_m,\sigma_m,\mu_\chi,\sigma_\chi)$ with $\mu_\chi\ge0$.

We hold the cluster model at
$(s,r_h,\fgc)=(\result{FiducialS},\result{FiducialRh}\,\mathrm{pc},0.004)$,
rebuild each trial field template by Monte Carlo, profile over $R_{\rm field}$,
and grid-scan the four shape parameters. This prevents a gradient-free optimizer
from using a spuriously broad field Gaussian to absorb $15$--$25\,\Msun$ events
already supplied by clusters. The likelihood is maximized for
\begin{equation}
  m_1\sim\mathcal{N}(\result{FieldTunedMone},\,\result{FieldTunedMoneSig})\,\Msun,\qquad
  \chieff\sim\mathcal{N}(+\result{FieldTunedChi},\,\result{FieldTunedChiSig}),
\end{equation}
a \emph{sharp} low-mass primary peak and a mildly spin-\emph{aligned} effective-spin
distribution---the alignment the dynamical channel structurally cannot produce.
The
tuned mixture improves on the cluster-only fit by
$\Delta\ln\mathcal{L}\simeq+\result{FieldTunedDlnL}$ on the tuning objective, which is not directly comparable to
the fixed-shape $+\result{FieldSplitDlnL}$ of Sec.~\ref{sec:results:split},
matches the detected count, and assigns the field a $\result{FieldDetFrac}$
detected fraction
concentrated below $20\,\Msun$ (the decomposition of Fig.~\ref{fig:split}, with
the tuned template). The widths are
\emph{intrinsic}---the per-event likelihood folds in each event's measurement
uncertainty---and are \emph{measurement-limited upper bounds}: the likelihood is flat
for $\sigma_m\lesssim1\,\Msun$ and $\sigma_\chi\lesssim0.05$, below the $\simeq2\,
\Msun$ and $\simeq0.08$ measurement errors, because the observed low-mass events
cluster at $m_1\simeq\result{FieldTunedMone}\,\Msun$ more tightly than their individual
uncertainties. The field
is therefore as sharp as the data can resolve; we adopt these floor values, and
propagating the (one-sided) shape uncertainty by sampling the four parameters jointly
with the formation parameters is a straightforward extension.

\section{Amplitude posteriors: $\fgc$ and the local rate budget}
\label{app:amplitudes}
\begin{figure*}
  \centering
  \includegraphics[width=\textwidth]{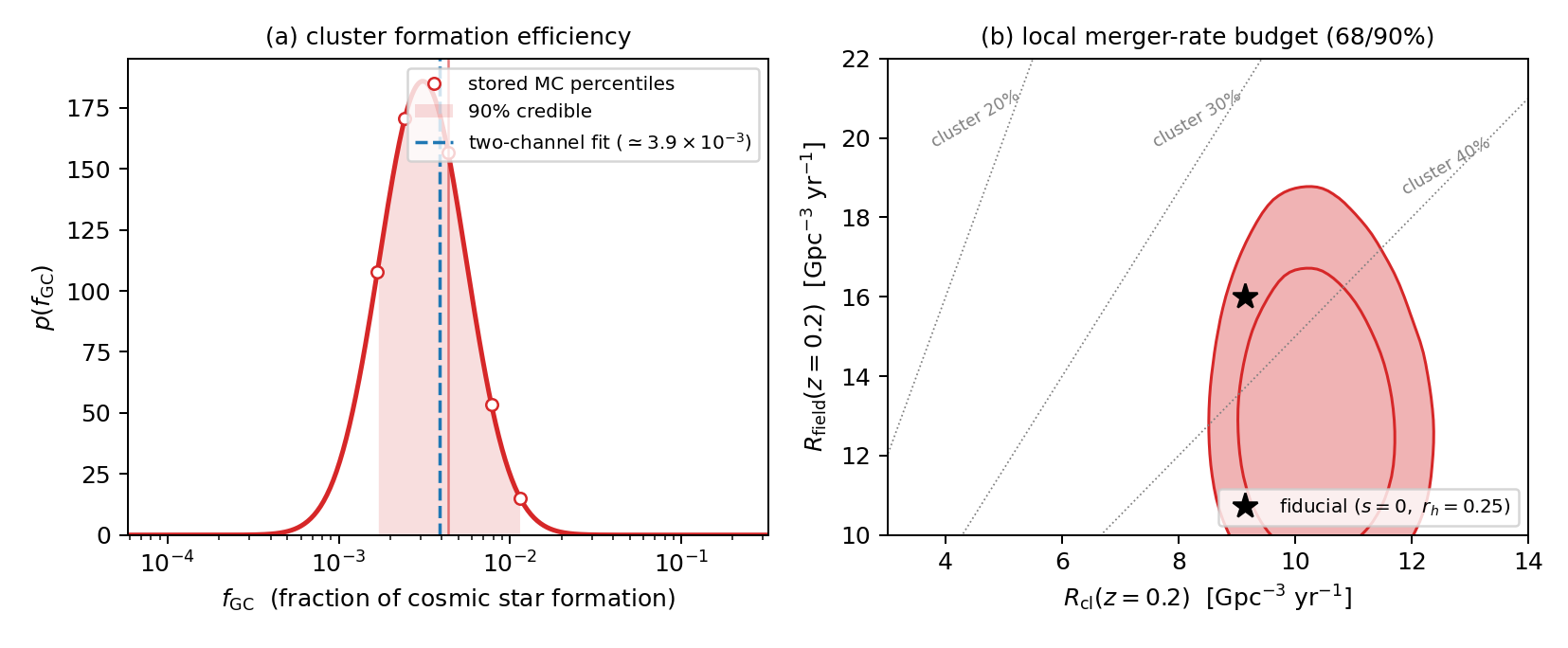}
  \caption{\textbf{Amplitude posteriors.} (a)~The cluster formation efficiency
  $\fgc$, marginalized over the $(s,r_h)$ posterior, Monte-Carlo seed scatter, and
  the dominant LVK high-mass-rate normalization [Eq.~\eqref{eq:ratefac}]; open
  circles are the stored posterior-predictive Monte-Carlo percentiles, and the dashed
  line marks the two-channel fit, in which the field carries the low-mass events.
  (b)~The joint local ($z{=}0.2$) rate budget $(R_{\rm cl},R_{\rm field})$ ($68\%$
  and $90\%$ contours), with lines of constant cluster fraction of the total rate:
  statistical (fitted-amplitude) and compactness uncertainty; normalization
  systematics shared by both channels are common-mode and not shown.}
  \label{fig:fgc_rate_corner}
\end{figure*}

Figure~\ref{fig:fgc_rate_corner}(a) shows $p(\fgc)$ marginalized over
$(s,r_h)$, seed scatter, and the LVK high-mass-rate normalization. Its stored
percentiles are log-symmetric within $2\%$ and are represented by a matched
lognormal ($\sigma_{\ln}\simeq0.52$); the dashed line marks the two-channel
$\hatfgc\simeq\result{FieldSplitFgc}$. Panel~(b) shows the joint $z=0.2$
posterior for $(R_{\rm cl},R_{\rm field})$, including fitted-amplitude and
compactness uncertainty. Common-mode selection-calibration uncertainty is
omitted. Every supported point is field dominated, with cluster fraction
$\simeq0.3$--$0.45$ ($90\%$).

\section{Measurement smearing for detected-space comparisons}
\label{app:smear}
Detected-space data curves use NAL likelihood centroids where available and
catalog medians otherwise. Their median mass errors grow from
$\result{SmearSigmaMoneLow}\,\Msun$ below $15\,\Msun$ to
$\result{SmearSigmaMoneHigh}\,\Msun$ above $45\,\Msun$, while median
$\sigma_{\chieff}=\result{SmearSigmaChiMed}$. Point-estimate histograms are
therefore measurement broadened, whereas raw model densities are intrinsic.
This affects display only; Eq.~\eqref{eq:nal_kernel} already incorporates the
uncertainties in the likelihood.

We therefore adopt one convention for every such comparison. Model detected
samples are \emph{measurement-smeared} before plotting: each sample of mass
$m_1$ is assigned the $(\sigma_{m_1},\sigma_{\chieff})$ \emph{pair} of one of
its $K{=}15$ nearest catalog events in $\log_{10}m_1$ (chosen uniformly among
them, preserving both the strong $\sigma_{m_1}(m_1)$ trend and the per-event
pairing of the two widths), and Gaussian noise with those widths is added
($\chieff$ clipped to its physical range). Observed curves remain point-estimate
constructions; in the $\chieff$--$m_1$ plane (Fig.~\ref{fig:chieff_m1}) we
instead draw each event as its $1\sigma$ NAL marginal ellipse, making the
per-event noise visible directly. The smear uses a fixed seed: it is a
deterministic display transform, applied identically to every component and to
the posterior-predictive band totals (Sec.~\ref{sec:methods:rate}), and never
enters a fit.

The convention is backtested at the fiducial two-channel fit by comparing
detected-population dispersions. In effective spin, the observed centroid
scatter is $\result{SmearChiStdObs}$; the intrinsic model width is
$\result{SmearChiStdIntr}$, and smearing brings it to
$\result{SmearChiStdSmear}$---the measurement noise supplies most of the
missing breadth.
In high-mass log-mass ($m_1>20\,\Msun$), the observed
scatter is $\result{SmearLgmStdObs}$~dex against an intrinsic
$\result{SmearLgmStdIntr}$~dex, smeared to $\result{SmearLgmStdSmear}$~dex.
Residual differences between the smeared model and the observed curves are
then genuine shape information, not noise artifacts.

\section{The intermediate-mass effective-spin distribution}
\label{app:chieff_iso}
\begin{figure}
  \centering
  \includegraphics[width=\columnwidth]{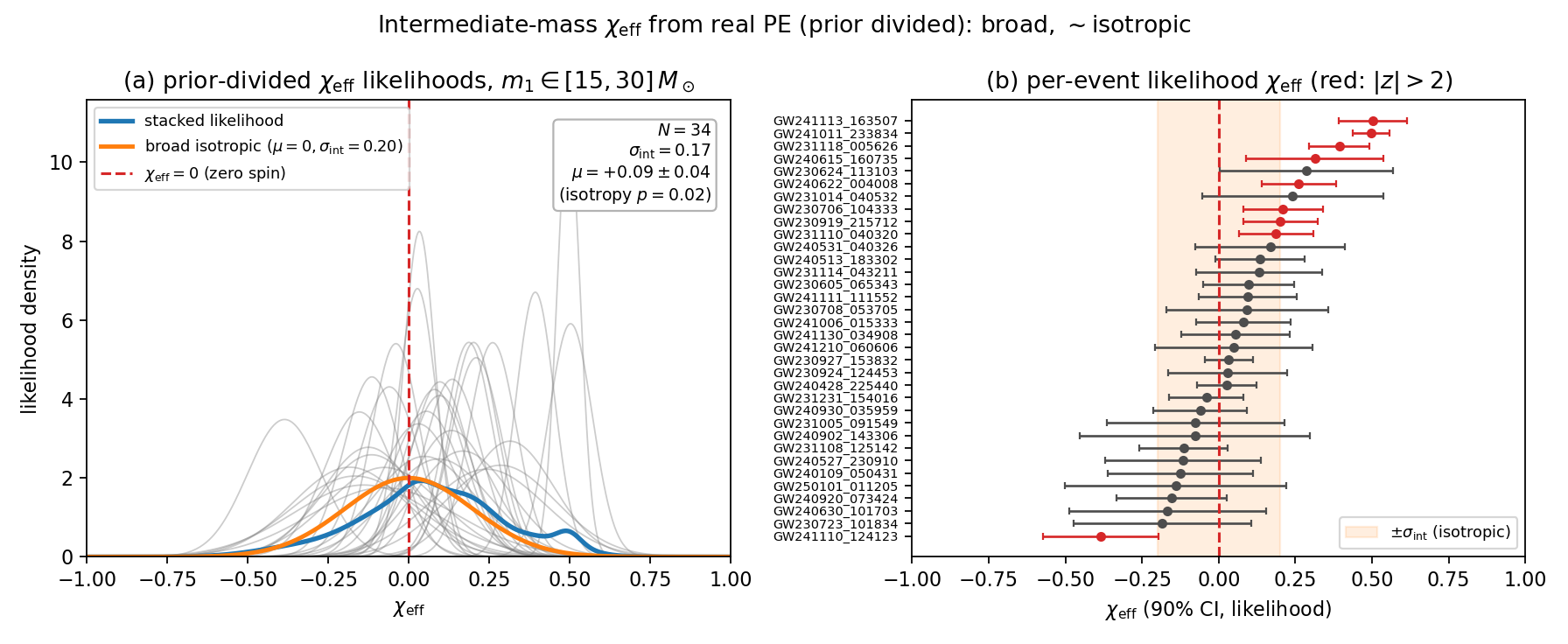}
  \caption{Effective spins of the intermediate-mass ($m_1\in[15,30]\,\Msun$) events,
  from \emph{proper prior-divided likelihoods} built from the real GWTC-4.0/5.0 PE
  samples. (a) Per-event $\chieff$ likelihoods (gray), their stack (blue), the best-fit
  broad-isotropic model ($\mu=0$, $\sigma_{\rm int}\simeq\result{ChiIsoSigmaIntIso}$; orange), and the
  zero-spin delta (red). The inset instead reports the free-mean fit
  ($\sigma_{\rm int}=\result{ChiIsoSigmaInt}$, $\mu=\result{ChiIsoMu}$), which is the pair
  quoted in the text. (b) Per-event likelihood $\chieff$ ($90\%$), sorted; $|z|>2$
  events (red) occur on \emph{both} sides of zero---$\result{ChiIsoNsigPos}$ positive, of
  which GW241011 at $\result{ChiIsoLoudChi}\pm\result{ChiIsoLoudSig}$ is the most extreme,
  and $\result{ChiIsoNsigNeg}$ negative, GW241110 at
  $\result{ChiIsoChiMin}\pm\result{ChiIsoChiMinSig}$.
  The distribution is broad and two-sided with a small positive mean
  ($\mu=\result{ChiIsoMu}\pm\result{ChiIsoMuErr}$), so it is neither a spike at zero
  (rejected at $p\sim\result{ChiIsoP}$) nor the strictly one-sided excess an aligned
  population would give; it is also not isotropic in the strict sense
  ($\mu=0$ disfavored at $p\simeq\result{ChiIsoPiso}$). Dividing out the zero-peaked PE
  prior \emph{broadens} the likelihood, so this is conservative relative to the
  summary-table Gaussian. This figure uses event-level prior-divided PE likelihoods
  (data/\texttt{o4\_nal.json} and the pooled GWTC-4.0/5.0 samples), not any LVK
  population product.}
  \label{fig:chieff_consistency}
\end{figure}

At $m_1\in[15,30]\,\Msun$, cluster mergers are mostly first generation and
would have $\chieff=0$ for zero natal spin. The fitted mixture also contains an
aligned field and a hierarchical tail, so an all-zero test is only a floor test,
not a test of the full model. We use prior-divided GWTC-4.0/5.0 event
likelihoods for the $\result{ChiIsoN}$ events in this window
(Fig.~\ref{fig:chieff_consistency}). Dividing out the zero-peaked PE prior
broadens the likelihood; nevertheless, it agrees with the summary-CI estimate:
$\sigma_{\rm int}=\result{ChiIsoSigmaInt}$ versus
$\result{ChiIsoSigmaIntSumm}$ for the same free-mean fit.

That a strictly zero-spin black-hole population is inconsistent with the GW catalog
is well established at the \emph{population} level: hierarchical analyses of the
full catalog find that most merging black holes carry small but nonzero spin, with
the fraction of a possible zero-spin subpopulation
debated but bounded well below
unity~\cite{gwastro-Galaudage-SpinSubpop2021,gwastro-Tong-SpinPop-2022,
gwastro-Callister-NoZeroSpin2022,MouldGerosaTaylor2022}, and $\chieff$--mass
correlations reported across the catalog~\cite{gwastro-Callister-ChiEffQ2021}.
Our test is deliberately narrower in scope: it isolates the $15$--$30\,\Msun$
window where the \emph{cluster's} non-hierarchical first generation dominates,
so it bears specifically on the natal spin of this channel's black holes---the
quantity that controls remnant retention (Sec.~\ref{sec:compactness})---rather
than on the catalog-wide mixture those analyses constrain.
The strong ``all identically zero'' hypothesis is decisively rejected
($T=\sum_i(\chieff^{(i)}/\sigma_i)^2=\result{ChiIsoT}$ over the $\result{ChiIsoN}$ events with released PE,
$p\sim\result{ChiIsoP}$), driven by well-measured events that scatter far from zero---most
strikingly GW241011, whose \emph{real} posterior gives
$\chieff=\result{ChiIsoLoudChi}\pm\result{ChiIsoLoudSig}$, a
genuine, tightly measured high-aligned-spin BBH and not a summary-table or
broad-posterior artifact. The large-$|\chieff|$ events appear with \emph{both signs}
(maximum $\result{ChiIsoChiMax}$, minimum $\result{ChiIsoChiMin}$). A broad-population fit $\chieff^{(i)}\sim
\mathcal{N}(\mu,\sigma_{\rm int}^2+\sigma_i^2)$ gives an intrinsic width
$\sigma_{\rm int}\simeq\result{ChiIsoSigmaInt}$ with a small positive mean
($\mu=\result{ChiIsoMu}\pm\result{ChiIsoMuErr}$, i.e.\ $\mu=0$ disfavored at
$p\simeq\result{ChiIsoPiso}$); constraining $\mu=0$ instead---the
strictly symmetric case---widens the fit to
$\sigma_{\rm int}=\result{ChiIsoSigmaIntIso}$.

Three separate statements can be read off this fit, and they are not equally strong; we
keep them apart because it is easy to run them together.

\emph{(i) Orientations must be misaligned---supported, but resting on one event.}
$\chieff$ is the mass-weighted projection of the component spins on the orbital angular
momentum, so appreciable support at $\chieff<0$ requires a spin--orbit tilt beyond
$90^\circ$ for the dominant component. No joint spin-magnitude/tilt inference is needed
for that inference, and no preferentially-aligned population produces it. In this window
the empirical basis is nonetheless narrow: of the $\result{ChiIsoN}$ events,
$\result{ChiIsoNsigNeg}$ is significantly negative---GW241110, at
$\chieff=\result{ChiIsoChiMin}\pm\result{ChiIsoChiMinSig}$---against
$\result{ChiIsoNsigPos}$ significantly positive. That supports ``not strictly one-sided'';
it does not by itself establish randomized orientations.

\emph{(ii) The distribution is not isotropic in the strict sense.} With
$\result{ChiIsoNpos}$ of $\result{ChiIsoN}$ central values positive and
$\mu=\result{ChiIsoMu}\pm\result{ChiIsoMuErr}$, exact symmetry about zero is itself
disfavored at $p\simeq\result{ChiIsoPiso}$. The accurate description is a \emph{broad,
two-sided, predominantly misaligned} $\chieff$ distribution with a small positive mean---
not a ``nearly isotropic'' one, and not the narrow one-sided excess a purely aligned
low-mass population would give.

\emph{(iii) Attributing the residual to a dynamical channel is model-conditional.} The
fitted mean is statistically indistinguishable from the aligned field template's own spin
centre ($\mu_\chi=+\result{FieldTunedChi}$; App.~\ref{app:field}), so this one-dimensional fit does
\emph{not} discriminate a dynamical population from an aligned field with extra width plus
a misaligned outlier. What favours the dynamical reading here is (i) together with the
structure of our model, in which the field is the only aligned channel; it is not
delivered by the $\chieff$ fit alone. We note too that a mild positive spin trend at
intermediate-to-high mass has been read by others as evidence for an \emph{aligned}
high-mass population rather than a dynamical one
(cf.~\citet{2026PhRvL.137b1407L}; \citet{2026arXiv260700565F}, specifically on the
mass-dependence of BBH spin properties), so our $\mu=\result{ChiIsoMu}$ is consistent with
that literature rather than in tension with it; what our window adds is the negative-side
event that an aligned population cannot make.

Reproducing this width from the cluster's \emph{own} 1G mergers, however, would require
a natal spin $a\gtrsim0.3$, and such spins enhance the GW recoil enough to eject most
merger remnants. The tradeoff is one-sided: the widest Beta-distributed natal spin we
scan reaches only $\sigma\simeq\result{SpinTradeoffSigmaMax}$ in the 1G $\chieff$---still
short of the observed spread---while already dropping the hierarchical fraction and the
$m_1>45\,\Msun$ rate by $\simeq\result{SpinTradeoffSuppression}\times$ (from
$\simeq\result{SpinTradeoffFhighZero}$ to $\simeq\result{SpinTradeoffFhighSpin}$ of
mergers), gutting the high-mass tail the cluster channel exists to explain. \emph{No}
single natal spin reproduces both the high-mass population (which prefers $s\simeq0$ for
remnant retention) and the intermediate-mass spin width. We therefore read the genuine
high-$|\chieff|$ intermediate-mass events as a \emph{separate, truly-spinning
sub-population}. Their magnitudes are large and their signs are two-sided, which requires
at least some \emph{misaligned} tilts rather than a purely aligned population: the natural
candidates are
dynamically re-paired merger remnants (the reprocessed component of
Sec.~\ref{sec:results:fieldfed}) or a distinct formation
path~\cite{dcc-Tong-Hierarchical-2025}, whereas an aligned field channel reaching
up to $\simeq20\,\Msun$ would have to produce the negative-$\chieff$ event some other
way. We do not claim the tilts are \emph{randomized}: as set out above, that would need
more than the single significantly-negative event this window contains, and the fitted
mean is small but positive. Either way the sub-population sits atop a low-spin dynamical bulk that remains
$\chieff\!\approx\!0$ (the waist of Fig.~\ref{fig:chieff_m1}) and continues to build the
high-mass, isotropic population through hierarchical assembly. The zero-spin
ensemble is thus retained as the cluster model: it gives the best high-mass fit, and the
intermediate-mass spin scatter is not a cluster natal-spin signal but the imprint of a
second population carrying large spins.

A dynamical origin for the scatter itself was tested with the cluster-compactness
\emph{mixture} model (Sec.~\ref{sec:methods:fit}, model~c; a free distribution
$g(r_h)$ summing to unity), since denser clusters make more hierarchical
(2G$+$1G) mergers---which at zero natal spin carry a broad, two-sided $\chieff$ from the
$\simeq0.7$ remnant spin. Fit over $\result{RhMixNbins}$ available bins from
$r_h=\result{RhMixDenseRh}$ to $\result{RhMixDiffuseRh}\,$pc, the
mixture prefers a \emph{bimodal} compactness distribution---$\simeq\result{RhMixDiffuseFrac}$ ordinary,
diffuse clusters at $r_h=\result{RhMixDiffuseRh}\,$pc (supplying the low-mass,
$\chieff\!\approx\!0$ bulk) plus $\simeq\result{RhMixDenseFrac}$
dense cores at $r_h=\result{RhMixDenseRh}\,$pc (supplying the
high-mass tail and the hierarchical
$\chieff$ spread). This is worth emphasising in its own right, because it is the
\emph{ensemble} answer to the compactness question of Sec.~\ref{sec:compactness}: the
population the data prefer is dominated by ordinary, diffuse clusters, with a dense
minority doing the hierarchical work---not a uniformly very compact population. It also
helps with the spin width, raising the predicted intermediate-mass value above its
single-$r_h$ level to $\simeq\result{RhMixSigInt}$ and improving the fit---but it
\emph{plateaus} well below the observed $\result{ChiIsoSigmaInt}$: the intermediate-mass hierarchical
fraction peaks at a moderate compactness ($r_h\sim0.1\,$pc) and \emph{falls} at higher
density, where runaway growth carries the 2G$+$ products straight past $30\,\Msun$ (we
verified this is physical, not a simulation-time artifact). The dynamical channel
therefore supplies part of the intermediate-mass spin scatter but not all of it,
reinforcing the need for the separate high-spin sub-population above.

\section{Delete-truncation pair-instability variant}
\label{app:pisn_truncate}
\begin{figure}
  \centering
  \includegraphics[width=\columnwidth]{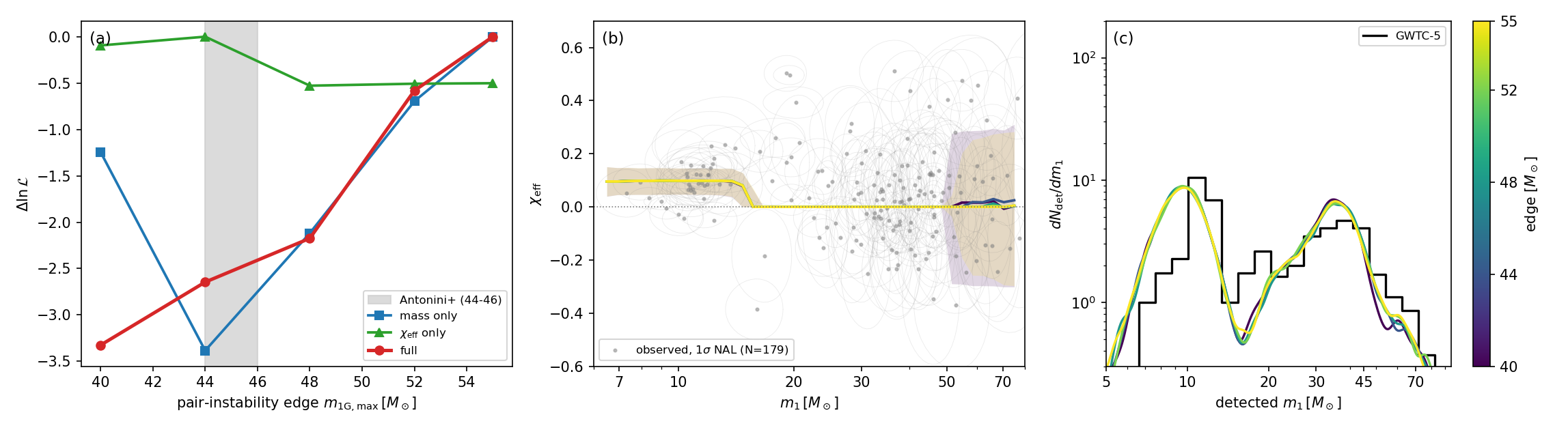}
  \caption{As Fig.~\ref{fig:pisn_edge}, but for the \emph{delete-truncation} gap (\Rapster\
  native): would-be-gap remnants are removed rather than piled up. The likelihood
  (\emph{left}) rails monotonically toward the $\simeq56\,\Msun$ ceiling, and the best-fit
  detected models (\emph{right}) are nearly edge-independent---deletion erases the edge's
  mass signature, weakening the constraint relative to the pile-up fiducial.}
  \label{fig:pisn_edge_truncate}
\end{figure}

The fiducial scan piles pulsational pair-instability remnants at the edge;
native \Rapster{} instead deletes them. Repeating the scan with deletion gives
the same qualitative result (Fig.~\ref{fig:pisn_edge_truncate}): $\Delta\ln\mathcal{L}\simeq3.5$ toward the
$\simeq56\,\Msun$ ceiling. The constraint is weaker and purely monotonic because
deletion makes the detected high-mass and hierarchical fractions nearly
edge-independent. We therefore retain the physically motivated pile-up, whose
mass signature lets the turnover track the maximum first-generation mass.

\bibliography{LIGO-publications,gw-astronomy-mergers,gw-astronomy-mergers-approximations,gw-astronomy-mergers-ns,gw-astronomy-mergers-ns-gw170817,gw-astronomy-mergers-emcounterparts,gw-astronomy-mergers-ns-nuclearphysics,gw-astronomy-mergers-wd,popsyn,popsyn_gw-merger-rates,popsyn-constraints,gw-astronomy-mergers-dynamical-agndisk,gw-astronomy,gw-astronomy-mergers-nr,gw-astronomy-mergers-eccentric,mm-statistics,gw-systematics,claude_refs,refs}

\end{document}